\documentclass[a4paper,twocolumn,prx]{revtex4-2}
\usepackage[english]{babel}
\usepackage[utf8]{inputenc}
\usepackage{amsmath}
\usepackage{amsfonts}

\usepackage{bm}
\usepackage{siunitx}
\usepackage{xspace}

\newcommand{\braket}[2]{\langle #1 \vert #2 \rangle}
\newcommand{\ket}[1]{\vert #1 \rangle}
\newcommand{\mel}[3]{\langle #1 \vert #2 \vert #3 \rangle}
\newcommand{\expval}[1]{\langle #1 \rangle}

\newcommand{\mat}[1]{\begin{pmatrix} #1 \end{pmatrix}}
\renewcommand{\vec}[1]{\ensuremath{\boldsymbol{#1}}\xspace}

\usepackage[colorlinks=true,linkcolor=black,citecolor=black]{hyperref}
\usepackage[capitalise]{cleveref}

\usepackage{graphicx}
\graphicspath{ {figures/} }

\usepackage{natbib}

\newcommand{\affA}{Department of Physics and Astronomy, Aarhus University, DK-8000 Aarhus C, Denmark}
\newcommand{\affB}{Aarhus Institute of Advanced Studies, Aarhus University, DK-8000 Aarhus C, Denmark}

\date{\today}
\begin{document}
	\title{Lattice gauge theory and dynamical quantum phase transitions using noisy intermediate scale quantum devices} 

	\author{Simon Panyella Pedersen}
	\email{spp@phys.au.dk}
	\affiliation{\affA}
	\author{Nikolaj Thomas Zinner}
	\email{zinner@phys.au.dk}
	\affiliation{\affA}
	\affiliation{\affB}

\begin{abstract}
	Lattice gauge theories are a fascinating and rich class of theories relating to the most fundamental models of particle physics, and as experimental control on the quantum level increases there is a growing interest in non-equilibrium effects such as dynamical quantum phase transitions. To demonstrate how these physical theories can be accessed in near-term quantum devices, we study the dynamics of a (1+1)D U(1) quantum link model following quenches of its mass-term. We find that the system undergoes dynamical quantum phase transitions for all system sizes considered, even the smallest where the dynamics can be solved analytically. We devise a gauge invariant string order parameter whose zeros correlates with the structure of the Loschmidt amplitude, making the order parameter useful for experimental study in near-term devices. The zeros of the Loschmidt amplitude as well as the zeros of our order parameter are revealed by vortices in their phases, which can be counted by a topologically invariant winding number.	With noisy intermediate scale quantum devices in mind, we propose a class of superconducting circuits for the general implementation of U(1) quantum link models. The principles of these circuits can be generalized to implement other, more complicated gauge symmetries. Furthermore, the circuit can be modularly scaled to any lattice configuration. Simulating the circuit dynamics with realistic circuit parameters we find that it implements the target dynamics with a steady average fidelity of $ 99.5\% $ or higher. Finally, we consider readout of the circuit using a method that yields information about all the degrees of freedom with resonators coupled dispersively to only a subset of them. This constitutes a direct and relatively straightforward protocol to access both Loschmidt amplitudes and the order parameter. 
\end{abstract}

\maketitle

\section{Introduction}\label{sec:intro}
There is an increasing study of non-equilibrium quantum dynamics as improving experimental quantum control makes it accessible \cite{Georgescu2014}. Quantum simulators have been realized with cold atoms in optical lattices, ions, and superconducting quantum circuits (SQCs) among others, and have already been used to study exciting dynamical phenomena like time crystals \cite{Zhang2017,Choi2017}, many-body localization \cite{Schreiber2015,Smith2016}, prethermalization and thermalization \cite{Gring2012,Neyenhuis2017,Neill2016}, and particle-antiparticle creation and annihilation \cite{Martinez2016}. In particular an interest in dynamical quantum phase transitions (DQPTs) is emerging \cite{Heyl2013,Heyl2015,Canovi2014,Pekker2014,Vosk2014,Schmitt2015b,Zvyagin2017,Huang2018,Lacki2019,Goes2020,Zvyagin2016,Heyl2018,Heyl2019}. These occur when the Loschmidt amplitude $ \mathcal{G}(t) = \braket{\psi(0)}{\psi(t)} $, which is the overlap between the initial state and the state at time $ t $, becomes zero or shows non-analytic behaviour. In the context of DQPTs, $ \mathcal{G}(t) $ formally takes the place of a partition function. These points of vanishing or non-analytic behaviour happen in the proper time evolution of the system, and heralds a transition between dynamical phases. This is to be compared with equilibrium phase transitions which are heralded by non-analytical behaviour occurring as system properties are externally changed \cite{Sachdev2011}. DQPTs have been studied experimentally \cite{Jurcevic2017,Guo2019,Xu2019,Xu2020}, and offer a broad spectrum of fascinating physics, like a connection to topology \cite{Hu2019,Vajna2015,Hagymasi2019,Zache2019}, allowing for the definition of dynamical topological order parameters \cite{Budich2016,Xu2020,Ding2020}, vortex dynamics \cite{Flaschner2018}, scaling and universality \cite{Heyl2015}, and a showing both a connection to underlying equilibrium phase transitions \cite{Heyl2013,Vajna2015,Huang2016}, as well as being completely independent of them \cite{Fagotti2013,Vajna2014,Andraschko2014,Porta2020}, the latter showing their truly non-equilibrium nature. An interesting type of system for the study of dynamics is gauge theories, specifically lattice gauge theories (LGTs) \cite{Wilson1974,Kogut1975,Smit2002}. Gauge theories are at the basis of our understanding of particle physics, and are notoriously difficult to handle both analytically and numerically. They are thus ideally suited for analogue simulation \cite{Banerjee2012,Kuhn2014,Kuno2015,Kasper2016,Dehkharghani2017,Kuno2017,Mil2019,Banuls2020}. Recently, a method for extracting defining information about a quantum field theory from experimental data was proposed \cite{SanchezPalencia2020,Zache2020}. Furthermore, a direct observation of U(1) gauge invariance and an equilibrium phase transition in a 71-site ultracold atom system was reported in Ref. \cite{Yang2020}. DQPTs in gauge theories, on the other hand, have been studied numerically \cite{Huang2018,Zache2019}, as well as analytically in the non-interacting limit in Ref. \cite{Zache2019}, but have yet to be observed experimentally to the best of our knowledge.

In this work we show how to obtain LGTs, in the form of quantum link models (QLM) \cite{Horn1981,Orland1990,Chandrasekharan1997,Hauke2013,Banerjee2012,Huang2018}, in SQCs in a fully consistent way, showing that we get the desired Hamiltonians with very high fidelity. We use the example of U(1) to demonstrate this, and we show that exploring the new field of DQPTs is possible with NISQ-era devices \cite{Preskill2018}. We study a particular U(1) symmetric system, the massive Schwinger model \cite{Schwinger1962,Coleman1975,Martinez2016,Ercolessi2018,Zache2019}, exhibiting DQPTs after a quench for all system sizes considered. A recent study by Zache et al. has considered DQPTs in continuum and LGT models. In Ref. \cite{Zache2019}, DQPTs were found through the study of vorticity in an appropriate order parameter, implying that the transitions have a topological nature. Here, we consider a QLM over a larger range of parameters and write down a gauge invariant string order parameter that is accessible even in small systems of relevance in near-term quantum devices. We show how the zeros of both this order parameter as well as those of the Loschmidt amplitude can be found by looking for vortices in their respective phases. The zeros of the order parameter correlate for all system sizes with the low points of the Loschmidt amplitude

In the second part of this paper we propose a superconducting quantum circuit, which through use of the eigenmodes of the capacitive network \cite{Bergeal2010,Kounalakis2018,Roy2017,Roy2018,Devoret2017} the circuit implements three spin-$ 1/2 $'s interacting via Z-type couplings (i.e. couplings consisting solely of products of Pauli-Z matrices) and a direct three-body XXX-coupling. Through appropriate tuning the XXX-coupling yields the desired U(1) invariant interaction necessary for the analogue simulation of a U(1) QLM. The Z-type couplings essentially just shift the energy levels of the system, and do not disturb the desired feature of the circuit, but merely make more complicated numerical tuning necessary. We find that with appropriately tuned parameters the circuit implements the desired dynamics with an average fidelity of about $ 99.5\% $ or higher, with most of the loss caused by leakage to higher levels, which could be further suppressed at the cost of slower dynamics. The circuit can be scaled in a modular way to construct any desired spin lattice configuration. Furthermore, we provide a readout scheme for how to observe this in concrete setups, inspired by that of Refs. \cite{Roy2017,Roy2018}. This makes it possible now to use NISQ devices to do precision studies of LGTs and DQPTs.

In \cref{sec:system} we go through the QLM, the quench of the mass, and our order parameter. In \cref{sec:counting} we go through our method for finding the zeros of the Loschmidt amplitude and our order parameter by looking at vortices in their phases. \cref{sec:sim} contains our numerical results concerning these quantities, showing how even at the smallest system size considered we see DQPTs and vortex dynamics. In \cref{sec:circuit} we introduce and analyse the circuit used to implement the direct XXX-coupling. We then discuss tuning of the circuit parameters in \cref{sec:opt}, showing an example in the Supplemental Material \cite{suppmat} of viable circuit parameters which yield good spin model parameters for the simulation of the quench dynamics discussed in the first part of the paper. In \cref{sec:readout} we then go through our proposal for readout of the circuit with intent to perform quantum state tomography. Finally, in \cref{sec:conc} we summarize and conclude on the paper.

\section{System and procedure}\label{sec:system}
\subsection{Hamiltonian}
In this work we study an example of an interacting LGT, specifically the (1+1)D U(1) gauge theory, on a periodic lattice. We represent the fermionic field with spinless, staggered mass fermions on the sites of the lattice, and transform these via the Jordan-Wigner transformation \cite{Jordan1928,Susskind1977,Marcos2013,Hauke2013} into spin-$ 1/2 $'s. We will be working in the quantum link model framework, where gauge fields, living on the links of the lattice, are represented by spin-$ 1/2 $'s. Thus the entire model is represented by a spin-$ 1/2 $ system. The Hamiltonian for this system is
\begin{align}
H &= \sum_{n=0}^{N-1}\left[-(-1)^{n}\frac{m}{2}\sigma_{n}^{z} + \frac{J}{2}\left(\sigma_{n}^{+}S_{n,n+1}^{+}\sigma_{n+1}^{-} + \textup{H.c.}\right)\right] \label{eq:gaugeH}
\end{align}
where $ N $ is the number of matter sites, which must be even to conserve the symmetry between particles and antiparticles, $ m $ is the staggered mass of the fermions, and $ J $ is the matter-gauge coupling strength. The $ J/2 $ coefficient is usually written as $ 1/2a $, where $ a $ is the lattice spacing. However, since we are not interested in the continuum limit $ a \rightarrow 0 $, we prefer to think of the coefficient as a coupling strength, $ J = 1/a $. $ \sigma_{n}^{\alpha} $ with $ \alpha = z,+,- $ are Pauli-Z, step-up and -down matrices pertaining to the matter field spin at site $ n $. Likewise $ S_{n,n+1}^{\alpha} $ for $ \alpha = z,+,- $ are spin operators for the gauge field spin on the link connecting site $ n $ and site $ n+1 $. The electric field energy term $ E_{n,n+1}^{2} \sim (S_{n,n+1}^{z})^{2} $ has been neglected because it is constant, when the gauge field is represented by spin-$ 1/2 $'s. The sign in front of $ (-1)^{n}\frac{m}{2}\sigma_{n}^{z} $ is to make $ (1,0)^{T} $ and $ (0,1)^{T} $ the ground and excited state respectively for even $ n $. Our $ \sigma^{\pm} $ are correspondingly defined. We will consider quenches of the sign of the mass $ m \rightarrow -m $, i.e. we will initialize in the ground state of the above Hamiltonian, and then perform time-evolution with the Hamiltonian where the sign of the mass has been switched. This corresponds to a maximal quench of the vacuum angle. The vacuum angle is a parameter that may be included in quantum chromodynamics as well as the Schwinger model, relating to the non-trivial structure of their vacua \cite{Callan1976,Jackiw1976}, and quantifying a CP-violating term. For more information see \cite{Coleman1975,Coleman1976,Callan1976,Jackiw1976,Schafer1998,Gabadadze2002,Peccei2008,Peccei2010,Kim2010}. We will, however, not be considering the vacuum angle itself, but will rather focus on the quench, and the subsequent DQPTs. Explicitly we will be initializing the system in the ground state of the pre-quench Hamiltonian $ H_{i} = H(m,J) $ at time $ t = 0 $, and then perform unitary time evolution according to the post-quench Hamiltonian $ H_{f} = H(-m,J) $. We study the post-quench dynamics by looking at the Loschmidt amplitude $ \mathcal{G}(t) $ or Loschmidt echo $ \mathcal{L}(t) = \vert\mathcal{G}(t)\vert^2 $, as well as an order parameter introduced below.

\begin{figure*}
	\includegraphics[width=0.8\textwidth]{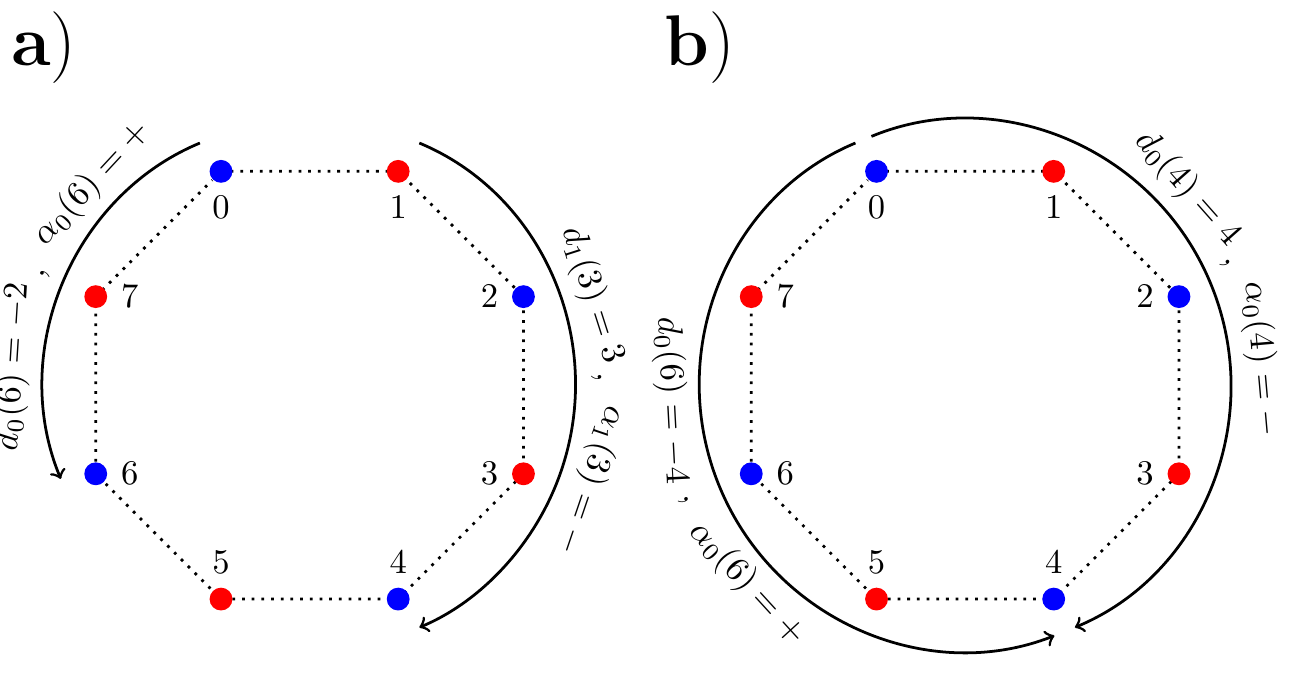}
	\caption{\label{fig:circOP} Some examples of the paths taken by the string operators which are summed over in the order parameter for a system size of $ N = 8 $ matter sites. a) Two examples, on originating from particle site $ 0 $ and the other from the antiparticle site $ 1 $, showing the sign conventions for clockwise and counter-clockwise paths. b) An example of the case where $ m-n = N/2 $, where the equidistant clockwise and counter-clockwise paths are both taken into account.}
\end{figure*}

\subsection{Symmetries}
The system has several symmetries which are conserved across the quench. We will therefore be simplifying numerical simulation by only working in the symmetry sector of the Hilbert space, to which the initial state belongs. The system of course has a local U(1) symmetry, generated by
\begin{align*}
G_{n} = S_{n-1,n}^{z} - S_{n,n+1}^{z} + \sigma_{n}^{z} - (-1)^{n}
\end{align*}
We will be working in the gauge sector of no background charges, i.e. the physical states satisfy $ G_{n}\ket{\textup{phys}} = 0 $ for all $ n $. Furthermore, the system has parity and charge conjugation symmetries, $ P $ and $ C $, implemented as
\begin{align*}
\sigma_{n}^{z} &\xrightarrow{P} \sigma_{-n}^{z} &\quad \sigma_{n}^{\pm} &\xrightarrow{P} \sigma_{-n}^{\pm}\\
S_{n,n+1}^{z} &\xrightarrow{P} -S_{-n-1,-n}^{z} &\quad S_{n,n+1}^{\pm} &\xrightarrow{P} S_{-n-1,-n}^{\mp}\\[1em]
\sigma_{n}^{z} &\xrightarrow{C} -\sigma_{n+1}^{z} &\quad \sigma_{n}^{\pm} &\xrightarrow{C} \sigma_{n+1}^{\mp}\\
S_{n,n+1}^{z} &\xrightarrow{C} -S_{n+1,n+2}^{z} &\quad S_{n,n+1}^{\pm} &\xrightarrow{C} S_{n+1,n+2}^{\mp}
\end{align*}
where the subscript index is calculated modulus $ N $. These symmetries satisfy $ P^{2} = C^{N} = 1 $. The fact that $ C $ is cyclic is a consequence of the periodic boundaries of our system. The symmetries make it possible to solve the dynamics analytically for small system sizes, $ N = 2,4 $. That is, the Hilbert space can be divided into symmetry sectors with different eigenvalues for each of the symmetries, with so few states in each of them that they can each be analytically diagonalized. The diagonalization boils down to solving a characteristic polynomial equation as is standard. As is well-known by Galois theory, this implies that for polynomial degree of $ 4 $ or smaller, the solution can be given in rational form.
For $ N = 6 $ the sector containing the ground state we initialize in already has $ 5 $ states. For the analytical solution of the $ N = 2,4 $ systems, see Supplemental Material \cite{suppmat}. For $ N = 2 $ it is even possible to analytically determine the zeros of $ \mathcal{G}(t) $, and thus the times of the DQPTs.

Because of these symmetries, the periodicity of the system, and the state we initialize in it can furthermore be shown that (see Supplemental Material \cite{suppmat})
\begin{align}
	\begin{split}
	\expval{\sigma_{n}^{z}} &= (-1)^{n}\expval{\sigma_{0}^{z}} \\
	\expval{S_{n,n+1}^{z}} &= (-1)^{n}\frac{\expval{\sigma_{0}^{z}} - 1}{2} \label{eq:regression}
	\end{split}
\end{align}
This is essentially a consequence of the fact that we have periodic boundaries, and so the system is symmetric under stepwise rotations, i.e. all particle-sites must have the same dynamics, and likewise for antiparticle-sites. Coupled together with the fact that are working in the sector of equally many particles and antiparticles (zero total charge), we get the above relation between the $ \expval{\sigma_{n}^{z}} $. The relation for the gauge link spins is a consequence of fact that the dynamics of the gauge fields are completely determined by the matter fields in a spin-$ 1/2 $ QLM. Furthermore, $ \expval{\sigma_{n}^{\beta}} = \expval{S_{n,n+1}^{\beta}} = 0 $ for $ \beta = x,y $ and all $ n $, which follows simply from the fact that $ \sigma_{n}^{\beta},S_{n,n+1}^{\beta} $ are gauge \emph{variant}, and so their expectation value within a certain gauge sector is zero. This means that a single measurement of $ \sigma^{z} $ for any spin in the system, reveals the equivalent quantity for all spins in the system, and only $ \sigma^{z} $ yields a non-zero measurement. This will be a decisive observation for the readout scheme in an experimental realization.

\subsection{Order parameter}
The order parameter we use to study the post-quench dynamics is $ g(k,t) = \mel{\psi(0)}{\mathfrak{g}(k)}{\psi(t)} $, where
\begin{align*}
\mathfrak{g}(k) = \sum_{m=0,1}\sum_{n=0}^{N-1}e^{-ikd_{m}(n)}\sigma_{m}^{-}\prod_{i=m}^{n-1}S_{i,i+1}^{\alpha_{m}(n)}\sigma_{n}^{+}
\end{align*}
This is a sum over two representative sites $ m = 0,1 $ (a particle site and an antiparticle site respectively), and over all sites of the lattice $ n = 0,...,N-1 $. The summand is a Fourier coefficient times a string operator, consisting of two matter site operators, one at the representative site $ m $ and the other at $ n $, connected by the gauge link operators between the two sites, making the total operator gauge invariant. Here the products of link operators are over the \emph{shortest} path between site $ m $ and site $ n $, i.e. either counter-clockwise or clockwise along the circular lattice, see \cref{fig:circOP}. Likewise $ d_{m}(n) $ is the distance from site $ m $ to $ n $ along the shortest path, with $ d_{m}(n) $ being positive for clockwise paths and negative for counter-clockwise paths. Similarly, $ \alpha_{m}(n) = - $ for clockwise paths and $ \alpha_{m}(n) = + $ for counter-clockwise paths, ensuring the gauge invariance of the summands. For sites on the exact opposite side of the circular lattice, i.e. $ m-n = N/2 $, the two paths around the lattice are equidistant and so both are included, see \cref{fig:circOP}b. Thus the operator $ \mathfrak{g}(k) $ is essentially the Fourier transform of the gauge invariant string operators connecting the sites $ 0 $ and $ 1 $ with all sites of the lattice. The order parameter $ g(k,t) $ is then the Fourier transform of the amplitudes of a matter excitation moving from either site $ 0 $ or $ 1 $ to site $ n $, via the shortest path, in the time between initialization and $ t $. The reason we have both a term for site $ 0 $ and one for site $ 1 $ is to make the operator symmetric with respect to particles and antiparticles. The Fourier parameter $ k $ takes its value in the Brillouin zone $ [-\pi/a,\pi/a] $, though the periodic boundaries of the system makes the order parameter symmetric around $ k = 0 $. We therefore only consider $ k \in [0,\pi/a] $. We define the phase $ \phi_{g} $ of the order parameter via $ g = \vert g\vert e^{i\phi_{g}} $. As we will see this novel order parameter has zeros along the troughs of the Loschmidt echo, $ \mathcal{L}(t) $, for all system sizes. These zeros are accompanied by vortices in $ \phi_{g} $, which can be counted by a winding number. The vortices show an interesting dynamics of creation and annihilation as the coupling constant is varied. Hence, even in the smallest interesting system this order parameter reveals the structure of $ \mathcal{L}(t) $, which repeats itself as the system size is increased, and exhibits non-trivial vortex dynamics. The order parameter that we employ here is related to the gauge-invariant time-ordered Green's function computed in the context of lattice gauge theory in \cite{Zache2019}, which was shown to reduce to the Loschmidt amplitude in the non-interacting case, and be correlated with the Loschmidt amplitude in the interacting case. Our work demonstrates how to transfer this to QLM models.

\begin{figure*}
	\includegraphics[width=\textwidth]{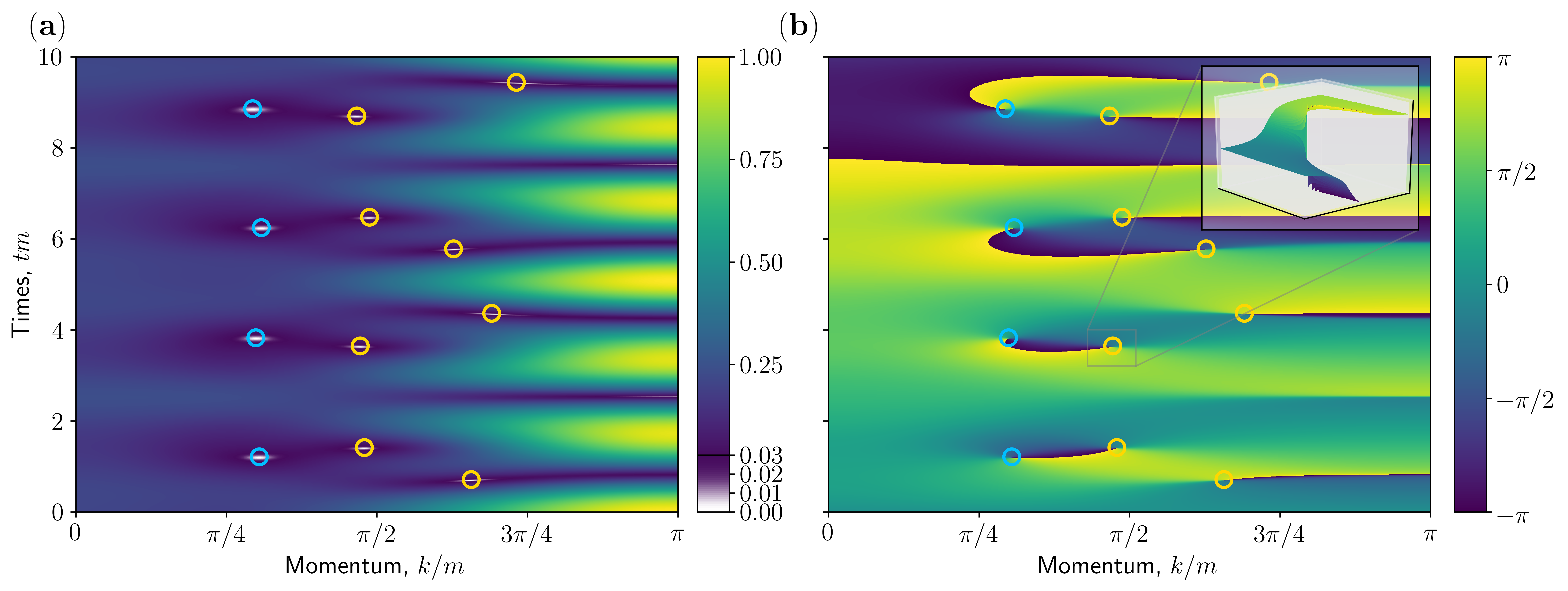}
	\caption{\label{fig:OP} (a) The modulus $ \vert g\vert $ and (b) the phase $ \phi_{g} $ of the order parameter $ g $ for $ N = 4 $ and $ J/m = 1.95 $. The vortices of the phase, corresponding to the zeros of $ g $, as found by the method described in the text, have been marked with circles coloured blue for right-winding vortices, and yellow for left-winding. In the inset a 3D zoom-in of a vortex can be seen, rotated to put the tear clearly into view.}
\end{figure*}
\begin{figure}
	\includegraphics[width=\columnwidth]{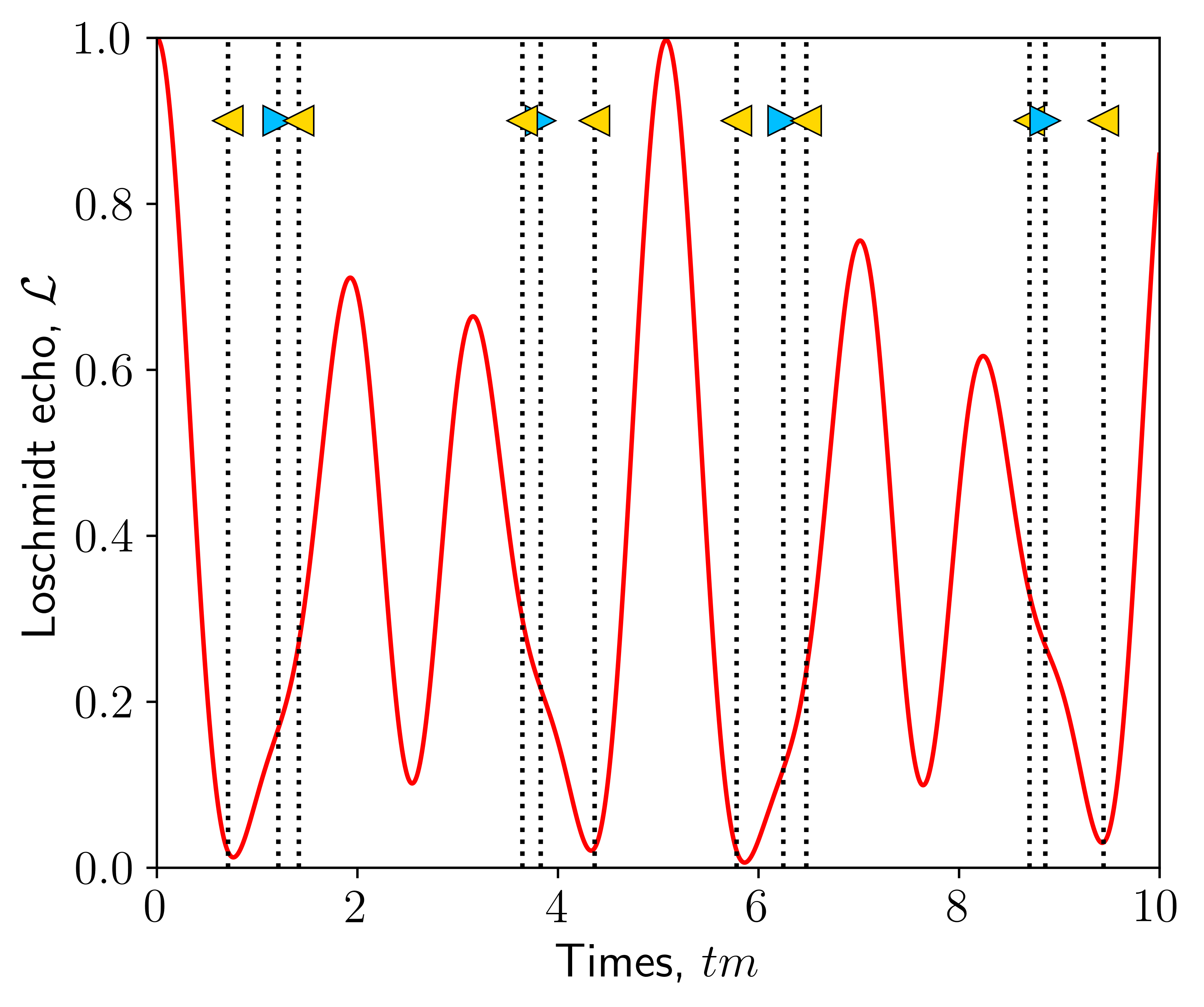}
	\caption{\label{fig:L} The Loschmidt echo $ \mathcal{L}(t) $ in a solid red line and its rate function $ \lambda(t) $ in a dashed blue, following the mass quench, $ m \rightarrow -m $ with the same system parameters as in \cref{fig:OP}. The vertical dotted lines indicate the times at which vortices occur in $ g $, and the triangular arrow heads indicate the orientation of the vortices, with arrow heads pointing right indicating right-winding vortices, and arrow heads pointing left for left-winding vortices. There is a clear correlation between the deepest minima of $ \mathcal{L}(t) $ (the sharpest peaks of $ \lambda(t) $) and the appearance of vortices, except when the vortices come in pairs of opposite orientation. For slightly larger $ J/m $, vortices appear at times corresponding to the two minima of $ \mathcal{L}(t) $ (two peaks of $ \lambda(t) $) at $ t\sim 2.5,7.5 $. }
\end{figure}

\section{Counting vortices}\label{sec:counting}
We study the dynamics of the Loschmidt amplitude and the order parameter introduced above over a range of the coupling strength $ J $ (keeping $ m $ fixed) and a range of the system size $ N $. We are in particular interested in the zeros of these quantities. Considering $ \mathcal{G}(t) $ as a complex function in the $ (J,t) $-plane and $ g $ a complex function in the $ (k,t) $-plane (with $ J $ fixed), we can find their zeros in a similar fashion, namely by looking at their phase. This method for numerically finding the vortices is an adaptation of the work in Ref. \cite{Fukui2005}, developed for computing Chern numbers in momentum space. When a complex function, say $ f = re^{i\varphi} $, becomes zero at some critical point, its phase $ \varphi $ is undefined at that point. For a complex function of two variables, $ f = f(x,y) $, this results in a vortex in $ \varphi $ surrounding the critical point, see the inset of \cref{fig:OP}b showing a vortex in the phase of $ g(k,t) $. The intuition of this is that while $ \varphi $ is undefined at the critical point it is otherwise smooth, up to discontinuities of $ 2\pi $. If there were no discontinuity around the critical point there would obviously be a meaningful, smooth extension of $ \varphi $ at the undefined point, which contradicts the fact that $ \varphi $ is undefinable. The phase must thus have a line of $ 2\pi $ discontinuity extending from the critical point. Starting a this discontinuity and going around the critical point, $ \varphi $ must then attain all possible values between $ -\pi $ and $ \pi $ in a smooth way, as there would otherwise again be a meaningful, smooth extension of $ \varphi $. Such vortices can be counted by a winding number 
\begin{align*}
\nu = \frac{1}{2\pi}\oint_{\mathcal{C}}d\vec{l}\cdot\nabla\varphi
\end{align*}
where $ \mathcal{C} $ is a closed curve. This number may then be considered a dynamical topological order parameter \cite{Budich2016,Xu2020}, as it is a parameter changing its value with time, taking on discrete values which only depend on the topology of $ \varphi $, i.e. its vortices, and whether $ \mathcal{C} $ encloses these vortices. Closing such a curve tightly around a vortex, the winding number essentially detects that a line of $ 2\pi $ discontinuity enters the area enclosed by the curve without exiting again. Together with the image of the vortices always being tailed by these lines, which we will refer to as tears (as in torn fabric), we see that to find the vortices we must simply be able to identify the tears and their ends. The tears can only end either at the edge of the considered parameter space or at a vortex. Hence, one only needs sufficient resolution (in a simulated or experimental data) to distinguish discontinuous jumps of $ 2\pi $ from the jumps between the data points on a coarse grid in order to find the zeros of the function $ f $. This makes it possible to find and study zeros and vortices with a minimal numerical computational effort, and we have used an algorithm based on this idea to do so in our system. Furthermore, one can be certain that these will be true zeros of the function, and not points where the complex function merely has a very small modulus. This is otherwise principally quite hard to do, as unavoidable numerical imprecision would usually make it necessary to set an arbitrary limit on when the modulus is small enough to indicate that the function has actually become zero. 

\begin{figure*}
	\includegraphics[width=.9\textwidth]{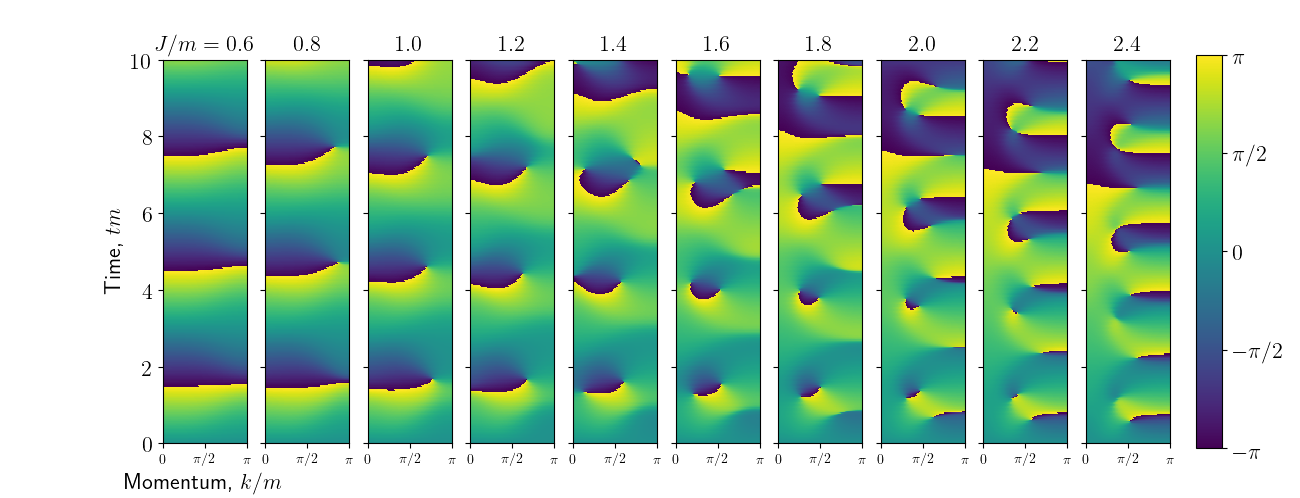}
	\caption{\label{fig:OPdyn} Ten plots of $ \phi_{g} $ showing the dynamics of its vortices. Following the three tears extended across the first plot as $ J/m $ increases, we see how they detach from the edges of the Brillouin zone via the creation of vortices, and how these vortices eventually annihilate with each other, with other vortices, or disappear at the edges. We can also see the creation of a pair of vortices at $ J/m = 1.4 $, $ tm \simeq 7 $, $ k/m \simeq 3\pi/4 $.}
\end{figure*}

\section{Simulation of dynamics}\label{sec:sim}
\subsection{Dynamics of $ g $}
In \cref{fig:OP} we show an example of the modulus and phase of $ g $ for a system size of $ N = 4 $ and a coupling strength of $ J/m = 1.95 $. The zeros of $ g $, as found by the approach described in the previous section, clearly do correspond uniquely to $ \vert g \vert $ becoming very small, confirming the method. Furthermore, looking at the plot of $ \phi_{g} $, the strength of this approach becomes clear. It is easy to see that the tears in $ \phi_{g} $ end in vortices or at the edge of the Brillouin zone, and it is therefore easy to find the zeros of the order parameter. The inset shows a 3D zoom-in of a vortex. This shows how the $ 2\pi $-discontinuity ends at a point where the surface plot of $ \phi_{g} $ is a vertical line, i.e where $ \phi_{g} $ is undefined, and surrounding this point the surface goes smoothly from $ -\pi $ to $ \pi $ in a helix structure. 

\cref{fig:L} shows the Loschmidt echo $ \mathcal{L}(t) $ for the parameters in \cref{fig:OP}. The time of the vortices in $ \phi_{g} $ are marked with vertical dashed lines and triangular, coloured arrow heads, indicating the orientation of the vortices. While we find that the vortices in $ \phi_{g} $ generally appear at values of $ (J,t) $ where $ \mathcal{L}(t) $ is small, we find that vortices of opposite winding which are close in the $ (k,t) $-plane, can be found at larger values of $ \mathcal{L}(t) $. Intuitively, one might think of the vortices as charges, and when two charges of opposite sign are near each other, they screen each other off. There are some additional dips at approximately $ tm = 2.5,7.5 $ not accompanied by vortices. However, for slightly larger $ J/m $, vortices will appear at these dips. This temporary discrepancy is caused by the fact that $ \mathcal{L}(t) $ varies smoothly as a function of $ J/m $, while the appearance and disappearance of vortices is sudden.

Looking at $ \phi_{g} $ for different $ J/m $ these vortices "move" in the $ (k,t) $-plane, being created or annihilated in pairs of opposite orientation, or appearing and disappearing at the edges of the Brillouin zone. This is how $ \phi_{g} $ changes its number of vortices in integer steps. In \cref{fig:OPdyn} are ten plots of $ \phi_{g} $ for different consecutive values of $ J/m $. At first we see three tears stretching across the entire plot, but as $ J/m $ is increased these tears detach from the right and left side of the Brillouin zone, via the creation of respectively left-winding and right-winding vortices at their ends. These vortices remain connected by the tear until they annihilate either with other, other vortices, or disappear again at the edges. At $ J/m = 1.4 $ a pair of vortices is created at $ tm \simeq 7 $ and $ k/m \simeq 3\pi/4 $. Thus we see how vortices appear and disappear at the edges or are created in pairs, how they annihilate with each other, and how they are always trailed by a tear, whose ends must be at the edges or at a vortex. Such vortex dynamics is known from other systems, for example Berezinskii-Kosterlitz-Thouless transitions, both in equilibrium and non-equilibrium \cite{Berezinskii1972,Kosterlitz1973,Mitra2012,Mathey2017}. 

\begin{figure}
	\includegraphics[width=\columnwidth]{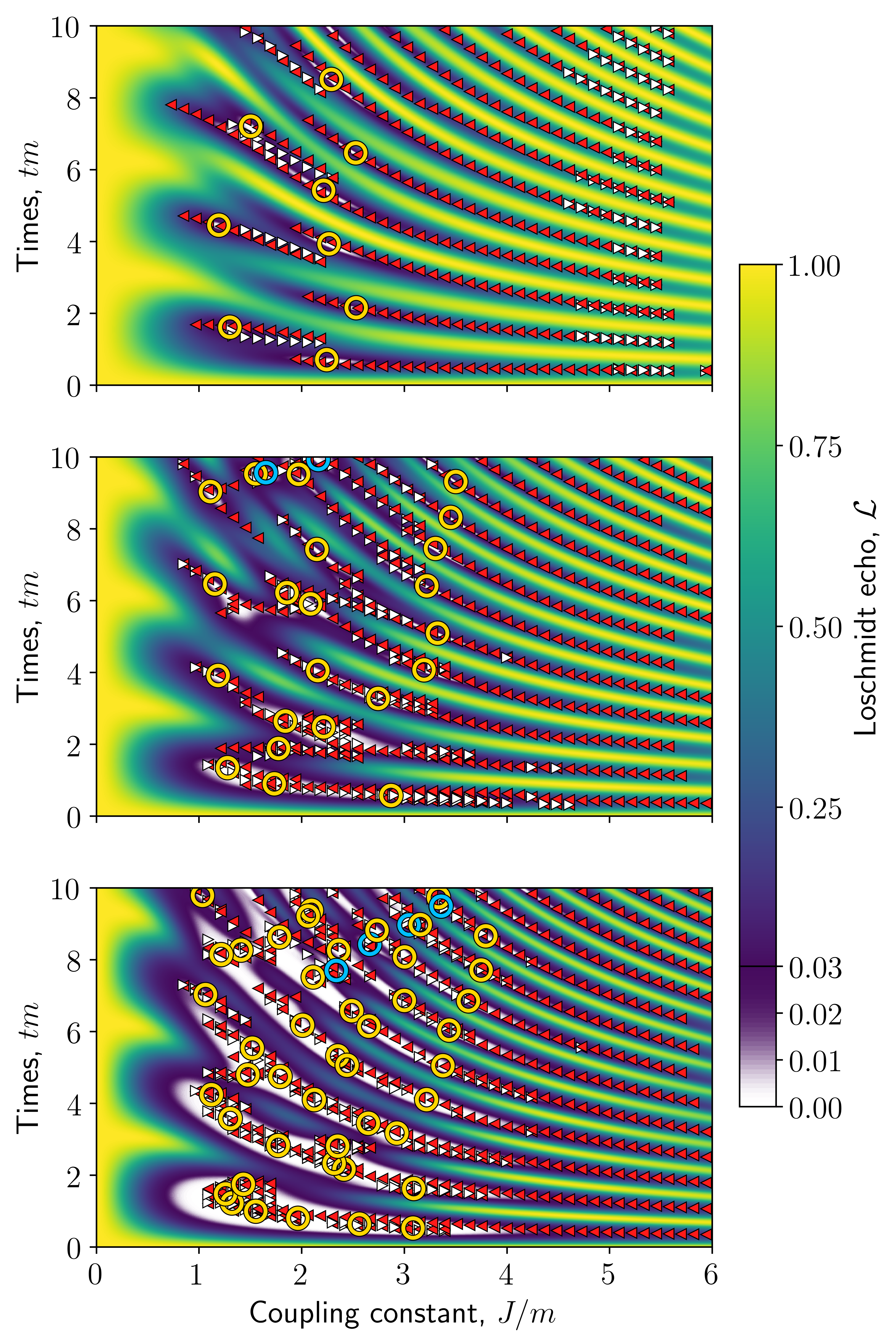}
	\caption{\label{fig:LA} Contour plots of $ \mathcal{L}(t) $ for the considered range of $ J/m $ and $ t $, for system sizes $ N = 4,8,16 $ from top to bottom. The zeros, as found by considering vortices in the phase of $ \mathcal{G}(t) $, are marked with circles coloured blue for right-winding vortices, and yellow for left-winding. A representative set of the vortices of the order parameter are plotted with arrow heads. White arrow heads indicate right-winding vortices, and red arrow heads indicate left-winding vortices.	Particularly for $ N = 4 $ it can be clearly seen how vortices of opposite orientation move around the $ (J,t) $-plane and annihilate. The structure of $ \mathcal{L}(t) $ and its zeros has a clear pattern that is present in all three plots. For $ N = 16 $ $ \mathcal{L}(t) $ is very close to zero in large areas, and its zeros, particularly at early times, trace out a curve following the center of these lobes.}
\end{figure}
\begin{figure}
	\includegraphics[width=\columnwidth]{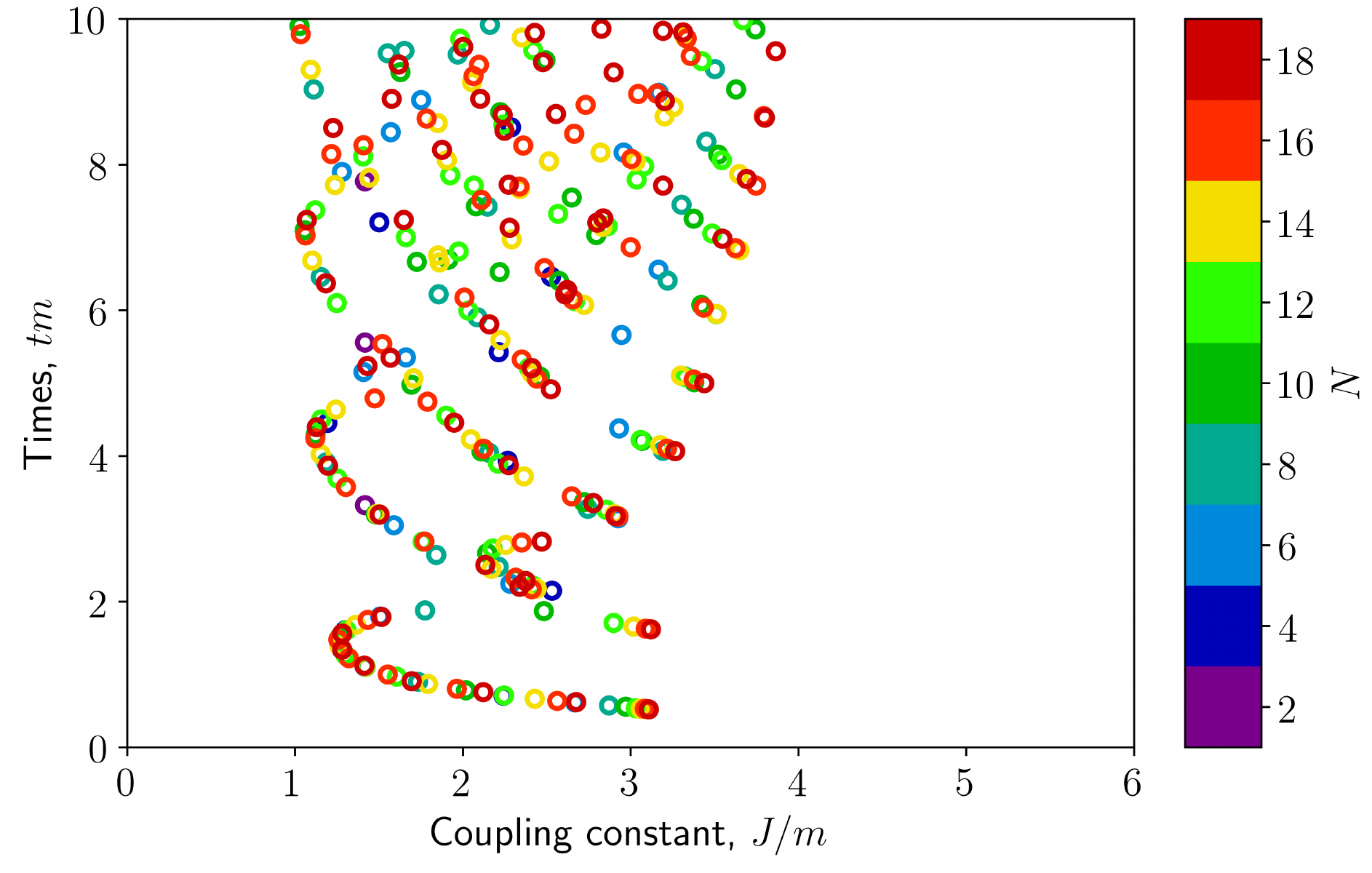}
	\caption{\label{fig:comp_Lvortices} All vortices of $ \mathcal{G}(t) $ for all considered system sizes plotted together as circles, with a colour code denoting which $ N $ each data point pertains to. Some patterns and curves emerge here, particularly for early times.}
\end{figure}

\subsection{Dynamics of $ \mathcal{L} $}
In \cref{fig:LA} $ \mathcal{L}(t) $ can be seen for the considered range of $ J/m $ and $ t $, for system sizes $ N = 4,8,16 $. The zeros of $ \mathcal{L}(t) $ are marked with circles using the same colour code as previously. Most of them are left-winding with a few right-winding at late times in the middle and lower frames of \cref{fig:LA}. It is unclear whether the orientation of the vortices in the phase of $ \mathcal{G}(t) $ has any physical significance, or whether it is simply a mathematical detail. In the phase of the order parameter $ \phi_{g} $ we saw dynamics of the vortices including annihilation of oppositely winding vortices as function of $ J/m $, but we can not speak of something similar in the phase of $ \mathcal{G}(t) $. The zeros of $ \mathcal{L}(t) $ are found by applying the method described in the previous section to the phase, $ \phi_{\mathcal{G}(t)} $, of $ \mathcal{G}(t) $. The zeros of the order parameter are also shown in \cref{fig:LA}, indicated with triangles and a different colour code for distinguishability. We can see how the zeros of the order parameter lie along the troughs of $ \mathcal{L}(t) $, and for $ N = 4 $ in the upper panel of \cref{fig:LA} it is particularly clear in for example the region $ 1 < J/m < 2 $, how vortices in $ \phi_{g} $ of opposite orientation move and annihilate as function of $ J/m $. There is a clear pattern of $ \mathcal{L}(t) $ being small in large lobes that start at around $ J/m = 0.5 $ and stretch towards higher $ J/m $ while $ \mathcal{L}(t) $ increases. The zeros of $ \mathcal{L}(t) $ occur near the center of these lobes, along lines which are traced out by the zeros of the order parameter. This pattern appears already at $ N = 4 $ and repeats itself for larger system sizes, showing see how the small system with $ N = 4 $ reproduces features of the much larger $ N = 16 $ system. This means that even a small experimental realization of this system could yield interesting results. We discuss this possibility in greater detail later in the paper. While the even smaller system with $ N = 2 $ does show zeros both in $ g $ and $ \mathcal{L}(t) $ the behaviour of these is considerably different and simpler than for larger $ N $, see Supplemental Material \cite{suppmat}. We explain this different behaviour by the fact that, as mentioned, the system is completely solvable for such a small size and the zeros of $ \mathcal{L}(t) $ can found analytically. 

For increasing system size $ \mathcal{L}(t) $ becomes very small in increasingly larger areas and has more zeros. The large areas of small value can be explained by the scaling of $ \mathcal{L}(t) = e^{-N\lambda(t)} $, where $ \lambda(t) $ is the rate function of the Loschmidt echo, which converges in the thermodynamic limit of $ N\rightarrow\infty $. That is, the Loschmidt echo is exponentially suppressed by the system size and thus naturally becomes very small for large systems. The zeros of $ \mathcal{L}(t) $ appear to arrange themselves along curves tracing out the center of the lobes where $ \mathcal{L}(t) $ is small. In \cref{fig:comp_Lvortices} the zeros of $ \mathcal{L}(t) $ are plotted together for all the considered $ N $. Especially for small $ t $ there is a clear hook-shaped curve along which the zeros arrange themselves. A similar pattern seems to repeat itself a few times for larger $ t $, defining branches of zeros. While the exact position on the branches of the individual zeros is dependent on $ N $, the branches do become more defined (at least for small $ t $) for increasing $ N $, where the number of zeros also increases. This condensation of discrete vortices is similar to how the zeros of a partition function are known to converge to lines in the thermodynamic limit \cite{Fisher1965,Heyl2013,Vajna2015,Schmitt2015b,Heyl2018}. A plot of $ \lambda(t) $ for each $ N $ considered is shown in the Supplemental Material \cite{suppmat}, including a zoom-in where the formation of the line of zeros at early times can be clearly seen. The occurrence and quick converge of these lines beginning already at $ N = 4 $ shows how the interesting dynamical effects of our system are not just finite size effects, but from early on reflect the dynamics of the thermodynamic limit.

\cref{fig:comp_Lvortices} furthermore clearly shows how the zeros of $ \mathcal{L}(t) $ occur within an interval of $ J/m $ that expands slowly for increasing $ tm $. For both $ J/m \ll 1 $ and $ J/m \gg 1 $, we expect $ \mathcal{L}(t) $ to go unity. This is because for $ J/m \ll 1 $ the mass term dominates and so the pre-quench ground state is still close to being an eigenstate of the post-quench Hamiltonian (the sign of its energy will simply have changed). For $ J/m \gg 1 $ the interaction term dominates and so the pre-quench ground state is still the ground state after the quench, as the interaction term does not change in the quench. For these reasons it makes sense that there is a limited interval of $ J/m $ in which the zeros occur. We expect this interval to expand for increasing $ tm $, simply because the weak or suppressed dynamics responsible for the zero will have more time to occur. For the same reasons we expect the order parameter to not show vortices in these limits, which is consistent with \cref{fig:LA}.

\subsection{Possibility of underlying equilibrium transition}
One of the most observed cases for what causes a DQPT is quenches across some underlying equilibrium phase transition \cite{Heyl2018}. However, given that the DQPTs we observe occur at unsystematic values of $ J/m $ and are dependent on the system size, it is difficult to see how an equilibrium phase transition should be the cause of the DQPTs. Furthermore, while equilibrium phase transitions are often discovered by looking at average magnetizations and similar quantities, in our periodic system we have found that $ \expval{\sigma_{n}^{z}} = (-1)^{n}\expval{\sigma_{0}^{z}} $ and $ \expval{S_{n,n+1}^{z}} = (-1)^{n}(\expval{\sigma_{0}^{z}} - 1)/2 $, meaning that any average of these is identically zero. Indeed, since all expectation values of $ \sigma^{z} $ for both sites and links are given by $ \expval{\sigma_{0}^{z}} $, and expectation values of $ \sigma^{x},\sigma^{y} $ for sites and links are zero (since they are gauge variant), the only local variable we could study is indeed $ \expval{\sigma_{0}^{z}} $. Alternatively, it might be possible to devise a gauge invariant, non-local order parameter, similar to $ g(k,t) $, which captures an equilibrium phase transition that is the cause of the DQPTs we have found here. We expect such an order parameter to be topological in nature, and thus the equilibrium phase transition to topologically non-trivial. In this context the Kibble-Zurek mechanism \cite{Kibble1976,Zurek1985} might be relevant to consider to relate a possible equilibrium transition to the DQPTs we see \cite{Shimizu2018,Roychowdhury2020}. Whether an underlying transition is the cause of the observed DQPTs and how such a transition should be revealed is an interesting direction to take for future research in our system.

\section{Circuit realization of QLM}\label{sec:circuit}
\begin{figure*}
	\includegraphics[width=\textwidth]{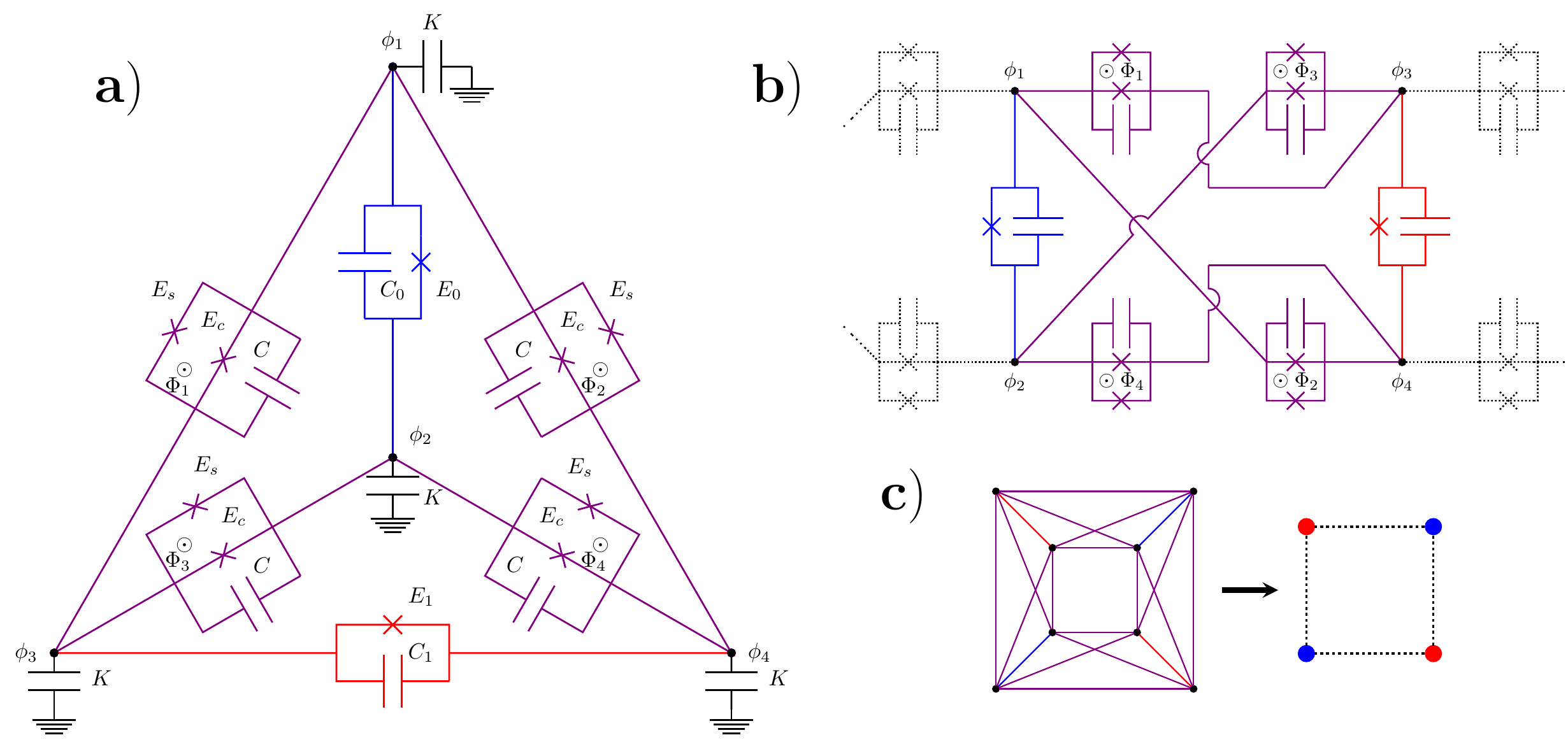}
	\caption{\label{fig:XXX} (a) The diagram of the circuit used to implement three spins interacting via a direct three-body XXX-coupling. Notice the identical circuit parameters. The identical grounding of each node has been included only for completeness of the study. We implement the spins via the eigenmodes of the capacitive network. The branches pertaining to each of these modes have been coloured separately. The blue and red branches will implement matter sites spins, while the purple branches implement the gauge link spin. (b) The same circuit but now folded to make the modular scalability of the circuit completely clear. Multiple copies of the circuit, sharing the matter site branches pairwise, will implement a chain of matter sites coupled via gauge links. (c) A simple diagram of how four copies of the circuit could be put together to implement the periodic $ N = 4 $ version of \cref{eq:gaugeH}.}
\end{figure*}
To realize a QLM you need discrete sites arranged in a lattice, with links connecting them in some configuration. On the sites live matter field degrees of freedom, which for a fermionic field is equivalent to a discrete two-level system, as discussed above. Likewise, on the links live finite degrees of freedom, a spin-$ S $, representing the gauge field coupling the matter fields. Compared to LGTs, where the gauge fields have an infinite number of degrees of freedom, the advantage of QLMs is thus a vastly reduced Hilbert space as a consequence of the finite number of states available to the gauge field. The QLM are valid and interesting gauge theories in themselves, but can also be seen as an approximation of LGTs, as an LGT is recovered by letting $ S\rightarrow\infty $.

We now go on to discuss a possible experimental implementation of U(1) symmetric spin-$ 1/2 $ systems using SQCs. We present a circuit which implements two matter site spins and a gauge link spin interacting via a direct three-body XXX-coupling, which through appropriate tuning yields the desired U(1) interaction in the rotating wave approximation (RWA). The circuit scales naturally in a modular fashion, and could be used to create 1D chains, 2D lattices, or any other configuration of matter sites interacting through gauge links. Hence, the circuit could be used to experimentally implement dynamics in 1D or 2D models, to study for example vacuum quenches similar to what we looked at above, or strong CP-breaking in gauge models. \cref{fig:XXX}c shows a diagrammatic implementation of a plaquette of four sites, hinting how a 2D configuration would have to be made. Early work was done to indicate that circuits could be used for simulating LGTs \cite{Marcos2013,Marcos2014,Mezzacapo2015,LeHur2016,Hartmann2016,Brennen2016,Alaeian2018,Klco2018,Roushan2017,Yang2016}, but this work did not consider concrete cases in detail, nor any checks whether the circuits actually realize the right dynamics with high fidelity. Here we do those things for the first time. In the next section we directly compare the time evolution operator implemented by our circuit with the desired time evolution operator of the U(1) spin-$ 1/2 $ QLM presented in previous sections (see \cref{eq:gaugeH}) using average fidelity \cite{Nielsen2002}. Average fidelity is a measure of how well a certain process implements a desired operation. In our case the process is the time evolution of the circuit and the operation we compare this with is the time evolution according to the target Hamiltonian. The average is over all possible initial states.

\subsection{The circuit}
The circuit can be seen in \cref{fig:XXX}a. It consists of four all-to-all connected nodes, and we may divide the branches in two groups. The blue and red branches have the same circuit elements but each their own circuit parameters. They will be shown to each implement a spin representing a matter field. The purple branches likewise have the same circuit elements, with identical parameters, and they connect the blue and red branches to each other. These branches will altogether implement a spin representing a gauge link. A similar circuit was designed and experimentally tested in Ref. \cite{Gyenis2019} to realize a qubit with a very long life time. Each node in our circuit has been coupled capacitively to ground via an identical capacitance $ K $. Ideally this capacitance is zero, but has been included to study the effects of coupling to ground as well as capacitive coupling to control or readout devices. There are four external fluxes, $ \Phi_{i} $ for $ i = 1,2,3,4 $, threaded through loops, each consisting of two Josephson junctions in parallel, in the purple branches. The Josephson junctions themselves are imagined to be implemented with SQUIDs such that the Josephson energy of each has an increased interval of possible values and high tunability. In \cref{fig:XXX}b the layout of the circuit is chosen such that the $ \Phi_{i} $ only pass through the circuit loops with the Josephson junctions pertaining to $ E_{s} $, using airbridges \cite{Abuwasib2013,Chen2014,Jin2018,Dunsworth2018,Hukai2019,Schmitt2015a}, and the modular scalability of the circuit has also been made explicit. The idea is to make copies of the circuit in sequence, while using the same branches for the matter site spins. If the circuit parameters are chosen with the same symmetries as in \cref{fig:XXX}a, the eigenmodes of the capacitance network will not be affected by the additional copies. The shared branch of two connected blocks will precisely represent a matter connected to two gauge links, and only the desired interactions will be present. The circuit can thus be quite intuitively scaled to a chain of matter sites interacting through gauge links. A matter site branch could potentially also be shared by more than two copies of the circuit, making it possible to realize more complicated configurations. This would however result in many wires connecting to the same branch and require many bridges. The ability to cross conductors via airbridges makes SQCs a suitable platform to implement periodic boundary conditions. Bridges make it possible to access all points in a complicated circuit, while keeping it planar. In order to simulate the periodic system considered in the first part of this paper for the simplest interesting case, i.e. $ N = 4 $, we would have four copies of the circuit put together in this way, forming a square. A simple diagram of such a circuit and the resulting spin system can be seen in \cref{fig:XXX}c. In \cref{sec:readout} we will consider readout of the circuit by dispersively coupling resonators to just the matter site branches, i.e. readout of the square would be done by coupling resonators to its corner branches. 

\subsection{Hamiltonian}
In the following we will consider only a single copy of the circuit. Putting several together to form a square or some larger system, the circuit parameters would have to be re-tuned. However, only the spin model parameters of the modes on the branches which are shared with the new copies will be affected by them, i.e. the matter field modes. These do, however, have their own circuit parameters, which affect only the matter modes. These are the parameters with a subscript in \cref{fig:XXX}a on the red and blue branches. Hence, the effect of the new copies could be compensated for by changing these parameters correspondingly. 

We define the node fluxes of the circuit $ \vec{\phi} = (\phi_{1},\phi_{2},\phi_{3},\phi_{4})^{T} $ \cite{Devoret2017}, but will be working in the eigenmodes of the capacitive network \cite{Bergeal2010,Kounalakis2018,Roy2017,Roy2018,Pedersen2019}, $ \vec{\psi} = (\psi_{CM},\psi_{0},\psi_{g},\psi_{1})^{T} $, defined through
\begin{align*}
\vec{\phi} = \mat{1 & 1 & \frac{1}{2} & 0 \\ 1 & -1 & \frac{1}{2} & 0 \\ 1 & 0 & -\frac{1}{2} & 1 \\ 1 & 0 & -\frac{1}{2} & -1}\vec{\psi}
\end{align*}
This results in no interactions through the capacitors, greatly reducing the complexity of the interactions in the system. We will se how the modes $ \psi_{0} $ and $ \psi_{1} $ will represent two matter sites, and $ \psi_{g} $ will represent the gauge link between them. We will furthermore introduce a new set of external fluxes, $ \Psi_{0}, \Psi_{g} $ and $ \Psi_{1} $, of which the $ \Phi_{i} $ are certain simple, linear combinations. We set these new external fluxes to be constant $ \Psi_{j} = -\pi/2 $ for $ j = 0,g,1 $. For details on the external fluxes and the derivation of the Hamiltonian see Supplemental Material \cite{suppmat}. In these coordinates and with these choices of external fluxes the circuit Hamiltonian becomes
\begin{align}
\begin{split}
H_{c} &= \frac{\mathcal{K}_{00}^{-1}}{2}q_{0}^2 + \frac{\mathcal{K}_{gg}^{-1}}{2}q_{g}^2 + \frac{\mathcal{K}_{11}^{-1}}{2}q_{1}^2\\
	&\hspace{11.5pt} - E_{0}\cos\psi_{0} - E_{1}\cos\psi_{1} \\
	&\hspace{11.5pt} - 4E_{c}\cos\psi_{0}\cos\psi_{g}\cos\psi_{1}\\
	&\hspace{11.5pt} - 4E_{s}\sin\psi_{0}\sin\psi_{g}\sin\psi_{1} \label{eq:circH}
\end{split}
\end{align}
where the $ q_{j} $ are momentum variables conjugate to the $ \psi_{j} $, and $ \mathcal{K}_{jj}^{-1} $ are the diagonal entries of the inverse capacitance matrix in the basis of the $ \psi_{j} $ coordinates. This Hamiltonian describes three transmon-like \cite{Transmon2007} anharmonic oscillator modes, interacting only through the interesting triple cosine and sine interactions. These are direct, completely even and completely odd respectively, three-body interactions. The sine functions come about as a consequence of setting $ \Psi_{j} = -\pi/2 $. Recasting each of the $ \psi_{j} $ and $ q_{j} $ variables in terms of harmonic oscillator operators, i.e. bosonic creation and annihilation operators, $ a_{j}^{\dagger} $ and $ a_{j} $, and truncating the system to the two lowest levels of each, yields the following spin Hamiltonian
\begin{align}
\begin{split}
H_{s} &= -\frac{1}{2}\Omega_{0}\sigma_{0}^{z} - \frac{1}{2}\Omega_{g}\sigma_{g}^{z} - \frac{1}{2}\Omega_{1}\sigma_{1}^{z}\\
&\hspace{11.5pt} + J_{0g}^{z}\sigma_{0}^{z}\sigma_{g}^{z} + J_{01}^{z}\sigma_{0}^{z}\sigma_{1}^{z} +  J_{g1}^{z}\sigma_{g}^{z}\sigma_{1}^{z}\\
&\hspace{11.5pt}  + J_{0g1}^{z}\sigma_{0}^{z}\sigma_{g}^{z}\sigma_{1}^{z} + J_{0g1}^{x}\sigma_{0}^{x}\sigma_{g}^{x}\sigma_{1}^{x} \label{eq:spinH}
\end{split}
\end{align}
where the $ \Omega $'s and $ J $'s are spin model parameters. To calculate the parameters we use a method introduced by the authors in Ref. \cite{Pedersen2019}, which avoids approximating the trigonometric functions via a Taylor expansion, but instead takes their full effect into account. This gives more accurate parameters, when truncating the flux Hamiltonian, and can be used for any sine or cosine of a linear combination of the flux coordinates. The exact dependence of the spin model parameters on the circuit parameters, and details on their derivation can be seen in the Supplemental Material \cite{suppmat}. We note however that the XXX-coupling strength is proportional to $ E_{s} $, and $ E_{s} $ does not appear anywhere else in the spin model parameters, making the XXX-coupling separately tunable. The circuit has thus resulted in three spins interacting through several Z-type couplings, and a direct XXX-coupling. As we detail below, the XXX-coupling can be tuned to $ \sigma_{0}^{+}\sigma_{g}^{+}\sigma_{1}^{-} + \textup{H.c.} $, which is exactly the U(1) gauge coupling in \cref{eq:gaugeH}, $ \sigma_{0}^{+}S_{0,1}^{+}\sigma_{1}^{-} + \textup{H.c.} $. This shows how the $ \psi_{0} $- and $ \psi_{1} $-modes will represent matter site spins, while the $ \psi_{g} $-mode has the role of gauge link spin. The circuit could be simplified a great deal, by removing the $ E_{c} $ junctions, which would seemingly not disturb any of the desired terms in the Hamiltonian. However, this is not quite true as the triple cosine term is the only source of anharmonicity for the $ \psi_{g} $-mode. The anharmonicities $ \alpha_{j} $ of the three modes, which justify the truncation to the two lowest level of each anharmonic oscillator, can be seen in the Supplemental Material \cite{suppmat}.

\subsection{Higher levels}
Let us briefly consider the effect of interactions between the spin-$ 1/2 $ subspace and the higher levels of the circuit. There will generally be even interactions like $ a_{j}^{\dagger}a_{j}^{\dagger}a_{j'}a_{j'} + \textup{H.c.} $ or $ a_{j}^{\dagger}a_{j}^{\dagger}a_{j'}^{\dagger}a_{j'}^{\dagger} + \textup{H.c.} $, where there is an even number of creation and annihilation operators for each mode, as well as odd interactions where is an odd number of creation and annihilation operators. The even interactions will generally be suppressed if their coupling strengths are much smaller than the spin transition energies, while the odd interactions will generally be suppressed if their coupling strengths are much smaller than the anharmonicities. As the $ J^{z} $-coupling strengths and $ J_{0g1}^{x} $ represent the couplings strengths of higher order even and odd couplings respectively, this means that we must have $ J^{z} \ll \Omega_{j} $ and $ J_{0g1}^{x} \ll \alpha_{j} $ in our system. However, even in this regime, where the dynamics mainly take place in the spin-$ 1/2 $ subspace, there will still be the important question of the exact effect of the higher levels in the anharmonic oscillator degrees of freedom, both for effective interactions and leakage \cite{Krantz2019,Kounalakis2018,Rol2019,Zhao2020,Transmon2007,Loft2018,Baekkegaard2019,Rasmussen2019,Barfknecht2019,Pedersen2019}. Below we will include higher levels in our numerics to judge the impact directly and show the regimes necessary to reduce these effects. In particular, the system will undergo virtual excitations and de-excitations, which effectively renormalize the spin model parameters. This results in having to tune effective parameters and not the explicit ones which appear in \cref{eq:spinH}. This is what we have done numerically, and we will detail on it in the next section.

\subsection{Connecting to the QLM Hamiltonian}
In order to achieve a staggered mass for the matter site spins, and no mass for the gauge link spin, we use an approach from Ref. \cite{Marcos2013}. Let $ H_{s} = H_{0} + H_{\textup{int}} $, where $ H_{0} $ contains all the Z-type terms and $ H_{\textup{int}} $ is just the XXX-coupling. Consider then $ H_{s} $ in a frame rotating with respect to $ H_{m} = H_{0} + \frac{1}{2}m(\sigma_{0}^{z} - \sigma_{1}^{z}) $
\begin{align}
\begin{split}
H_{R} &= e^{iH_{m}t}\left[H_{s} - H_{m}\right]e^{-iH_{m}t}\\
&= -\frac{1}{2}m\sigma_{0}^{z} + \frac{1}{2}m\sigma_{1}^{z} + J_{0g1}^{x}\sum_{p,r,s\ \in\ \{+,-\}}e^{-i\omega_{prs}t}\sigma_{0}^{p}\sigma_{g}^{r}\sigma_{1}^{s} \label{eq:rotH}
\end{split}
\end{align}
where the sum is over all eight combinations of the three $ \sigma_{i}^{\pm} $, and the frequency of their phase is given by
\begin{align*}
\omega_{prs} &= p(\Omega_{0} - m) + r\Omega_{g} + s(\Omega_{1} + m) + 2prsJ_{0g1}^{z}
\end{align*}
If the system is now to tuned such that for example $ \omega_{++-} = -\omega_{--+} = 0 $, then the operator $ \sigma_{0}^{+}\sigma_{g}^{+}\sigma_{1}^{-} + \textup{H.c.} $ will be resonant, as desired. All other combinations will be off-resonant, and would disappear in a RWA, as long as the spin transition frequencies and their differences are much larger than the $ J^{z} $, which is already something we must fulfil to justify the truncation to the spin-$ \frac{1}{2} $ subspace. Furthermore, we have recovered the staggered mass of \cref{eq:gaugeH} via the terms $ -\frac{1}{2}m\sigma_{0}^{z} + \frac{1}{2}m\sigma_{1}^{z} $. Thus in an appropriately rotating frame, the Hamiltonian in \cref{eq:spinH} implemented by the circuit indeed recreates the one-dimensional U(1) quantum link model of \cref{eq:gaugeH} for two matter sites and the link between them. 

The circuit design principles we have used here, i.e. looking at the eigenmodes of the capacitive network in a symmetric circuit to achieve multi-body couplings and suppressing as many undesired interactions as possible, could be used to achieve other interesting gauge invariant systems. It would be an obvious next step to work towards higher gauge symmetries, like SU(2), or to attempt to implement gauge link operators with three levels. The latter would allow for the study of confinement, and might be implemented by using two spin-$ 1/2 $'s to represent one gauge field.

\section{Optimizing circuit parameters and average fidelity}\label{sec:opt}
\subsection{Initial considerations}
We now go through some considerations necessary for the optimization of circuit parameters. As mentioned, higher order contributions from interactions with states outside the spin-$ \frac{1}{2} $ subspace mean that we must consider effective spin model parameters when optimizing the circuit parameters. We are in particular interested in the effective detuning, $ \Delta_{\textup{eff}} $, between $ \ket{\uparrow_{0}\uparrow_{g}\downarrow_{1}} $ and $ \ket{\downarrow_{0}\downarrow_{g}\uparrow_{1}} $ (with $ \downarrow_{i} $ and $ \uparrow_{i} $ referring to the ground and excited state of the $ i $'th spin) corresponding to $ \omega_{++-} $ in the above, and the effective coupling strength, $ J_{\textup{eff}} $, of the XXX-coupling corresponding to $ J_{0g1}^{x} $ in the above. Furthermore, we want the numerical value of the anharmonicities of the modes to be about $ \SI{100}{\times2\pi\mega\hertz} $ or larger \cite{Krantz2019}. For details on how to numerically determine $ \Delta_{\textup{eff}} $ and $ J_{\textup{eff}} $, see Supplemental Material \cite{suppmat}. We want to show that this circuit can be used to realize the quench dynamics studied in the first part of this paper. In this case, the effective detuning should not be zero, but rather it defines the staggered mass $ m $. Considering the rotated Hamiltonian $ H_{R} $ in \cref{eq:rotH}, it can be seen that we get the desired mass term when $ \Delta_{\textup{eff}} = 2m $. Likewise, if $ J $ is the desired strength of the matter-gauge coupling in \cref{eq:gaugeH}, then we must have $ J_{\textup{eff}} = J/2 $, because of the factor $ 1/2 $ in the interaction term in \cref{eq:gaugeH}. We will thus be tuning $ J/m = 4J_{\textup{eff}}/\Delta_{\textup{eff}} $.

Since only the Josephson energies of a SQC can be tuned \emph{in situ}, it is difficult to actually perform the appropriate quench of the circuit. Instead we intend for the circuit to be constructed with the post-quench parameters. The quench will then be implemented by initializing the system in the ground state of the pre-quench Hamiltonian. Whether we have the system in its pre-quench setup, go into its ground state, and then quench to the post-setup, or simply start with the system in the post-quench setup, and then quickly initialize in the ground state of the pre-quench Hamiltonian, we will see the same resulting dynamics. This moves the difficulty from performing a fast quench to performing a fast initialization. 

\subsection{Optimizing with respect to average fidelity}
With all this in mind we tune the circuit parameters to yield a negative $ J/m $ (as the quench is to a negative mass, $ m \rightarrow -m $), corresponding to post-quench parameters. After finding appropriate circuit parameters we do a check of the overall behaviour of the circuit, ensuring that it works as intended, including no disturbing interactions with higher levels. To do this we use average fidelity \cite{Nielsen2002}. As mentioned average fidelity is a measure of how well a certain process implements a desired operation. It is calculated as the fidelity of the process' implementation of the operation, averaged over initial states. In our case the process is the time evolution of the circuit, determined by $ H_{c} $ in \cref{eq:circH}, in a frame rotating such that the resulting Hamiltonian is $ H_{R} $ in \cref{eq:rotH}, where $ m $ is set to $ \Delta_{\textup{eff}}/2 $ and the bare coupling strength $ J_{0g1}^{x} $ is replaced with $ J_{\textup{eff}} $. To take contributions from higher level interactions into account we truncate to the lowest four levels. We only rotate the spin-$ 1/2 $ states, i.e. we use a four-level version of $ H_{0} + \frac{1}{2}m(\sigma_{0} - \sigma_{1}) $, where all entries pertaining to levels higher than the spin-$ 1/2 $ states are just zero. Furthermore, we must use an effective version of $ H_{0} $, where contributions to the energy levels from virtual interactions are included. In an experimental setting this $ H_{0,\textup{eff}} $ could be determined in a separately, by setting $ E_{s} = 0 $ via flux tuning, thus turning off the XXX-coupling, and then initializing in each of the spin-$ 1/2 $ states, which would then be very close to eigenstates of the system, such that their phase over time would yield their effective energy. 

The operation we compare this with is the time evolution according to the target Hamiltonian, i.e. $ H $ from \cref{eq:gaugeH} with two matter sites and a gauge link between them. We compare only the dynamics of the spin-$ 1/2 $ states, i.e. time-evolution of the circuit takes place with four levels included for each mode, and the result is then projected down to the spin-$ 1/2 $ subspace, before comparing with the time-evolution of $ H $. For details on how the average fidelity is calculated see Supplemental Material \cite{suppmat}. 

The average fidelity will thus be comparing the very time evolution operators themselves for the circuit and the target system. The mass and coupling strength of the target Hamiltonian are chosen to be $ \Delta_{\textup{eff}}/2 $ and $ 2J_{\textup{eff}} $. The fidelity would thus \emph{a priori} be expected to be quite high, but since this is all done with four levels included in each anharmonic mode, the fidelity will be a measure of how much the higher levels affect the dynamics of the circuit beyond just the renormalization of the mass and coupling strength. In particular, some population will be lost to the higher levels, and just as virtual processes contribute to the strength of the XXX-coupling, they will also to some extent induce other effective interactions. These will disturb the desired dynamics and might be gauge variant, resulting in population moving outside of the $ G_{n} = 0 $ gauge sector of the spin-$ 1/2 $ subspace. 

\subsection{Result}
In the Supplemental Material \cite{suppmat} we show an example of a realistic set of circuit parameters satisfying our demands, yielding $ J/m = 4J_{\textup{eff}}/\Delta_{\textup{eff}} = -2.0 $, which according to \cref{fig:LA,fig:comp_Lvortices} would result in interesting dynamics of the order parameter and Loschmidt amplitude within a time of $ tm = 2 $, corresponding here to $ t = \SI{49.5}{\nano\second} $ (remember that we optimize for post-quench $ J/m $, i.e. negative values, while \cref{fig:LA,fig:comp_Lvortices} shows pre-quench values of $ J/m $). In our work with tuning the circuit we have found that it is well capable of implementing the interval of $ J/m $ considered in the first part of this paper. If a circuit is made that implements an interesting value of $ J/m $, other nearby values could be achieved by varying just the Josephson energies, making it possible to use the same circuit to study different values of $ J/m $.

\begin{figure}
	\includegraphics[width=\columnwidth]{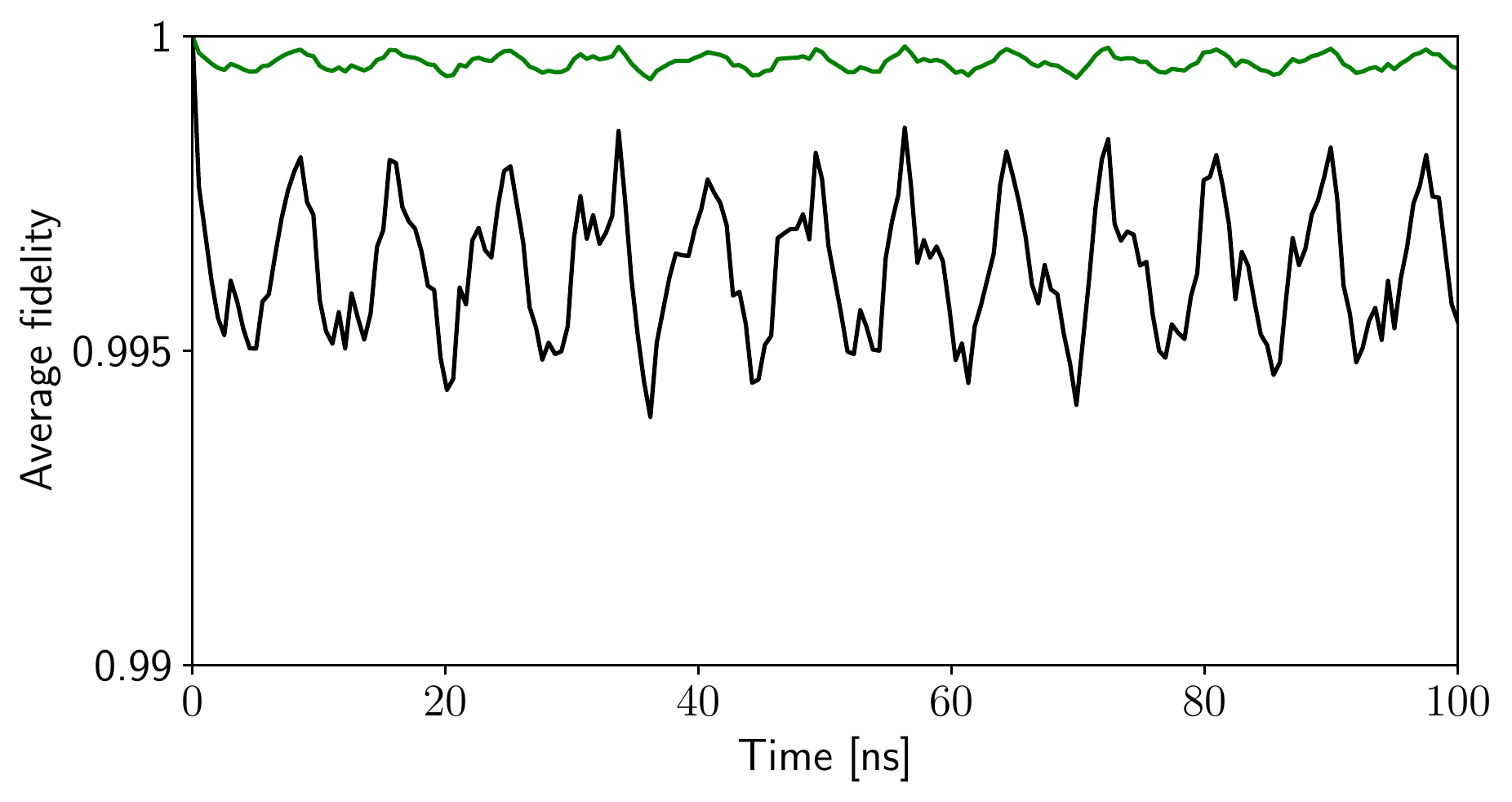}
	\caption{\label{fig:fids} In black: The average fidelity of the circuit's implementation of the target dynamics in a rotating frame. The fidelities are close to or above $ 99.5\% $ and keep steady over long times, with an oscillation on a short time scale. In green: The fidelity without taking leakage to the higher levels into account. From this we see that the largest part, $ 0.2-0.45\% $, of the lost fidelity is due to population immediately leaking into the higher levels and then partly oscillating back and forth. A smaller contribution, about $ 0.05\% $, is from the effective interactions induced by virtual processes involving the higher levels.}
\end{figure}

All calculations of dynamics are performed without including noise. The effect of different types of noise in analogue simulation of LGTs and how to suppress it has already been studied extensively \cite{Kasamatsu2013,Kuhn2014,Lamm2020,Halimeh2020,Halimeh2020_2,Halimeh2020_3,Halimeh2020_4,Halimeh2020_5}. Our results here instead highlight the basic high quality of the presented circuit implementation of a QLM. Furthermore, with present superconducting qubit life times \cite{Wang2019,Touzard2019,Gyenis2019} we do not believe noise would significantly disturb the results presented here. While we have explained how the XXX-coupling in a RWA yields the desired U(1) interaction term, we do not actually use the resulting approximate Hamiltonian in our numerics, but instead retain all terms to show directly that they do indeed not disturb the desired dynamics significantly. In \cref{fig:fids} the calculated average fidelity of the circuit's implementation of the target dynamics using the circuit parameters presented in the Supplemental Material \cite{suppmat} can be seen in black. The fidelity is about or above $ 99.5\% $ at all times, and while it oscillates on a short timescale, it seems to keep steady over the plotted interval. Hence, the implementation of the desired dynamics is good, and stable in the sense that we are not accumulating error or continuously losing population to the higher levels. We seem to lose a small fraction of the population immediately, which then partly oscillates back and forth. In green is plotted the same average fidelity plus the leakage to higher levels, i.e. this plot shows the fidelity if we do not take leakage into account. Hence, we can see that about $ 0.2-0.45\% $ fidelity is lost because of population leaking to the higher levels of the circuit, while about $ 0.05\% $ is lost due to effective interactions induced by virtual processes involving the higher levels. These high and steady fidelities show directly how our superconducting circuit truly implements the desired dynamics, with circuit parameters available to experiments. Hence, the circuit is a strong candidate for studying the U(1) QLM with present, NISQ-era devices.

\section{Readout for state tomography}\label{sec:readout}
\subsection{Reading out $ \mathcal{G}(t) $ and $ g(k,t) $}
\begin{figure*}
	\includegraphics[width=\textwidth]{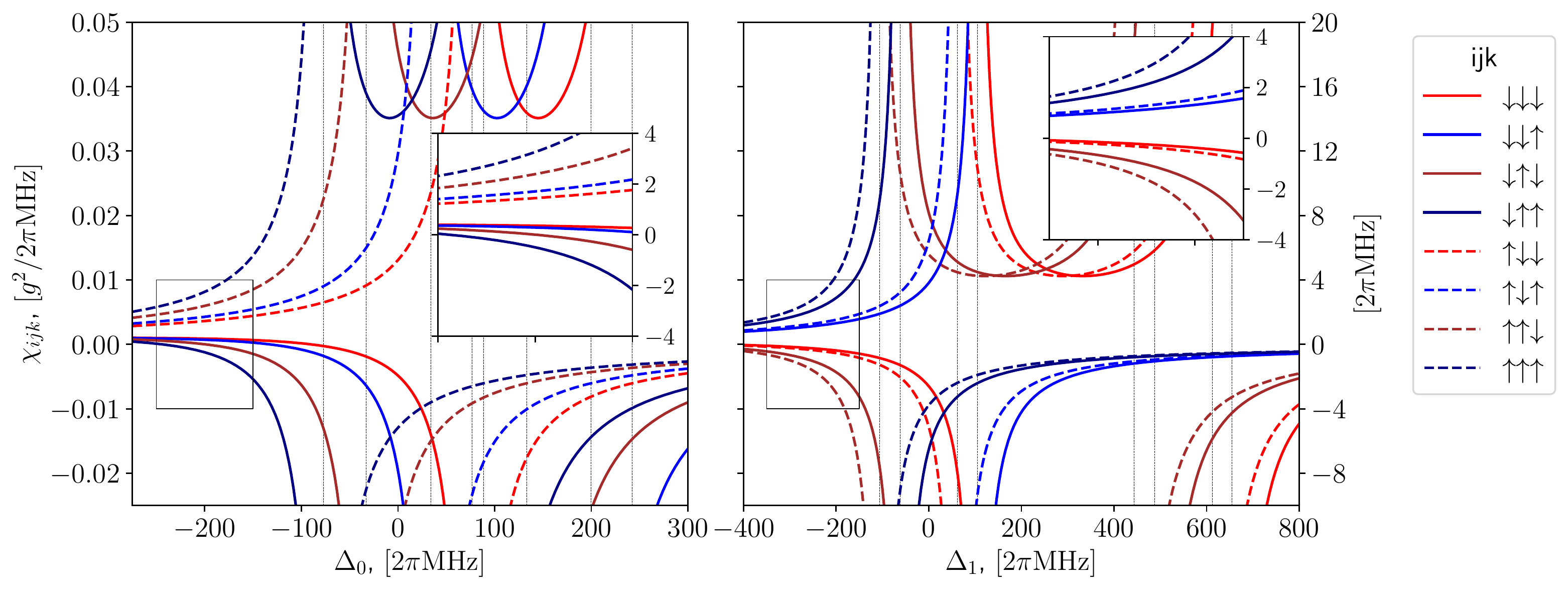}
	\caption{\label{fig:readout} Left and right, the shift $ \chi_{ijk} $ of the resonance frequency of a resonator dispersively coupled to spin $ 0 $ or spin $ 1 $, respectively, for all eight spin-$ 1/2 $ states of the circuit. The shifts are plotted as functions of the bare detuning $ \Delta_{n} = \Omega_{n} - \omega_{r} $ for $ n = 0,1 $ respectively. The indices $ ijk = \downarrow,\uparrow $ denote the state of each of the three spins. Assuming a resonator coupling strength of $ g_{r} = \SI{20}{\times2\pi\mega\hertz} $, we get the energy scale shown on the right $ y $-axis. The insets show intervals of the bare detunings where the dispersive shifts of each state are distinct enough to distinguish between the eight spin state with just these two measurements, particularly when the information from each measurement is compared. To be in the dispersive regime we must have $ \chi_{ijk} \ll g_{r} $, i.e. we must stay well away from the points where $ \chi_{ijk} $ diverges, which are marked with vertical dashed lines. }
\end{figure*}
An important thing to note is that the Loschmidt amplitude in the circuit's own frame will not be the same as in the rotating frame. This is because under unitary transformations like $ e^{-iH_{c}t} \rightarrow Ue^{-iH_{c}t} $, where the operators are simultaneously transformed as $ A \rightarrow UAU^{\dagger} $, quantities like  $ \mathcal{G}(t) = \mel{\psi(0)}{e^{-iH_{c}t}}{\psi(0)} $ or the order parameter $ g(k,t) = \mel{\psi(0)}{\mathfrak{g}(k)e^{-iH_{c}t}}{\psi(0)} $ get an odd number of $ U $ operators. Hence, these operators can not cancel and the quantities are not invariant. The Loschmidt amplitude and our order parameter are therefore not invariant under rotations like the one we performed to find $ H_{R} $. However, the Loschmidt amplitude is often measured by performing state tomography \cite{Qin2017,Baur2012,Filipp2009,Dicarlo2009,Steffen2006} and then calculating $ \mathcal{G}(t) $ from the results \cite{Heyl2018,Xu2019,Guo2019,Xu2020,Jurcevic2017,Flaschner2018}. With full information about the state of the circuit at any time, the Loschmidt amplitude can easily be calculated. 

\subsection{Dispersive readout of coupled qubits}
To perform readout of the circuit we would use a method inspired by Refs. \cite{Roy2017,Roy2018}. Here they perform quantum state tomography of two qubits by measuring the dispersive shift of a resonator coupled to just one of them. The idea is that strong ZZ-couplings shift the energy of one qubit conditioned on the state of the other sufficiently such that it can be seen in the dispersive shift of the resonator. Hence, where normally one observes two shifts of the resonator corresponding to the two eigenvalues of $ \sigma^{z} $, one would see four shifts corresponding to the four combinations of eigenvalues from the two qubits. Single qubit rotations are used to measure the qubits along their $ x $- and $ y $-axes allowing for full state tomography. For details see Refs. \cite{Roy2017,Roy2018} and their supplemental materials. Similarly, here we imagine doing readout of just the matter site spins. A resonator coupled through identical capacitors to the two nodes pertaining to a matter site mode will couple to just that mode. Hence, usual dispersive readout of the spin can be performed. The ZZ-couplings between this mode and its neighbouring gauge and matter modes will make it possible to derive some information about them as well. In particular, we want to extract information about the gauge modes by coupling to just the matter modes. It is easier to couple to the matter modes, as they live on a single branch between two nodes, while the gauge modes live on four branches between four nodes. We therefore propose measuring on all matter modes and comparing the data to extract information about the whole system. For a single module of our circuit measurements on either of the matter modes, gives information on both matter modes and the gauge mode between them. For a 1D chain of modules, measurement on a matter mode would give information on that matter mode, and the two matter modes and the two gauge modes to which it is coupled.

\subsection{Effective dispersive shifts}
We now consider the shifts one would measure for a single module of our circuit. The effective resonance frequency of a resonator dispersively coupled to a qubit is \cite{Transmon2007}
\begin{align}
\omega_{r}' = \omega_{r} - \frac{g_{r}^2}{\Delta + \alpha} - \left(\frac{g_{r}^2}{\Delta} - \frac{g_{r}^2}{\Delta + \alpha}\right)\sigma^{z} \label{eq:shift}
\end{align}
Here $ \omega_{r} $ is the bare resonance frequency, $ \sigma_{z} $ pertains to the qubit, $ \Delta = \omega_{q} - \omega_{r} $ is the detuning between the transition frequency $ \omega_{q} $ of the qubit and $ \omega_{r} $, $ \alpha $ is the anharmonicity of the qubit, and finally $ g_{r} $ is the strength of the dispersive coupling. Here $ \sigma_{z} $ is not to be understood as an operator, but the appropriate eigenvalue of the state that the qubit has collapsed to. The transition frequency $ \omega_{q} $ is in our case a combination of the bare spin transition frequency and ZZ-coupling strengths. If we consider coupling a resonator to spin $ 0 $ in $ H_{s} $ of \cref{eq:spinH}, then we can see that
\begin{align*}
\omega_{q} = \Omega_{0} - 2\left(J_{0g}^{z}\sigma_{g}^{z} + J_{01}^{z}\sigma_{1}^{z} + J_{0g1}^{z}\sigma_{g}^{z}\sigma_{1}^{z}\right)
\end{align*}
where the operators, like the $ \sigma^{z} $ in \cref{eq:shift}, are to be understood as some specific eigenvalue corresponding to the state of the circuit, which has collapsed as we measured it. We can now consider the dispersive shift $ \chi_{ijk} = \omega_{r}' - \omega_{r} $ of the resonator frequency as a function of the bare detuning $ \Delta_{0} = \Omega_{0} - \omega_{r} $, where $ i,j,k = \downarrow,\uparrow $ refers to whether spin $ 0,g,1 $, respectively, has collapsed to $ \ket{\downarrow} $ or $ \ket{\uparrow} $. In \cref{fig:readout} the eight shifts, corresponding to the eight spin-$ 1/2 $ states of the circuit, for both coupling to spin $ 0 $ and $ 1 $ are plotted, using the same circuit parameters is in our previous plots. We now want to find values of $ \Delta_{0} $ and $ \Delta_{1} $ such that the eight shifts are as distinct as possible, and where comparing shifts from both of the spins helps to determine the state of the gauge link spin. In order to remain in the dispersive region we must satisfy $ g_{r}/\vert\Delta\vert,g_{r}/\vert\Delta + \alpha\vert \ll 1 $ where $ \Delta $ now has a different value for each of the spin-$ 1/2 $ states. Looking at \cref{eq:shift}, we can see that $ \chi_{ijk} \sim g_{r}^2/\vert\Delta\vert, g_{r}^2/\vert\Delta + \alpha\vert $, and thus the conditions for the dispersive regime can be written as $ \chi_{ijk} \ll g_{r} $. This essentially means we must stay well away from the regions where $ \chi_{ijk} $ diverges. These are marked with vertical dashed lines in \cref{fig:readout}. If for the sake of example we consider a resonator coupling strength of $ g_{r} = \SI{20}{\times2\pi\mega\hertz} $, we get the energy scale shown on the right $ y $-axis of \cref{fig:readout}. In the insets of \cref{fig:readout} can be seen zoomed in regions which are between $ \chi_{ijk} = \pm\SI{4}{\times2\pi\mega\hertz} $. If we choose a bare detuning within these regions, we could use the shift of the first resonator (the left inset) to distinguish between the dashed and solid lines, i.e. the state of spin $ 0 $, and use the second resonator (the right inset) to distinguish between the blue/navy and the red/brown lines, i.e. the state of spin $ 1 $. This is similar to dispersive measurement of qubits, but we can further use this information to distinguish between the shifts caused by the state of the gauge link spin, $ g $. With a resolution of $ \SI{1}{\times2\pi\mega\hertz} $ in a measurement of the dispersive shift, which is experimentally feasible \cite{Roy2017,Roy2018,Guo2019,Heinsoo2018}, it would be possible to distinguish the states of the circuit with this or even a smaller choice of $ g_{r} $. 

The above analysis is approximate, as it uses only the bare spin model parameters, and the formula \cref{eq:shift} is derived for a single qubit with some specific bare transition frequency coupled to a resonator. In our case it is clear that the higher levels of the circuit would affect these calculations, and it is in fact a resonator, or several, coupled to the circuit, a system of multiple interacting qubits or spins. A more accurate analysis using numerical methods could be carried out to find the actual shifts of the resonance frequency of the resonator dependent on the circuit state. However, the quantitative results would be the same, namely that the different states would result in different shifts. It would then be a matter of determining whether those shifts would be sufficient to distinguish the states in a measurement, using the comparative method outlined above. Whether or not this is the case is in the end a consequence of the chosen circuit parameters, so one could optimize the circuit parameters with respect to these considerations in addition to the conditions we outlined previously. These dispersive readouts could then be used to perform a full quantum state tomography, yielding all information about the Loschmidt amplitude or the order parameter, we introduced in the first part of this paper.

An alternative to performing full state tomography is to have multiple copies of the circuit, perform the quench experiment in just one of them while initializing the other in the appropriate initial state. The circuits are then connected using some appropriate scheme to make their states interfere, potentially yielding information about the quantities we are interested in. Such an approach is used in the context of atoms in an optical lattice in Refs. \cite{Daley2012,Pichler2013} to measure the Rényi entropy. This has in fact been experimentally probed \cite{Islam2015,Linke2018,Brydges2019}.

\section{Conclusion}\label{sec:conc}
We have shown how to realize lattice gauge theories through quantum link models in superconducting quantum circuits. Specifically, we have provided a method for general circuits to implement quantum link models with a high average fidelity. This opens up the possibility for experimental study of quantum link models in NISQ-era devices. As a demonstration of the principles in our work we have studied a periodic (1+1)D spin-$ 1/2 $ quantum link model with local U(1) gauge symmetry, corresponding to the Schwinger model in the continuum. Even with the smallest lattice considered the system undergoes dynamical quantum phase transitions after a quench of the sign of the mass. With this in mind we have proposed a superconducting circuit, which realizes three spin-$ 1/2 $'s interacting via a direct XXX-coupling, which through appropriate tuning becomes the matter-gauge interaction necessary for a U(1) quantum link model. The circuit can be modularly scaled in an intuitive way and for a single module of the circuit realizes the desired U(1) QLM dynamics with an average fidelity of about $ 99.5\% $ or above, using realistic circuit parameters. From this we expect that the dynamical quantum phase transitions we have found should be observable in an experimental realization of the circuit. 

We studied an order parameter, which is essentially the Fourier transform of the gauge invariant string order parameters connecting a representative particle and antiparticle site to all other sites of the system. This order parameter had zeros that correlated with the minima of the Loschmidt amplitude and its zeros. The zeros of both the Loschmidt amplitude and our complex order parameter were found by looking for vortices in their phases. These vortices, which appear exactly when the function goes to zero, are much easier to find numerically, as they are extended structures, and their center point can be found by looking at the line of discontinuity, which extends from it. The vortices are topological in nature, in particular they can be counted by a winding number which then constitutes a dynamical topological order parameter of the system. The vortices of the order parameter showed dynamics of creation and annihilation. We found that the structure of the Loschmidt amplitude as well as its zeros in the parameter space of matter-gauge coupling strength and time, quickly converges to a certain pattern, with the zeros lying along lines. Hence, even the smallest system size considered would reveal the tendencies of the larger systems in an experimental realization. Finally, we considered readout of the circuit, using a method of resonators coupled dispersively to a subset of the circuit spins, but which nonetheless gave information about all the spins, by exploiting their pairwise ZZ-couplings. To use this in the context of the quench dynamics we have studied, we imagine that quantum state tomography of the circuit is performed, to extract the data necessary to calculate the Loschmidt amplitude and our order parameter.

In future work it would be interesting to study dynamics of more complicated lattice configurations and gauge theories. For example periodic 2D, i.e. toric, QLMs could be considered, which would likely show interesting topological aspects. Additional degrees of freedom could be added to the link operators, indeed simply promoting them to spin-$ 1 $'s would allow for the study of confinement and pair production. It would be natural to work towards a superconducting circuit realization of such models using the same design principles we have presented here. A similarly modular circuit realizing SU(2) symmetric interactions between fermions, or some other interesting gauge symmetry like $ \mathbb{Z}_{n} $ would be interesting to develop. Further work could also be done on the specific system studied here. It would be interesting to link the dynamical quantum phase transitions found here to a potential underlying equilibrium phase transition or entropy production \cite{Jurcevic2017,Goes2020,Heyl2018}.

\section*{Acknowledgements}
The authors would like to thank Torsten Zache for insightful discussions. The authors would also like to thank K. S. Christensen, N. J. S. Loft, S. E. Rasmussen, L. B. Kristensen and T. Bækkegaard for general discussions pertaining to this work. The authors acknowledge support from the Independent Research Fund Denmark, the Carlsberg Foundation, and AUFF through the Jens Chr. Skou fellowship program.

\end{document}


\title{Supplemental Material: Lattice gauge theory and dynamical quantum phase transitions using noisy intermediate scale quantum devices} 

\author{Simon Panyella Pedersen}
\email{spp@phys.au.dk}
\affiliation{\affA}
\author{Nikolaj Thomas Zinner}
\email{zinner@phys.au.dk}
\affiliation{\affA}
\affiliation{\affB}

\maketitle
\onecolumngrid 

\section*{Appendix}\label{app}
In \cref{app:N2} we go through the analytical solution for the $ N = 2 $ QLM, and in \cref{app:N4} we do the same for the $ N = 4 $ case. \cref{app:regression} derives and discusses the relations between expectation values that we found in the system. In \cref{app:allLA} we show data for the Loschmidt echo rate function for all system sizes considered. \cref{app:circuitH} derives the Hamiltonian of our circuit in terms of the capacitive eigenmodes, harmonic oscillator modes, and finally spin-$ 1/2 $'s. Our method for exact truncation is also explained. \cref{app:effparam} goes through our numerical determination of effective spin model parameters in the circuit. Finally, in \cref{app:rotFrame} we prove the expression for the time evolution operator in a rotating frame.

\subsection{Analytical solution, N = 2} \label{app:N2}
We can solve analytically the two matter sites periodic chain with the U(1) staggered mass Hamiltonian
\begin{align*}
H &= \sum_{n=0}^{N-1}\left[-(-1)^{n}\frac{m}{2}\sigma_{n}^{z} + \frac{J}{2}\left(\sigma_{n}^{+}S_{n,n+1}^{+}\sigma_{n+1}^{-} + \textup{H.c.}\right)\right]
\end{align*}
We will only be looking at the subspace that we use in the main text, i.e. we first reduce the Hilbert space to the states with $ G_{n}\ket{\textup{phys}} = 0 $. We furthermore group the remaining states according to the parity symmetry of the system. 

\subsubsection{Eigenstates and -energies}
\begin{figure}
	\centering
	\includegraphics[width=0.7\textwidth]{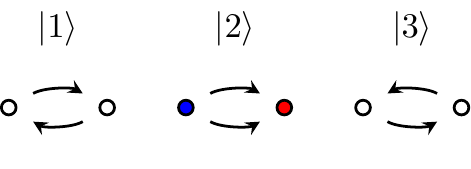}
	\caption{\label{fig:periodicChainHilbertSpace2} The three states of the two matter sites periodic chain that satisfy $ G_{n}\ket{\textup{phys}} = 0 $. They have been arranged such that the Hamiltonian couples each state with the states to its left and right.}
\end{figure}

For $ N = 3 $ there are only three states in the $ G_{n}\ket{\textup{phys}} = 0 $ subspace, see \cref{fig:periodicChainHilbertSpace2}. Numbering them according to the scheme of \cref{fig:periodicChainHilbertSpace2}, we can write the Hamiltonian as
\begin{align*}
H &= m\left(\op{2}{2} - \op{1}{1} - \op{3}{3}\right)\\
&\hspace{11.5pt} + \frac{J}{2}\left(\op{1}{2} + \op{2}{3} + \text{H.c.}\right)
\end{align*}
and the parity symmetry operator can be written as
\begin{align*}
P = \op{1}{3} + \op{3}{1} + \op{2}{2}
\end{align*}
To find the eigenvectors of $ H $, we choose to work with the eigenvectors of the parity symmetry (which can be chosen such that they are also eigenvectors of $ H $)
\begin{align*}
P = +1:& \ket{\psi_{i}} = a_{i}(\ket{1} + \ket{3}) + b_{i}\ket{3}\\
P = -1:& \ket{\psi_{-}} = \frac{1}{\sqrt{2}}\left(\ket{1} - \ket{3}\right)\\
\end{align*}
We know the coefficients to be real, because both the symmetry operator and the Hamiltonian are purely real. Since there are thus two degrees of freedom in the first set of the above eigenvectors, we expect there to be two possible eigenvectors in it, making a total of three. Thus $ i = 1,2 $. To find the values of the coefficients such that they are indeed eigenvectors of $ H $, we will act with $ H $ on them and demand that the result is a constant times the original vector. 

In this small system this is rather simple. $ \ket{\psi_{-}} $ is itself already an eigenvector with energy $ -m $, i.e. $ H\ket{\psi_{-}} = -m\ket{\psi_{-}} $. We find for $ \ket{\psi_{i}} $ (ignoring subscripts for ease of reading) that
\begin{align*}
H&\left[a(\ket{1} + \ket{3}) + b\ket{2}\right]\\
&= m\left(b\ket{2} - a(\ket{1} + \ket{3})\right) + \frac{J}{2}\left(2a\ket{2} + b(\ket{1} + \ket{3})\right)\\
&= \left(-m + \frac{Jb}{2a}\right)a(\ket{1} + \ket{3}) + \left(m + \frac{Ja}{b}\right)b\ket{2}
\end{align*}
We now demand that 
\begin{align*}
-m + \frac{Jb}{2a} = m + \frac{Ja}{b}
\end{align*}
Together with normalization
\begin{align*}
2a^2 + b^2 = 1
\end{align*}
this gives us two equations with two unknowns. We introduce $ x = J/m $ and $ y = m/J = 1/x $, and rewrite our equations as
\begin{align*}
2 + \frac{xa}{b} - \frac{xb}{2a} &= 0\\
2a^2 + b^2 &= 1
\end{align*}
Introducing $ p = b/a $, the first can be rewritten as
\begin{align*}
p^2 - 4yp - 2 = 0
\end{align*}
This has the solutions
\begin{align*}
p_{i} = \frac{4y \pm \sqrt{16y^2 + 8}}{2} = 2y \pm \sqrt{4y^2 + 2}
\end{align*}
These are two real solutions, as we wanted. With the second equation (normalization) we find
\begin{align*}
a_{i} = \frac{1}{\sqrt{2 - p_{i}^2}}
\end{align*}
which together with $ b_{i} = p_{i}a_{i} $, this gives us the coefficients. Furthermore, we can find the energies
\begin{align*}
E_{i} = -m + \frac{Jb}{2a} = J\left(\frac{p_{i}}{2} - y\right) = \begin{cases}
J\sqrt{y^2 + \frac{1}{2}}, & i = 1\\
-J\sqrt{y^2 + \frac{1}{2}} & i = 2
\end{cases}
\end{align*}
We thus have the energies and the coefficients of all eigenstates.

\subsubsection{Loschmidt amplitude}
We can now consider the Loschmidt amplitude. Of the energies $ E_{-} = -m $, $ E_{1} = \sqrt{m^2 + \frac{1}{2}J^2} $, and $ E_{2} = -\sqrt{m^2 + \frac{1}{2}J^2} $, $ E_{2} $ is clearly the always smallest. The ground state before the quench is therefore $ \ket{\psi_{2}} $. After the quench this state still only couples to the two $ P = +1 $ states. Let us note that the quench simply has the effect $ y \rightarrow -y $ in the expressions for the coefficients and energies. We can therefore see that the post-quench coefficients, denoted with a prime, relate to the pre-quench coefficients in the following way
\begin{align*}
a_{1}' &= a_{2}\\
a_{2}' &= a_{1}\\
b_{1}' &= -b_{2}\\
b_{2}' &= -b_{1}
\end{align*}
and the energies are unaffected by the quench. We can now calculate the Loschmidt amplitude in terms of the pre-quench coefficients. We find
\begin{align*}
\mathcal{G}(t) &= \mel{\psi_{2}}{e^{-iH't}}{\psi_{2}}\\
	&= (2a_{2}^2 - b_{2}^2)^2e^{-iE_{2}t} + (2a_{1}a_{2} - b_{1}b_{2})^2e^{-iE_{1}t}
\end{align*}
We find the following expressions for the two parentheses
\begin{align*}
2a_{2}^2 - b_{2}^2 &= -\frac{1}{1 + (2y^2 - y\sqrt{4y^2 + 2})^{-1}}\\
2a_{1}a_{2} - b_{1}b_{2} &= \frac{1}{\sqrt{2y^2 + 1}}
\end{align*}
We look at the amplitude again, and find with some algebra
\begin{align*}
\mathcal{G}(t) = (2a_{2}^2 - b_{2}^2)^2e^{-iE_{2}t}\left(1 + \left(x\frac{1 + \frac{x^2}{2 - \sqrt{4 + 2x^2}}}{\sqrt{2 + x^2}}\right)^2e^{-i(E_{1}-E_{2})t}\right)
\end{align*}
For this to be zero, we must have
\begin{align*}
\left(x\frac{1 + \frac{x^2}{2 - \sqrt{4 + 2x^2}}}{\sqrt{2 + x^2}}\right)^2e^{-i(E_{1}-E_{2})t} = -1
\end{align*}
Hence, $ e^{-i(E_{1}-E_{2})t} = e^{-2iE_{1}t} $ must be real and negative, i.e. $ e^{-2iE_{1}t} = -1 $. This occurs only when $ 2E_{1}t = (2n+1)\pi $. From this we find the times, $ t_{n} $, where $ \mathcal{G}(t) $ goes to zero
\begin{align*}
t_{n} = \frac{((2n + 1)\pi)\pi}{2E_{1}} = \frac{(2n + 1)\pi}{J\sqrt{4y^2 + 2}} = \frac{(2n + 1)\pi}{\sqrt{4m^2 + 2J^2}}
\end{align*}
for $ n = 1,2,3,... $. Furthermore, we must have
\begin{align*}
\left(x\frac{1 + \frac{x^2}{2 - \sqrt{4 + 2x^2}}}{\sqrt{2 + x^2}}\right)^2 = 1
\end{align*}
Again with some tedious algebra we rather pleasingly find that this has only one solution, namely $ x = \sqrt{2} $. With this we can reduce the expression for the critical times to $ t_{n} = \frac{(2n + 1)\pi}{\sqrt{8}m} $. Hence we have the coordinates $ \left(J/m=\sqrt{2},t_{n} = \frac{(2n + 1)\pi}{\sqrt{8}m}\right) $ at the vortices of the Loschmidt amplitude. We find this to be consistent with our numerical simulations.

\subsubsection{Order parameter}
Let us now consider the order parameter. In this small version of the system the order parameter operator, $ \mathfrak{g} $, has the form
\begin{align*}
\mathfrak{g} &= \sigma_{0}^{+}\sigma_{0}^{-} + e^{-ik}\sigma_{1}^{+}S_{0,1}^{-}\sigma_{0}^{-} + e^{ik}\sigma_{1}^{+}\Sigma_{0,1}^{+}\sigma_{0}^{-} \\
&\hspace{11.5pt} + \sigma_{1}^{+}\sigma_{1}^{-} + e^{-ik}\sigma_{0}^{+}\Sigma_{0,1}^{-}\sigma_{1}^{-} + e^{ik}\sigma_{0}^{+}S_{0,1}^{+}\sigma_{1}^{-}
\end{align*}
where we have temporarily used a $ \Sigma $ to denote the second gauge link spin between the two matter sites, i.e. the lower link in the diagrams in \cref{fig:periodicChainHilbertSpace2}. Written in terms of the states in the gauge sector, see \cref{fig:periodicChainHilbertSpace2}, we have
\begin{align*}
\mathfrak{g} &= \op{2}{2} + e^{-ik}\op{3}{2} + e^{ik}\op{1}{2}\\
&\hspace{11.5pt} + \op{1}{1} + \op{3}{3} + e^{-ik}\op{2}{1} + e^{ik}\op{2}{3}
\end{align*}
We can reorder the result a bit
\begin{align*}
\mathfrak{g} &= 1 + e^{-ik}\op{2}{1} + (e^{ik}\ket{1} + e^{-ik}\ket{3})\bra{2} + e^{ik}\op{2}{3}
\end{align*}
We can now calculate the order parameter, and after some algebra we find
\begin{align*}
g(t) &= \mel{\psi_{2}}{\mathfrak{g}e^{-iH't}}{\psi_{2}}\\
	&= (2a_{2}^2 - b_{2}^2)^2e^{-iE_{2}t}\left[1 + \frac{\left((2a_{1}a_{2} - b_{1}b_{2})^2 + (2a_{1}a_{2} - b_{1}b_{2})(2(a_{1}b_{2} - a_{2}b_{1})\cos k)\right)}{(2a_{2}^2 - b_{2}^2)^2}e^{-i(E_{1} - E_{2})t}\right]
\end{align*}
For this to be zero we must again have $ e^{-i(E_{1} - E_{2})t} = \pm 1 $, yielding the critical times $ t_{n} = \frac{n\pi}{\sqrt{4m^2 + 2J^2}} $. However, we must simultaneously have
\begin{align*}
\frac{\left((2a_{1}a_{2} - b_{1}b_{2})^2 + (2a_{1}a_{2} - b_{1}b_{2})(2(a_{1}b_{2} - a_{2}b_{1})\cos k)\right)}{(2a_{2}^2 - b_{2}^2)^2} &= \mp 1
\end{align*}
From before we have
\begin{align*}
2a_{2}^2 - b_{2}^2 &= -\frac{1}{1 + (2y^2 - y\sqrt{4y^2 + 2})^{-1}}\\
2a_{1}a_{2} - b_{1}b_{2} &= \frac{1}{\sqrt{2y^2 + 1}}
\end{align*}
and we can calculate
\begin{align*}
a_{1}b_{2} - a_{2}b_{1} = -\frac{1}{\sqrt{2}}
\end{align*}
With this we must solve
\begin{align*}
\left(\frac{1}{2y^2 + 1} - \sqrt{2}\frac{1}{\sqrt{2y^2 + 1}}\cos k\right)\left(1 + \frac{1}{2y^2 - y\sqrt{4y^2 + 2}}\right)^2 = \mp 1
\end{align*}
This can be reduced with some algebra to
\begin{align}
\cos k = \frac{x^2 \pm 2}{x\sqrt{4 + 2x^2}}
\end{align}
The vortices of course occur only when $ k_{c} $, the critical momentum satisfying the above condition, is real, i.e. $ -1 \leq \frac{x^2 \pm 2}{x\sqrt{4 + 2x^2}} \leq 1 $. Hence, we may divide the vortices into two sets, one for a $ + $ in the above with $ n $ odd, and one for a $ - $ with $ n $ even. One or both the sets of vortices may not appear for all values of $ x = J/m $. It can be found that $ k_{c,+} $ is real for $ x \geq \sqrt{2} $, and $ k_{c,-} $ is real for $ x \geq \sqrt{2\sqrt{5} - 4} $. 

Hence, for critical times $ t_{n} = \frac{n\pi}{\sqrt{4m^2 + 2J^2}} $, the critical momenta are $ k_{c,\pm} = \arccos\left[\frac{x^2 \pm 2}{x\sqrt{4 + 2x^2}}\right] $, where a $ - $ means $ n $ is even, and a $ + $ means $ n $ is odd. This is again consistent with our numerics, as is the boundary of $ J/m $ where the two sets of vortices appear. The critical times of the order parameter coincide with the critical times of the Loschmidt amplitude for $ n $ odd. These are all left-winding vortices. For even $ n $ the (right-winding) vortices of the order parameter do not follow the troughs of the Loschmidt echo, as is otherwise the case for higher $ N $ systems, except remarkably $ N = 6 $ where we likewise find that only the left-winding vortices follow the troughs of the Loschmidt echo, while the right-winding do not. For all other considered system sizes both left- and right-winding vortices appear at low values of the Loschmidt echo.

\subsection{Analytical solution, N = 4} \label{app:N4}
\begin{figure}
	\centering
	\includegraphics[width=0.9\textwidth]{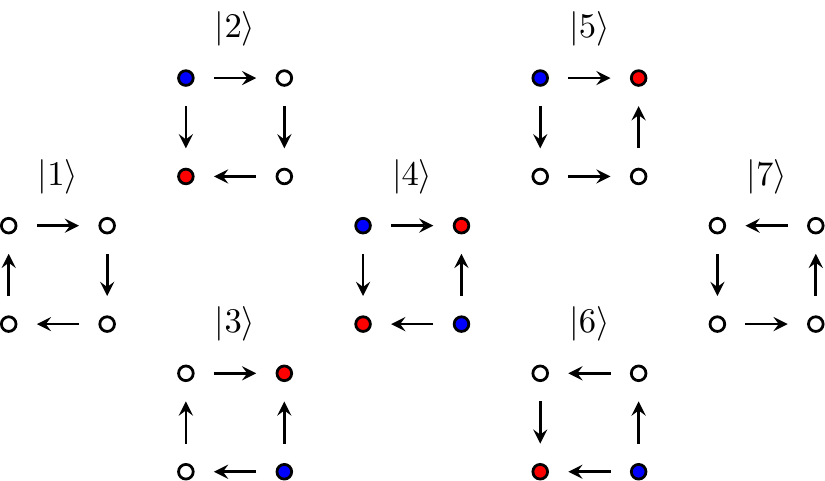}
	\caption{\label{fig:periodicChainHilbertSpace4} The seven states of the four matter sites periodic chain that satisfy $ G_{n}\ket{\textup{phys}} = 0 $. They have been arranged such that the Hamiltonian couples each state with just the states to its left and right (but not above or below).}
\end{figure}
We can also solve the $ N = 4 $ system analytically as well. Again we first reduce the Hilbert space to the states with $ G_{n}\ket{\textup{phys}} = 0 $, and then group the remaining states according to the parity symmetry of the system. The states with $ G_{n}\ket{\textup{phys}} = 0 $ can be most easily found by applying the U(1) assisted hopping term to the two vacua, see \cref{fig:periodicChainHilbertSpace4}. There are a total of seven of these states, and we have numbered them according to the scheme of \cref{fig:periodicChainHilbertSpace4}. In such terms the Hamiltonian can be written as
\begin{align*}
H &= 2m\left(\op{4}{4} - \op{1}{1} - \op{7}{7}\right)\\
&\hspace{11.5pt} + \frac{J}{2}\left(\op{1}{2} + \op{1}{3} + \op{2}{4} + \op{3}{4} + \op{4}{5} + \op{4}{6} + \op{5}{7} + \op{6}{7} + \text{H.c.}\right)
\end{align*}
and the parity symmetry operator can be written as
\begin{align*}
P = (\op{1}{7} + \op{2}{5} + \op{3}{6} + \text{H.c.}) + \op{4}{4}
\end{align*}
To find the eigenvectors of $ H $, we look at the eigenvectors of the parity symmetry
\begin{align*}
P = +1:& \ket{\psi_{i}} = a_{i}(\ket{1} + \ket{7}) + b_{i}(\ket{2} + \ket{3} + \ket{5} + \ket{6}) + c_{i}\ket{4}\\
P = +1:& \ket{\psi_{+}} = \frac{1}{2}\left(\ket{2} - \ket{3} + \ket{5} - \ket{6}\right)\\
P = -1:& \ket{\psi_{-}} = \frac{1}{2}\left(\ket{2} - \ket{3} - (\ket{5} - \ket{6})\right)\\
P = -1:& \ket{\psi_{j}'} = a_{j}'(\ket{1} - \ket{7}) + b_{j}'(\ket{2} + \ket{3} - \ket{5} - \ket{6})
\end{align*}
We know the coefficients to be real, because both the symmetry operator and the Hamiltonian are purely real. Since there are thus three and two degrees of freedom in the first and last set of the above eigenvectors, we expect there to be three and two possible eigenvectors in these sets, making a total of seven. Thus $ i = 1,2,3 $ and likewise $ j = 1,2 $. To find the values of the coefficients such that they are indeed eigenvectors of $ H $, we will act with $ H $ on them and demand that the result is a constant times the original vector. 

For $ \ket{\psi_{\pm}} $, we simply have $ H\ket{\psi_{\pm}} = 0 $, meaning that they are indeed eigenvectors of $ H $, with energies $ 0 $. For $ \ket{\psi_{i}} $ we find (ignoring subscripts for ease of reading)
\begin{align*}
H&\left[a(\ket{1} + \ket{7}) + b(\ket{2} + \ket{3} + \ket{5} + \ket{6}) + c\ket{4}\right]\\
&= 2m\left(-a(\ket{1} + \ket{7}) + c\ket{4}\right) + \frac{J}{2}\left(a(\ket{2} + \ket{3} + \ket{5} + \ket{6}) + b(2\ket{1} + 4\ket{4} + 2\ket{7}) + c(\ket{2} + \ket{3} + \ket{5} + \ket{6})\right)\\
&= \left(-2m + \frac{Jb}{a}\right)a(\ket{1} + \ket{7}) + \frac{J(a+c)}{2b}b(\ket{2} + \ket{3} + \ket{5} + \ket{6}) + \left(2m + \frac{2Jb}{c}\right)c\ket{4}
\end{align*}
We must now demand that this is a constant times the original vector, i.e. the explicit coefficients must be identical
\begin{align*}
-2m + \frac{Jb}{a} = \frac{J(a+c)}{2b} = 2m + \frac{2Jb}{c}
\end{align*}
There are two equations here, plus one which follows from the other two. We furthermore have normalization, which takes the form
\begin{align*}
2a^2 + 4b^2 + c^2 = 1
\end{align*}
Hence, we have three equations with three variables, which should be solvable. We introduce $ x = J/m $ and $ y = m/J = 1/x $, and write our equations as
\begin{align*}
x\left(\frac{b}{a} - \frac{a+c}{2b}\right) - 2 &= 0\\
x\left(\frac{b}{a} - \frac{2b}{c}\right) - 4 &= 0\\
2a^2 + 4b^2 + c^2 &= 1
\end{align*}
From the first two we derive
\begin{align*}
c &= \frac{2b^2}{a} - a - \frac{4b}{x}\\
\frac{1}{c} &= \frac{1}{2a} - \frac{2}{xb}
\end{align*}
We use this to get an equation with just $ a $ and $ b $
\begin{align*}
1 = \left(\frac{2b^2}{a} - a - \frac{4b}{x}\right)\left(\frac{1}{2a} - \frac{2}{xb}\right)
\end{align*}
We now change to $ y $ instead of $ x $ and we introduce $ p = b/a $. The equation then becomes
\begin{align*}
1 = \left(2p^2 - 1 - 4yp\right)\left(\frac{1}{2} - \frac{2y}{p}\right) = p^2 - 6yp + \frac{2y}{p} - \frac{1}{2} + 8y^2
\end{align*}
This can again be rewritten to
\begin{align*}
p^3 - 6yp^2 +\left(8y^2 - \frac{3}{2}\right)p + 2y = 0
\end{align*}
Such a third degree equation, $ p^3 + a_{2}p^2 + a_{1}p + a_{0} = 0 $, has three solutions, though two of them may be complex. To write the solutions we introduce
\begin{align*}
Q &= \frac{3a_{1} - a_{2}^2}{9} = -\frac{4}{3}y^2 - \frac{1}{2}\\
R &= \frac{9a_{2}a_{1} - 27a_{0} - 2a_{2}^3}{54} = \frac{1}{2}y\\
D &= Q^3 + R^2 = -\frac{64}{27}y^6 - \frac{8}{3}y^4 - \frac{3}{4}y^2 - \frac{1}{8}
\end{align*}
Here $ D $ is a manner of discriminant for the equation, and if it is negative then there are three real and unequal solutions to the equation. This is clearly the case, meaning that we will indeed find three different eigenvectors of $ H $ as we want. We additionally introduce the complex quantity
\begin{align*}
S = \sqrt[3]{R + \sqrt{D}}
\end{align*}
The solutions are then
\begin{align*}
p_{1} &= -\frac{1}{3}a_{2} + 2\Re(S) = 2y + 2\Re(S)\\
p_{2} &= -\frac{1}{3}a_{2} - \Re(S) - \sqrt{3}\Im(S) = 2y - \Re(S) - \sqrt{3}\Im(S)\\
p_{3} &= -\frac{1}{3}a_{2} - \Re(S) + \sqrt{3}\Im(S) = 2y - \Re(S) + \sqrt{3}\Im(S)
\end{align*}
which are all real. Remembering the results from before (and reintroducing the subscripts) we now have
\begin{align*}
b_{i} &= p_{i}a_{i}\\
c_{i} &= \left(2p_{i}^2 - 4yp_{i} - 1\right)a_{i}\\
2a_{i}^2 + 4b_{i}^2 + c_{i}^2 &= 1
\end{align*}
From this last one, we can derive
\begin{align*}
a_{i} = \frac{1}{\sqrt{4p_{i}^4 - 16yp_{i}^3 + 16y^2p_{i}^2 + 8yp_{i} + 3}}
\end{align*}
Hence, we have a set of $ a_{i},b_{i},c_{i} $ for each of the three $ p_{i} $. They have an energy of
\begin{align*}
E_{i} = -2m + \frac{Jb_{i}}{a_{i}} = J\left(p_{i} - 2y\right) = \begin{cases}
2J\Re(S), & i = 1\\
J\left(-\Re(S) - \sqrt{3}\Im(S)\right), & i = 2\\
J\left(-\Re(S) + \sqrt{3}\Im(S)\right), & i = 3
\end{cases}
\end{align*}
This gives us the energies and coefficients of the three eigenvectors of $ H $, which are on the form of $ \ket{\psi_{i}} $.

Let us now do the same with $ \ket{\psi_{j}'} $. Here we find
\begin{align*}
H&\left[a'(\ket{1} - \ket{7}) + b'(\ket{2} + \ket{3} - \ket{5} - \ket{6})\right]\\
&= 2m\left(-a'(\ket{1} - \ket{7})\right) + \frac{J}{2}\left(a'(\ket{2} + \ket{3} - \ket{5} - \ket{6} + b'(2\ket{1} + 2\ket{7}))\right)\\
&= \left(-2m + \frac{Jb'}{a'}\right)a'(\ket{1} - \ket{7}) + \frac{Ja'}{2b'}b'(\ket{2} + \ket{3} - \ket{5} - \ket{6})
\end{align*}
We thus simply get the equation
\begin{align*}
\frac{Jb'}{a'} - 2m = \frac{Ja'}{2b'}
\end{align*}
as well as normalization
\begin{align*}
2(a')^2 + 4(b')^2 = 1
\end{align*}
Introducing again $ p' = b'/a' $, the first equation becomes
\begin{align*}
(p')^2 - 2yp' - \frac{1}{2} = 0
\end{align*}
This has the solutions
\begin{align*}
p_{j}' = \frac{2y \pm \sqrt{4y^2 + 2}}{2} = y \pm \sqrt{y^2 + \frac{1}{2}}
\end{align*}
These are two real numbers, as we would want. Using the normalization equation we can now find
\begin{align*}
a_{j}' = \frac{1}{\sqrt{4p_{j}^2 + 2}}
\end{align*}
Thus we again have a set of $ a_{j}',b_{j}' $ for the two values of $ p_{j}' $. The corresponding eigenvectors have the energies
\begin{align*}
E_{j}' = -2m + Jp_{j}' = J(p_{j}' - 2y) = \begin{cases}
J\left(- y + \sqrt{y^2 + \frac{1}{2}}\right), & j = 1\\
J\left(- y - \sqrt{y^2 + \frac{1}{2}}\right), & j = 2
\end{cases}
\end{align*}

We thus have seven eigenvectors of the Hamiltonian arranged in sets according to their eigenvalue with respect to the parity symmetry. These eigenvectors are for the Hamiltonian both before and after the quench, though of course the coefficients change as $ J/m $ changes sign. Hence, we could now determine numerically which vector is the ground state for the different values of positive $ x $ (or $ y $) and then calculate its overlap with the eigenvectors after the quench, $ x\rightarrow -x $. Notice, that the quench does not affect the parity symmetry, and hence a vector from one of the sets will have zero overlap with vectors from others sets even after the quench. Thus, time evolution becomes quite simple as there will be at most three vectors relevant after the quench. We could again write down the Loschmidt amplitude and our order parameter, but in this case we do not believe the zeros of these to be analytically derivable.

\subsection{Regression formulae} \label{app:regression}
It follows from the charge conjugation symmetry, $ C $, of the system, which has the effect $ \sigma_{n}^{z} \xrightarrow{C} -\sigma_{n+1}^{z} $, that
\begin{align*}
\expval{\sigma_{n}^{z}} = -\expval{\sigma_{n+1}^{z}}
\end{align*}
or indeed
\begin{align*}
\expval{\sigma_{n}^{z}} = (-1)^{n}\expval{\sigma_{0}^{z}}
\end{align*}
This can be found in a different way, which reveals some other attributes of the system as well. The system has a rotation symmetry which is simply the square of the conjugation symmetry, $ R = C^{2} $, which acts as
\begin{align*}
\sigma_{n}^{z} \leftrightarrow \sigma_{n+2}^{z} ,&\qquad \sigma_{n}^{\pm} \leftrightarrow \sigma_{n+2}^{\pm}\\
S_{n,n+1}^{z} \leftrightarrow S_{n+2,n+3}^{z} ,&\qquad S_{n,n+1}^{\pm} \leftrightarrow S_{n+2,n+3}^{\pm}
\end{align*}
Because of this rotation symmetry, if we initialize in a likewise symmetric state (which is the case for our quench protocol), there will be the relation $ \expval{\sigma_{n}^{z}} = \expval{\sigma_{n+2}^{z}} $. Hence, all particle sites appear to be in an identical state, in some sense, and likewise for all antiparticle sites. We furthermore have conservation of the difference in the number of particles and antiparticles, expressed as $ [D,H] = 0 $, where the difference in the number of particles and antiparticles $ D $ is
\begin{align*}
D = N_{P} - N_{A} = \sum_{n}(-1)^{n}\frac{1 - (-1)^{n}\sigma_{n}^{z}}{2} = -\sum_{n}\frac{\sigma_{n}^{z}}{2}
\end{align*}
We have here used that the number or particles at site $ n $ is $ \frac{1 - (-1)^{n}\sigma_{n}^{z}}{2} $. This follows from the Jordan-Wigner transformation, and the fact that we have staggered mass. In other words, the magnetization $ \sum_{n}\sigma_{n}^{z} $ is conserved. From $ G_{n} = \frac{e}{2}\left(S_{n-1,n}^{z} - S_{n,n+1}^{z} + \sigma_{n}^{z} - (-1)^{n}\right) = 0 $ in the subspace we work with, we have
\begin{align*}
0 = \sum_{n}G_{n} = \frac{e}{2}\sum_{n}\sigma_{n}^{z}
\end{align*}
which follows from the fact that we have periodic boundaries and thus the electrical fields have cancelled pairwise around the lattice. Thus the magnetization is generally conserved, and we are in fact looking at the sector with vanishing magnetization. Since $ \expval{\sigma_{n}^{z}} = \expval{\sigma_{n+2}^{z}} $ we now have
\begin{align*}
0 = \sum_{n}\expval{\sigma_{n}^{z}} = \frac{N}{2}(\expval{\sigma_{0}^{z}} + \expval{\sigma_{1}^{z}})
\end{align*}
from which it follows $ \expval{\sigma_{n}^{z}} = -\expval{\sigma_{n+1}^{z}} $, or simply $ \expval{\sigma_{0}^{z}} = -\expval{\sigma_{1}^{z}} $, or indeed $ \expval{\sigma_{n}^{z}} = (-1)^{n}\expval{\sigma_{0}^{z}} $, i.e. particle and antiparticle sites have opposite expectation value of $ \sigma^{z} $, meaning that they in fact have the same expectation for the number of particles/antiparticles at the site. This means exactly that there is an equal number of particles and antiparticles. This is consistent with $ D = 0 $ being conserved, and is generally a consequence of working in the $ G_{n}\ket{\textup{phys}} = 0 $ subspace with periodic boundaries. Similarly we find that $ \expval{S_{n,n+1}^{z}} = -\expval{S_{n+1,n+2}^{z}} $, or $ \expval{S_{n,n+1}^{z}} = (-1)^{n}\expval{S_{0,1}^{z}} $. From this and Gauss' law it follows that
\begin{align*}
\expval{S_{n,n+1}^{z}} &= \frac{\expval{\sigma_{n}^{z}} - (-1)^{n}}{2} = (-1)^{n}\frac{\expval{\sigma_{0}^{z}} - 1}{2}
\end{align*}
Hence, the expectation values of Pauli-Z operators in the system are all simply related. Expectation value of X-, and Y-operators vanish, as they are completely gauge variant.

\subsection{All Loschmidt echo rate functions} \label{app:allLA}
\begin{figure}
	\includegraphics[height=0.5\textheight]{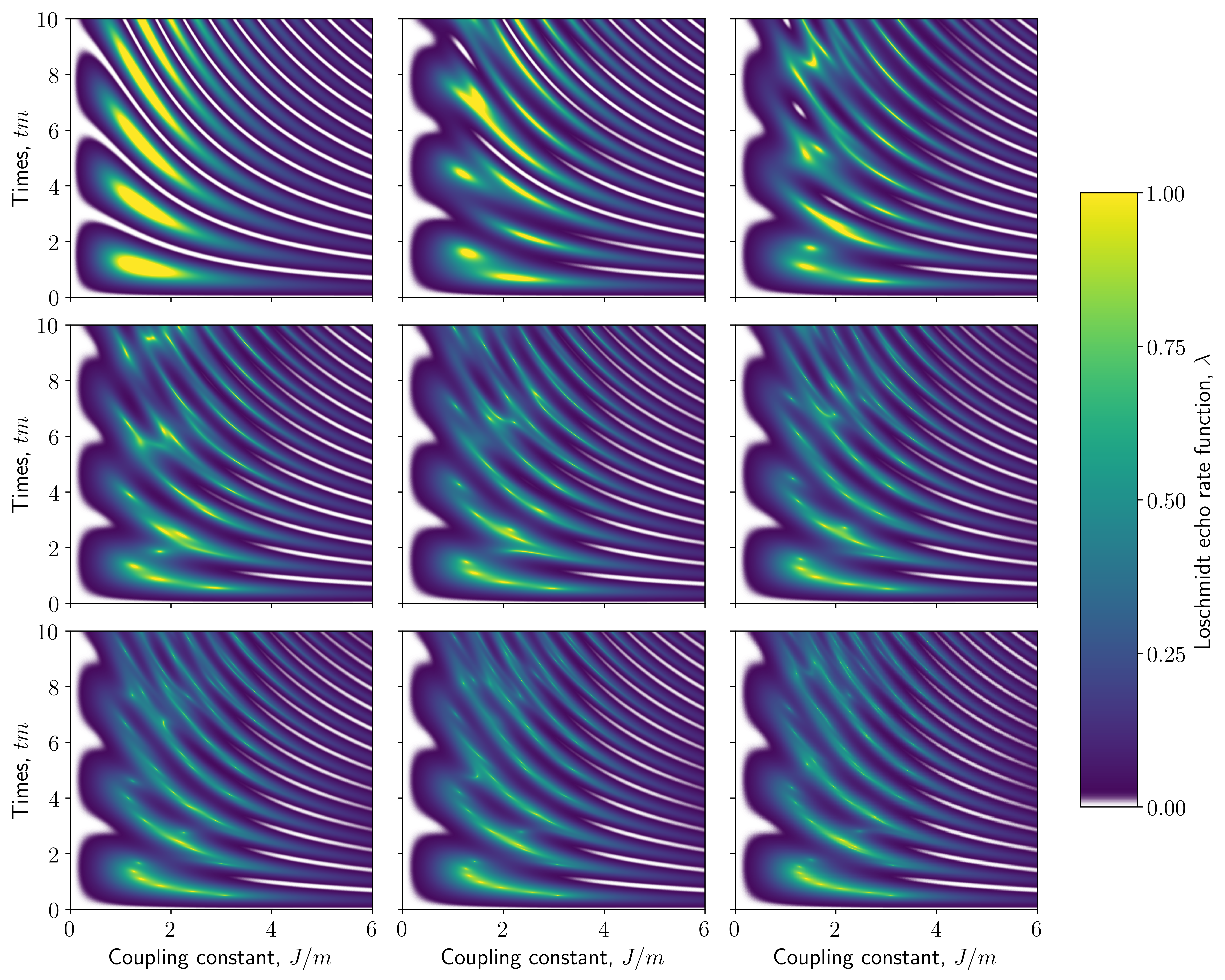}
	\caption{\label{fig:allLA} The Loschmidt echo rate function $ \lambda(t) = -\frac{1}{N}\log\left[\mathcal{L}(t)\right] $ for all systems sizes considered. $ N $ takes on the values $ 2,4,...,18 $ from the top left to the bottom right, increasing as you go from left to right and top to bottom.}
	\vfill
	\includegraphics[height=0.5\textheight]{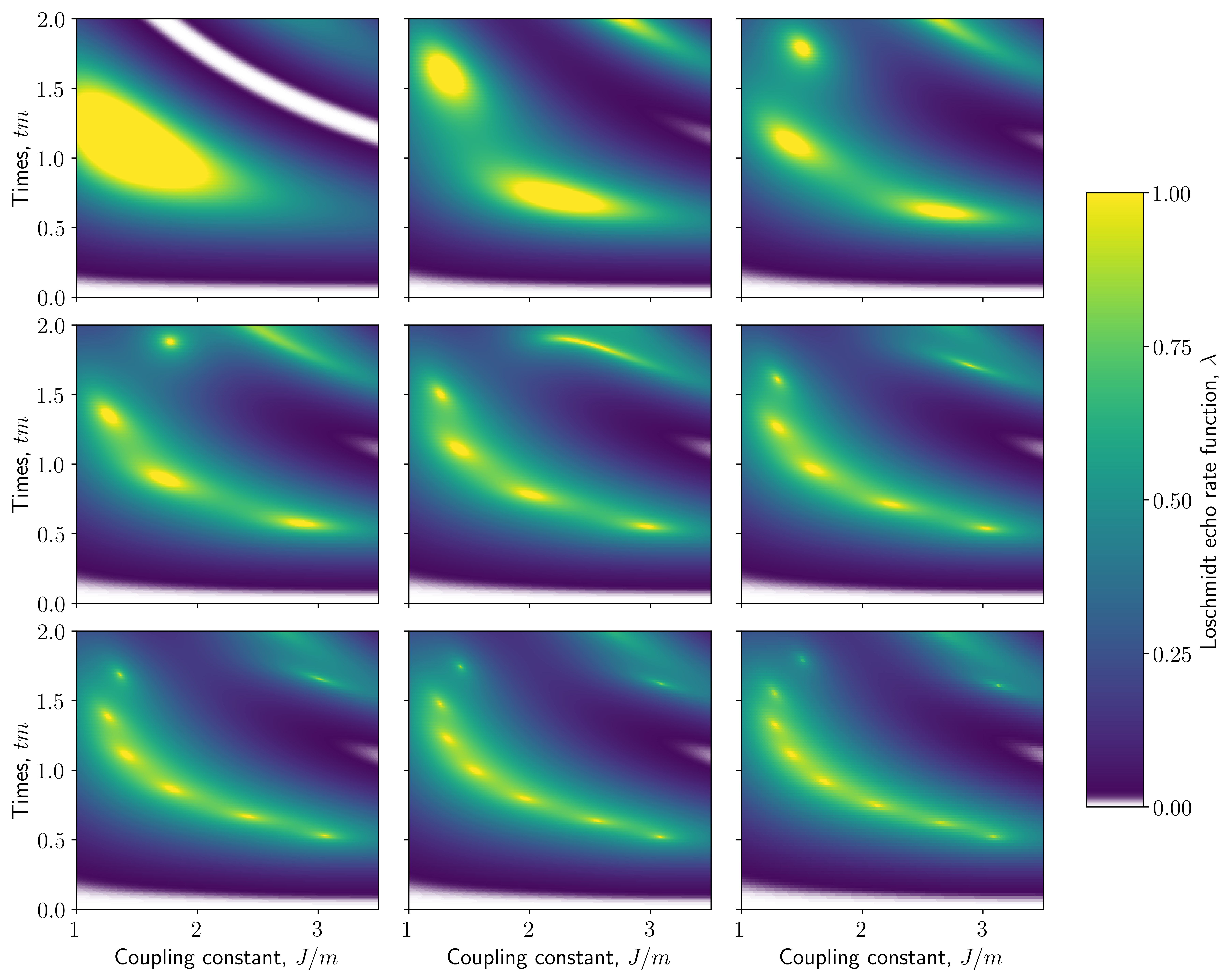}
	\caption{\label{fig:allLAzoom} Zoom-in of the area $ J/m \in [1,3.5] $, $ tm \in [0,2] $ for the data shown in \cref{fig:allLA}. The rate function has been cut off at $ 1 $, but of course attains higher values as it diverges near the zeros of $ \mathcal{L}(t) $.}
\end{figure}

In \cref{fig:allLA} can be seen the Loschmidt echo rate function $ \lambda(t) = -\frac{1}{N}\log\left[\mathcal{L}(t)\right] $ for all systems sizes considered, $ N = 2,4,...,18 $. The system size increases from left to right, and top to bottom. Clearly there is pattern quickly emerging and repeating itself. Furthermore, zooming in on the area $ J/m \in [1,3.5] $, $ tm \in [0,2] $, see \cref{fig:allLAzoom}, we can see the hook-shaped curve of zeros forming as $ N $ increases.
\clearpage

\subsection{Deriving the circuit Hamiltonian} \label{app:circuitH}
\begin{figure*}
	\includegraphics[width=0.7\textwidth]{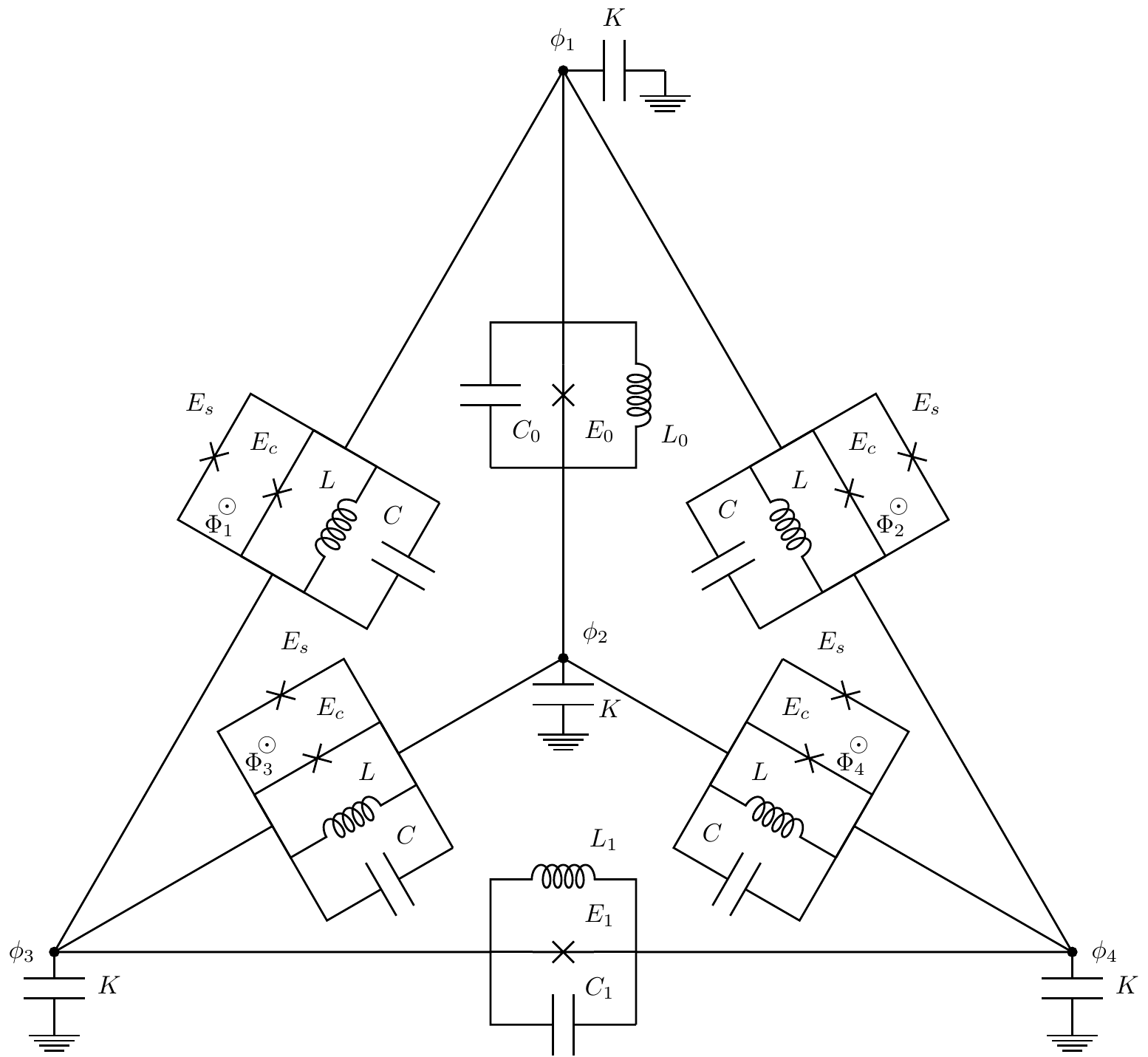}
	\caption{\label{fig:tetra} The circuit for implementing the XXX-coupling in a more general form than the used in the main text. We have here included linear inductors on all branches for the sake of generality.}
\end{figure*}
The only way (seemingly) to get a direct triple odd interaction like the XXX-coupling using capacitors, linear inductors, and Josephson junctions in a superconducting circuit, is to shift the position of the minimum of the potential away from the origin. Otherwise, all circuit elements contribute with terms that are even in the number of flux coordinates. This shifting is done by introducing (constant) drives of the potential terms, such that the different functions and their minima are shifted relative to each other. Shifting the minimum and expanding around it makes no difference for capacitive and linear inductive terms. Capacitive terms involve the derivative and so do not care about a constant shift, while linear inductive terms can only give first and second order terms (in the fluxes), and all first order terms disappear when we expand around an extremal point, like the potential minimum. Hence, we only get new and interesting terms from the Josephson junction. This in turn give us a myriad of messy terms. Consider the following cosine term, in the context of a larger system, where the coordinates are chosen such that the minimum of the full potential is at the origin
\begin{align*}
\cos(\phi + \alpha) = \cos(\alpha) - \sin(\alpha)\phi - \frac{1}{2}\cos(\alpha)\phi^2 + \frac{1}{6}\sin(\alpha)\phi^3 + \frac{1}{24}\cos(\alpha)\phi^4 + \mathcal{O}(\phi^{5})
\end{align*}
where $ \phi $ represent any combination of fluxes and $ \alpha $ is a shift combining different (constant) drives in the cosine, and shifts of the coordinates/potential. We are interested in the $ \phi^3 $ term for our XXX interaction. We seek a circuit whose symmetry ensures that all the undesired terms cancel. As we seek a three-body interaction we will work with a circuit of four nodes, all-to-all connected, and in the basis of the eigenvectors of the capacitance matrix, i.e. the eigenmodes of the capacitive network. Working in this basis removes interactions from capacitors, reducing the complexity by a lot, and also allows for more than two coordinates to be involved in some term.

In \cref{fig:tetra} can be seen our circuit in a more general form than was used in the main text. We have her included linear inductors on all branches for the sake of studying the most general case of the circuit. Notice the symmetry of the circuit parameters. Defining the original node flux coordinates $ \vec{\phi} = (\phi_{1},\phi_{2},\phi_{3},\phi_{4}) $, the circuit has the following capacitance matrix
\begin{align*}
\mathcal{C} = \mat{2C + C_{0} + K & -C_{0} & -C & -C \\ -C_{0} & 2C + C_{0} + K & -C & -C \\ -C & -C & 2C + C_{1} + K & -C_{1} \\ -C & -C & -C_{1} & 2C + C_{1} + K}
\end{align*}
We define new coordinates $ \vec{\psi} = (\psi_{CM},\psi_{1},\psi_{g},\psi_{2}) $, and make a change of coordinates according to
\begin{align*}
\vec{\phi} = \mat{1 & 1 & \frac{1}{2} & 0 \\ 1 & -1 & \frac{1}{2} & 0 \\ 1 & 0 & -\frac{1}{2} & 1 \\ 1 & 0 & -\frac{1}{2} & -1}\vec{\psi}
\end{align*}
Notice that the coordinates $ \psi_{1} $ and $ \psi_{2} $ are simply $ \phi_{2} - \phi_{3} $ and $ \phi_{1} - \phi_{4} $, making the terms corresponding to these branches affect only these modes individually, while $ \phi_{g} $ will be affected by the four branches with identical circuit parameters. This transformation diagonalizes the capacitance matrix to
\begin{align*}
\mathcal{K} = \mat{4K & 0 & 0 & 0 \\ 0 & 4C + 4C_{0} + 2K & 0 & 0 \\ 0 & 0 & 4C + K & 0 \\ 0 & 0 & 0 & 4C + 4C_{1} + 2K}
\end{align*}
It is important here that the capacitance to ground is identical for all nodes, as it will otherwise introduce interactions. 

Now let us consider the Josephson junctions on the branches pertaining to the $ \psi_{g} $ mode. Taking either of the four junctions with subscript $ c $ or $ s $, we find that they can be neatly summed as follows
\begin{align*}
&\cos(\phi_{1} - \phi_{2}) + \cos(\phi_{1} - \phi_{3}) + \cos(\phi_{2} - \phi_{4}) + \cos(\phi_{3} - \phi_{4})\\
&= \cos(-\psi_{1} + \psi_{g} + \psi_{2}) + \cos(\psi_{1} + \psi_{g} + \psi_{2}) + \cos(\psi_{1} - \psi_{g} + \psi_{2}) + \cos(-\psi_{1} - \psi_{g} + \psi_{2})\\
&= 2\cos(\psi_{1})\cos(\psi_{g} + \psi_{2}) + 2\cos(\psi_{1})\cos(\psi_{g} - \psi_{2})\\
&= 4\cos(\psi_{1})\cos(\psi_{g})\cos(\psi_{2})
\end{align*}
Hence, these four junctions result in some very neat interactions between the modes and contributions to their energies and anharmonicities. We choose the spanning tree and the exact layout of the circuit such that the external fluxes, $ \Phi_{i} $, $ i = 1,2,3,4 $, enter in the junctions with subscript $ s $. We then choose the $ \Phi_{i} $'s as sums of three other fluxes $ \Psi_{j} $, $ j = 1,g,2 $, where the sign of each $ \Psi_{j} $ in each cosine term must be the same as the corresponding $ \psi_{j} $. For example, if we include $ \Phi_{1} $ in the first cosine in the above calculation, $ \cos(\phi_{1} - \phi_{2} + \Phi_{1}) = \cos(-\psi_{1} + \psi_{g} + \psi_{2} + \Phi_{1}) $, then we can see that choosing $ \Phi_{1} = -\Psi_{1} + \Psi_{g} + \Psi_{2} $ results in the each $ \psi_{j} $ having a correspond $ \Psi_{j} $ with the same sign. Doing the same with all four cosines, we essentially shift $ \psi_{j} \rightarrow \psi_{j} + \Psi_{j} $. The results of the above calculation would then be
\begin{align*}
4\cos(\psi_{1} + \Psi_{1})\cos(\psi_{g} + \Psi_{g})\cos(\psi_{g} + \Psi_{2})
\end{align*}
Hence, we can use these to tune the interactions pertaining to $ E_{s} $. In particular choosing $ \Psi_{j} = -\frac{\pi}{2} $ for all $ j $, we get a triple sine interaction. The triple sine will have its minimum at $ \psi_{j} = 0 $, as all other terms already did, and so this shift will only affect this particular term of the Hamiltonian. Expanding to fourth order, it yields exactly a $ \psi_{1}\psi_{g}\psi_{2} $ interaction corresponding to the desired XXX coupling. Likewise for the four linear inductors we find
\begin{align*}
&(\phi_{1} - \phi_{2})^2 + (\phi_{1} - \phi_{3})^2 + (\phi_{2} - \phi_{4})^2 + (\phi_{3} - \phi_{4})^2\\
&= (-\psi_{1} + \psi_{g} + \psi_{2})^2 + (\psi_{1} + \psi_{g} + \psi_{2})^2 + (\psi_{1} - \psi_{g} + \psi_{2})^2 + (-\psi_{1} - \psi_{g} + \psi_{2})^2\\
&= 4(\psi_{1}^2 + \psi_{g}^2 + \psi_{2}^2)
\end{align*}
Thus they do no not introduce any interaction, as mentioned above, but merely add to the harmonicity and energy of the modes. The cosines pertaining to $ E_{c} $ will yield ZZ interaction, which we would rather be without, but they alone contribute to $ \psi_{g} $'s anharmonicity. If we instead had only the identical linear inductors, we would get similar contributions to the mode energies, but no interactions and also no contribution to anharmonicities. In this case $ \psi_{g} $ would be a pure harmonic oscillator, but we would have fewer interactions. The other two modes get their anharmonicities from the $ E_{1} $ and $ E_{2} $ junctions. The inductors turn out to not be necessary but could be used to tune the energies of $ \psi_{1} $ and $ \psi_{2} $ if needed. In order for the XXX interaction to work the modes need to be detuned in a proper way, as described in the main text. 

With these calculations and in these coordinates, the Hamiltonian of the circuit becomes
\begin{align*}
H &= \frac{1}{2}\frac{1}{4C + 4C_{1} + 2K}p_{1}^2 + \frac{1}{2}\frac{1}{4C + K}p_{2}^2 + \frac{1}{2}\frac{1}{4C + 4C_{2} + 2K}p_{2}^2\\
&\hspace{11.5pt} + \left(4E_{L_{1}} + 4E_{L}\right)\psi_{1}^2 - E_{1}\cos(2\psi_{1})  + 4E_{L}\psi_{g}^2 + \left(4E_{L_{2}} + 4E_{L}\right)\psi_{2}^2 - E_{2}\cos(2\psi_{2}) \\
&\hspace{11.5pt} - 4E_{c}\cos\psi_{1}\cos\psi_{g}\cos\psi_{2}\\
&\hspace{11.5pt} - 4E_{s}\sin\psi_{1}\sin\psi_{g}\sin\psi_{2}
\end{align*}
where $ E_{L} = \frac{1}{2L} $ and likewise for the others. We will not be performing the usual expansion to fourth order before recasting and truncation, but will rather use a method we developed for in a previous paper, with which the trigonometric functions can be truncated without performing any approximation. To do so we perform the recasting in terms of harmonic oscillator modes first. We introduce bosonic creation and annihilation operators, $ a_{j}^{\dagger},a_{j} $, via
\begin{align*}
\psi_{j} &= \frac{r_{j}}{\sqrt{2}}(a_{j}^{\dagger} + a_{j})\\
p_{j} &= i\frac{1}{\sqrt{2}r_{j}}(a_{j}^{\dagger} - a_{j})
\end{align*}
Here $ r_{j} $ is parameter which quantifies the "size" of the $ \psi_{j} $ coordinate. When performing the usual expansion to fourth order, one is in fact expanding to fourth order in $ r_{j}/\sqrt{2} $, and so the usual transmon regime (of circuit parameters) is when $ r_{j}/\sqrt{2} $ is very small. The $ r_{j} $ are defined as
\begin{align*}
r_{j} = \left(\frac{K_{p_{j}}}{K_{\psi_{j}}}\right)^{1/4}
\end{align*}
where $ K_{p_{j}} $ and $ K_{\psi_{j}} $ are the coefficients of the $ p_{j}^2 $ and the $ \psi_{j}^2 $ terms, respectively, in the Hamiltonian, when it is expanded in powers of $ \psi_{j} $. If we expand $ H $ to just second order to find these terms, we get
\begin{align*}
H &= \frac{1}{2}\frac{1}{4C + 4C_{1} + 2K}p_{1}^2 + \frac{1}{2}\frac{1}{4C + K}p_{2}^2 + \frac{1}{2}\frac{1}{4C + 4C_{2} + 2K}p_{2}^2\\
&\hspace{11.5pt} + \left(4E_{L_{1}} + 4E_{L}\right)\psi_{1}^2 - E_{1}\left(1 + 2\psi_{1}^2\right)  + 4E_{L}\psi_{g}^2 + \left(4E_{L_{2}} + 4E_{L}\right)\psi_{2}^2 - E_{2}\left(1 + 2\psi_{2}^2\right) \\
&\hspace{11.5pt} - 4E_{c}\left(1 + \frac{1}{2}\psi_{1}^2\right)\left(1 + \frac{1}{2}\psi_{g}^2\right)\left(1 + \frac{1}{2}\psi_{2}^2\right)\\
&\hspace{11.5pt} - 4E_{s}\psi_{1}\psi_{g}\psi_{2} + \mathcal{O}(\psi_{j}^3)\\
	&= \frac{1}{2}\frac{1}{4C + 4C_{1} + 2K}p_{1}^2 + \frac{1}{2}\frac{1}{4C + K}p_{2}^2 + \frac{1}{2}\frac{1}{4C + 4C_{2} + 2K}p_{2}^2\\
	&\hspace{11.5pt} + \left(4E_{L_{1}} + 4E_{L} - 2E_{1} - 2E_{c}\right)\psi_{1}^2  + \left(4E_{L} - 2E_{c}\right)\psi_{g}^2 + \left(4E_{L_{2}} + 4E_{L} - 2E_{2} - 2E_{c}\right)\psi_{2}^2 + \mathcal{O}(\psi_{j}^3)
\end{align*}
where we have ignored constant terms and all terms involving more than two $ \psi_{j} $ are hidden in $ \mathcal{O}(\psi_{j}^3) $. From this we can see that for example
\begin{align*}
K_{p_{1}} &= \frac{1}{2}\frac{1}{4C + 4C_{1} + 2K}\\
K_{\phi_{1}} &= 4E_{L_{1}} + 4E_{L} - 2E_{1} - 2E_{c}
\end{align*}
and similarly for the other coordinates. With these we can define
\begin{align*}
r_{1} &= \left[8(2C + 2C_{1} + K)\left(2E_{L_{1}} + 2E_{L} + E_{1} + E_{c}\right)\right]^{-1/4}\\
r_{g} &= \left[4(4C + K)(2E_{L} + E_{c})\right]^{-1/4}\\
r_{2} &= \left[8(2C + 2C_{2} + K)\left(2E_{L_{2}} + 2E_{L} + E_{2} + E_{c}\right)\right]^{-1/4}
\end{align*}
We will be writing the trigonometric functions in terms of complex exponentials, as this helps us truncate them later on. Thus, we can write the Hamiltonian in terms of the creation and annihilation operators as
\begin{align*}
H &= -\frac{1}{2}\frac{1}{4C + 4C_{1} + 2K}\frac{1}{2r_{1}^2}(a_{1}^{\dagger} - a_{1})^2 - \frac{1}{2}\frac{1}{4C + K}\frac{1}{2r_{g}^2}(a_{g}^{\dagger} - a_{g})^2 - \frac{1}{2}\frac{1}{4C + 4C_{2} + 2K}\frac{1}{2r_{2}^2}(a_{2}^{\dagger} - a_{2})^2\\
&\hspace{11.5pt} + 4\left(E_{L_{1}} + E_{L}\right)\frac{r_{1}^2}{2}(a_{1}^{\dagger} + a_{1})^2 - E_{1}\frac{1}{2}\left(e^{i\sqrt{2}r_{1}(a_{1}^{\dagger} + a_{1})} + e^{-i\sqrt{2}r_{1}(a_{1}^{\dagger} + a_{1})}\right)\\
&\hspace{11.5pt} + 4E_{L}\frac{r_{g}^2}{2}(a_{g}^{\dagger} + a_{g})^2\\
&\hspace{11.5pt} + 4\left(E_{L_{2}} + E_{L}\right)\frac{r_{2}^2}{2}(a_{2}^{\dagger} + a_{2})^2 - E_{2}\frac{1}{2}\left(e^{i\sqrt{2}r_{2}(a_{2}^{\dagger} + a_{2})} + e^{-i\sqrt{2}r_{2}(a_{2}^{\dagger} + a_{2})}\right) \\
&\hspace{11.5pt} - \frac{1}{2}E_{c}\left(e^{ir_{1}(a_{1}^{\dagger} + a_{1})/\sqrt{2}} + e^{-ir_{1}(a_{1}^{\dagger} + a_{1})/\sqrt{2}}\right)\left(e^{ir_{g}(a_{g}^{\dagger} + a_{g})/\sqrt{2}} + e^{-ir_{g}(a_{g}^{\dagger} + a_{g})/\sqrt{2}}\right)\left(e^{ir_{2}(a_{2}^{\dagger} + a_{2})/\sqrt{2}} + e^{-ir_{2}(a_{2}^{\dagger} + a_{2})/\sqrt{2}}\right)\\
&\hspace{11.5pt} - \frac{i}{2}E_{s}\left(e^{ir_{1}(a_{1}^{\dagger} + a_{1})/\sqrt{2}} - e^{-ir_{1}(a_{1}^{\dagger} + a_{1})/\sqrt{2}}\right)\left(e^{ir_{g}(a_{g}^{\dagger} + a_{g})/\sqrt{2}} - e^{-ir_{g}(a_{g}^{\dagger} + a_{g})/\sqrt{2}}\right)\left(e^{ir_{2}(a_{2}^{\dagger} + a_{2})/\sqrt{2}} - e^{-ir_{2}(a_{2}^{\dagger} + a_{2})/\sqrt{2}}\right)\\
\end{align*}
In order to truncate this Hamiltonian to any number of desired levels in each mode, we must calculate the corresponding matrix elements of each operator in the Hamiltonian. The matrix elements are calculated in the Fock basis, and so operators like $ (a_{j}^{\dagger} \pm a_{j})^2 $ are easily dealt with. Indeed, it is the trigonometric functions, now in terms of exponentials, that require some work. Here we recognize that the exponentials are on the form of a displacement operator, known from harmonic oscillators
\begin{align*}
D(\xi) = e^{\xi a^{\dagger} - \xi^{*}a}
\end{align*}
Here $ \xi $ is some complex number, and the fundamental property of the displacement operator is that $ D(\xi)\ket{0} = \ket{\xi} $, where $ \ket{0} $ is the ground state of a harmonic oscillator, and $ \ket{\xi} $ is a coherent state. The coherent states can in turn be written out as a series, specifically 
\begin{align*}
D(\xi)\ket{0} = e^{-\vert\xi\vert^2/2}\sum_{j = 0}^{\infty}\frac{\xi^{j}}{\sqrt{j!}}\ket{j} 
\end{align*}
Furthermore, we know the commutations relations of the displacement operator
\begin{align*}
D(\xi)a^{\dagger} = (a^{\dagger} - \xi^{*})D(\xi) 
\end{align*}
With these relations and the definition $ \ket{n} = \frac{1}{\sqrt{n!}}(a^{\dagger})^n\ket{0} $, we can calculate the needed matrix elements. Let us consider the general case of $ D(ik) = e^{ik(a^{\dagger} + a)} $. Let us also note that since the operator $ a^{\dagger} + a $ is symmetric, the operator $ D(ik) = e^{ik(a^{\dagger} + a)} $ is also symmetric, i.e. $ \mel{n}{e^{ik(a^{\dagger} + a)}}{m} = \mel{m}{e^{ik(a^{\dagger} + a)}}{n} $. We find
\begin{align*}
\begin{split}
\mel{0}{e^{ik(a^{\dagger} + a)}}{0} &= e^{-k^2/2}\\
\mel{1}{e^{ik(a^{\dagger} + a)}}{0} &= ike^{-k^2/2}\\
\mel{1}{e^{ik(a^{\dagger} + a)}}{1} &= (1 - k^2)e^{-k^2/2}
\end{split}
\end{align*}
With these we can write the operator $ e^{ik(a^{\dagger} + a)} $ truncated to two levels in terms of Pauli matrices and the identity
\begin{align*}
M_{2}\left[e^{ik(a^{\dagger} + a)}\right] = \left[\left(1 - \frac{k^2}{2}\right) + \frac{k^2}{2}\sigma^{z} + ik\sigma^{x}\right]e^{-k^2/2} 
\end{align*}
where $ M_{n}[\cdot] $ is the $ n\times n $ matrix representation of an operator. It is now just a matter of replacing exponentials with this expression and reducing the result, i.e. we don't need to perform the truncation calculation more than once, from now on it is just a mechanical method of inserting the right expressions into our equations. Likewise we can find the following two-level expressions for the remaining operators
\begin{align*}
M_{2}[a^{\dagger}a] &= \frac{1}{2}\left(1 - \sigma^{z}\right)\\
M_{2}[a^{\dagger} - a] &= -i\sigma^{y}\\
M_{2}[(a^{\dagger} - a)^2] &= -2 + \sigma^{z}\\
M_{2}[a^{\dagger} + a] &= \sigma^{x}\\
M_{2}[(a^{\dagger} + a)^2] &= 2 - \sigma^{z}
\end{align*}
This procedure is easily expanded to include more levels in the truncation. Performing this exact truncation of the trigonometric functions can reduce the numerical difficulty of simulating the dynamics of a circuit while including higher levels. Such a study is often relevant to check the actual effect of the higher levels of a circuit rather than only studying the approximate dynamics of system truncated to two levels in each mode. 

We can thus calculate the two level Hamiltonian exactly. We do so
\begin{align*}
H &= \frac{1}{2}\frac{1}{4C + 4C_{1} + 2K}p_{1}^2 + \frac{1}{2}\frac{1}{4C + K}p_{2}^2 + \frac{1}{2}\frac{1}{4C + 4C_{2} + 2K}p_{2}^2\\
&\hspace{11.5pt} + \left(\frac{2}{L_{1}} + \frac{2}{L}\right)\psi_{1}^2 - E_{1}\cos(2\psi_{1})  + \frac{2}{L}\psi_{g}^2 + \left(\frac{2}{L_{2}} + \frac{2}{L}\right)\psi_{2}^2 - E_{2}\cos(2\psi_{2}) \\
&\hspace{11.5pt} - 4E_{c}\cos\psi_{1}\cos\psi_{g}\cos\psi_{2}\\
&\hspace{11.5pt} - 4E_{s}\sin\psi_{1}\sin\psi_{g}\sin\psi_{2}\\
&= -\frac{1}{2}\frac{1}{4C + 4C_{1} + 2K}\frac{1}{2r_{1}^2}(a_{1}^{\dagger} - a_{1})^2 - \frac{1}{2}\frac{1}{4C + K}\frac{1}{2r_{g}^2}(a_{g}^{\dagger} - a_{g})^2 - \frac{1}{2}\frac{1}{4C + 4C_{2} + 2K}\frac{1}{2r_{2}^2}(a_{2}^{\dagger} - a_{2})^2\\
&\hspace{11.5pt} + 4\left(E_{L_{1}} + E_{L}\right)\frac{r_{1}^2}{2}(a_{1}^{\dagger} + a_{1})^2 - E_{1}\frac{1}{2}\left(e^{i\sqrt{2}r_{1}(a_{1}^{\dagger} + a_{1})} + e^{-i\sqrt{2}r_{1}(a_{1}^{\dagger} + a_{1})}\right)\\
&\hspace{11.5pt} + 4E_{L}\frac{r_{g}^2}{2}(a_{g}^{\dagger} + a_{g})^2\\
&\hspace{11.5pt} + 4\left(E_{L_{2}} + E_{L}\right)\frac{r_{2}^2}{2}(a_{2}^{\dagger} + a_{2})^2 - E_{2}\frac{1}{2}\left(e^{i\sqrt{2}r_{2}(a_{2}^{\dagger} + a_{2})} + e^{-i\sqrt{2}r_{2}(a_{2}^{\dagger} + a_{2})}\right) \\
&\hspace{11.5pt} - \frac{1}{2}E_{c}\left(e^{ir_{1}(a_{1}^{\dagger} + a_{1})/\sqrt{2}} + e^{-ir_{1}(a_{1}^{\dagger} + a_{1})/\sqrt{2}}\right)\left(e^{ir_{g}(a_{g}^{\dagger} + a_{g})/\sqrt{2}} + e^{-ir_{g}(a_{g}^{\dagger} + a_{g})/\sqrt{2}}\right)\left(e^{ir_{2}(a_{2}^{\dagger} + a_{2})/\sqrt{2}} + e^{-ir_{2}(a_{2}^{\dagger} + a_{2})/\sqrt{2}}\right)\\
&\hspace{11.5pt} - \frac{i}{2}E_{s}\left(e^{ir_{1}(a_{1}^{\dagger} + a_{1})/\sqrt{2}} - e^{-ir_{1}(a_{1}^{\dagger} + a_{1})/\sqrt{2}}\right)\left(e^{ir_{g}(a_{g}^{\dagger} + a_{g})/\sqrt{2}} - e^{-ir_{g}(a_{g}^{\dagger} + a_{g})/\sqrt{2}}\right)\left(e^{ir_{2}(a_{2}^{\dagger} + a_{2})/\sqrt{2}} - e^{-ir_{2}(a_{2}^{\dagger} + a_{2})/\sqrt{2}}\right)\\
&= \frac{1}{2}\frac{1}{4C + 4C_{1} + 2K}\frac{1}{2r_{1}^2}(2 - \sigma_{1}^{z}) + \frac{1}{2}\frac{1}{4C + K}\frac{1}{2r_{g}^2}(2 - \sigma_{g}^{z}) + \frac{1}{2}\frac{1}{4C + 4C_{2} + 2K}\frac{1}{2r_{2}^2}(2 - \sigma_{2}^{z})\\
&\hspace{11.5pt} + 2\left(E_{L_{1}} + E_{L}\right)r_{1}^2(2 - \sigma_{1}^{z}) - E_{1}\left[\left(1 - r_{1}^2\right) + r_{1}^2\sigma_{1}^{z}\right]e^{-r_{1}^2}\\
&\hspace{11.5pt} + 4E_{L}\frac{r_{g}^2}{2}(2 - \sigma_{g}^{z})\\
&\hspace{11.5pt} + 2\left(E_{L_{2}} + E_{L}\right)r_{2}^2(2 - \sigma_{2}^{z}) - E_{2}\left[\left(1 - r_{2}^2\right) + r_{2}^2\sigma_{2}^{z}\right]e^{-r_{2}^2} \\
&\hspace{11.5pt} - \frac{1}{2}E_{c}\left[\left(2 - \frac{r_{1}^2}{2}\right) + \frac{r_{1}^2}{2}\sigma_{1}^{z}\right]\left[\left(2 - \frac{r_{g}^2}{2}\right) + \frac{r_{g}^2}{2}\sigma_{g}^{z}\right]\left[\left(2 - \frac{r_{2}^2}{2}\right) + \frac{r_{2}^2}{2}\sigma_{2}^{z}\right]e^{-(r_{1}^2 + r_{g}^2 + r_{2}^2)/4}\\
&\hspace{11.5pt} - \frac{i}{2}E_{s}[\sqrt{2}ir_{1}\sigma_{1}^{x}][\sqrt{2}ir_{g}\sigma_{g}^{x}][\sqrt{2}ir_{2}\sigma_{2}^{x}]e^{-(r_{1}^2 + r_{g}^2 + r_{2}^2)/4}\\
&= -\frac{1}{2}\left[2(E_{1} + 4E_{L_{1}} + 4E_{L} + E_{c})r_{1}^2 + 2E_{1}r_{1}^2e^{-r_{1}^2} + \frac{E_{c}r_{1}^2}{2}\left(2 - \frac{r_{g}^2}{2}\right)\left(2 - \frac{r_{2}^2}{2}\right)e^{-\frac{r_{1}^2 + r_{g}^2 + r_{2}^2}{4}}\right]\sigma_{1}^{z}\\
&\hspace{7pt} - \frac{1}{2}\left[2(4E_{L} + E_{c})r_{g}^2 + \frac{E_{c}r_{g}^2}{2}\left(2 - \frac{r_{1}^2}{2}\right)\left(2 - \frac{r_{2}^2}{2}\right)e^{-\frac{r_{1}^2 + r_{g}^2 + r_{2}^2}{4}}\right]\sigma_{g}^{z}\\
&\hspace{7pt} - \frac{1}{2}\left[2(E_{2} + 4E_{L_{2}} + 4E_{L} + E_{c})r_{2}^2 + 2E_{2}r_{2}^2e^{-r_{2}^2} + \frac{E_{c}r_{2}^2}{2}\left(2 - \frac{r_{1}^2}{2}\right)\left(2 - \frac{r_{g}^2}{2}\right)e^{-\frac{r_{1}^2 + r_{g}^2 + r_{2}^2}{4}}\right]\sigma_{2}^{z}\\
&\hspace{7pt} - \frac{E_{c}r_{1}^2r_{g}^2}{8}\left(2 - \frac{r_{2}^2}{2}\right)e^{-\frac{r_{1}^2 + r_{g}^2 + r_{2}^2}{4}}\sigma_{1}^{z}\sigma_{g}^{z}\\
&\hspace{7pt} - \frac{E_{c}r_{1}^2r_{2}^2}{8}\left(2 - \frac{r_{g}^2}{2}\right)e^{-\frac{r_{1}^2 + r_{g}^2 + r_{2}^2}{4}}\sigma_{1}^{z}\sigma_{2}^{z}\\
&\hspace{7pt} - \frac{E_{c}r_{g}^2r_{2}^2}{8}\left(2 - \frac{r_{1}^2}{2}\right)e^{-\frac{r_{1}^2 + r_{g}^2 + r_{2}^2}{4}}\sigma_{g}^{z}\sigma_{2}^{z}\\
&\hspace{7pt} - \frac{E_{c}r_{1}^2r_{g}^2r_{2}^2}{16}e^{-\frac{r_{1}^2 + r_{g}^2 + r_{2}^2}{4}}\sigma_{1}^{z}\sigma_{g}^{z}\sigma_{2}^{z}\\
&\hspace{7pt} - \sqrt{2}E_{s}r_{1}r_{g}r_{2}e^{-\frac{r_{1}^2 + r_{g}^2 + r_{2}^2}{4}}\sigma_{1}^{x}\sigma_{g}^{x}\sigma_{2}^{x}
\end{align*}
Defining the spin-model parameters
\begin{align*}
\begin{split}
\Omega_{1} &= 2(E_{1} + 4E_{L_{1}} + 4E_{L} + E_{c})r_{1}^2 + 2E_{1}r_{1}^2e^{-r_{1}^2}\\
&\hspace{7pt} + \frac{E_{c}r_{1}^2}{2}\left(2 - \frac{r_{g}^2}{2}\right)\left(2 - \frac{r_{2}^2}{2}\right)e^{-\frac{r_{1}^2 + r_{g}^2 + r_{2}^2}{4}}
\end{split}\\
\Omega_{g} &= 2(4E_{L} + E_{c})r_{g}^2 + \frac{E_{c}r_{g}^2}{2}\left(2 - \frac{r_{1}^2}{2}\right)\left(2 - \frac{r_{2}^2}{2}\right)e^{-\frac{r_{1}^2 + r_{g}^2 + r_{2}^2}{4}}\\
\begin{split}
\Omega_{2} &= 2(E_{2} + 4E_{L_{2}} + 4E_{L} + E_{c})r_{2}^2 + 2E_{2}r_{2}^2e^{-r_{2}^2}\\
&\hspace{7pt} + \frac{E_{c}r_{2}^2}{2}\left(2 - \frac{r_{1}^2}{2}\right)\left(2 - \frac{r_{g}^2}{2}\right)e^{-\frac{r_{1}^2 + r_{g}^2 + r_{2}^2}{4}}
\end{split}\\
J_{1g}^{z} &= -\frac{E_{c}r_{1}^2r_{g}^2}{8}\left(2 - \frac{r_{2}^2}{2}\right)e^{-\frac{r_{1}^2 + r_{g}^2 + r_{2}^2}{4}}\\
J_{12}^{z} &= -\frac{E_{c}r_{1}^2r_{2}^2}{8}\left(2 - \frac{r_{g}^2}{2}\right)e^{-\frac{r_{1}^2 + r_{g}^2 + r_{2}^2}{4}}\\
J_{g2}^{z} &= -\frac{E_{c}r_{g}^2r_{2}^2}{8}\left(2 - \frac{r_{1}^2}{2}\right)e^{-\frac{r_{1}^2 + r_{g}^2 + r_{2}^2}{4}}\\
J_{1g2}^{z} &= -\frac{E_{c}r_{1}^2r_{g}^2r_{2}^2}{16}e^{-\frac{r_{1}^2 + r_{g}^2 + r_{2}^2}{4}}\\
J_{1g2}^{x} &= -\sqrt{2}E_{s}r_{1}r_{g}r_{2}e^{-\frac{r_{1}^2 + r_{g}^2 + r_{2}^2}{4}}
\end{align*}
we get the final Hamiltonian
\begin{align*}
H &= -\frac{1}{2}\Omega_{1}\sigma_{1}^{z} - \frac{1}{2}\Omega_{g}\sigma_{g}^{z} - \frac{1}{2}\Omega_{2}\sigma_{2}^{z}\\
&\hspace{7pt} + J_{1g}^{z}\sigma_{1}^{z}\sigma_{g}^{z} + J_{12}^{z}\sigma_{1}^{z}\sigma_{2}^{z} +  J_{g2}^{z}\sigma_{g}^{z}\sigma_{2}^{z}\\
&\hspace{7pt}  + J_{1g2}^{z}\sigma_{1}^{z}\sigma_{g}^{z}\sigma_{2}^{z} + J_{1g2}^{x}\sigma_{1}^{x}\sigma_{g}^{x}\sigma_{2}^{x}
\end{align*}
Furthermore, the anharmonicities can be calculated with the same exact method, by including a third level and defining the anharmonicities as $ \alpha = E_{12} - E_{01} $, where $ E_{ij} $ is the energy gap between level $ i $ and $ j $, and the energies are defined as the diagonal entries of the Hamiltonian. With this we get the following anharmonicities
\begin{align*}
\alpha_{1} &= -2E_{1}r_{1}^4e^{-r_{1}^2} - \frac{E_{c}r_{1}^4}{2}\left(1 - \frac{r_{g}^2}{4}\right)\left(1 - \frac{r_{2}^2}{4}\right)e^{-\frac{r_{1}^2+r_{g}^2+r_{2}^2}{4}},\\
\alpha_{g} &= -\frac{E_{c}r_{g}^4}{2}\left(1 - \frac{r_{1}^2}{4}\right)\left(1 - \frac{r_{2}^2}{4}\right)e^{-\frac{r_{1}^2+r_{g}^2+r_{2}^2}{4}},\\
\alpha_{2} &= -2E_{2}r_{2}^4e^{-r_{2}^2} - \frac{E_{c}r_{2}^4}{2}\left(1 - \frac{r_{1}^2}{4}\right)\left(1 - \frac{r_{g}^2}{4}\right)e^{-\frac{r_{1}^2+r_{g}^2+r_{2}^2}{4}}
\end{align*}
In the main text we did not include the linear inductors in the circuit. Removing them corresponds simply to setting $ E_{L},E_{L_{0}},E_{L_{1}} = 0 $ in the above expressions.

\subsection{Derive effective parameters} \label{app:effparam}
\begin{figure}
	\includegraphics[width=0.7\textwidth]{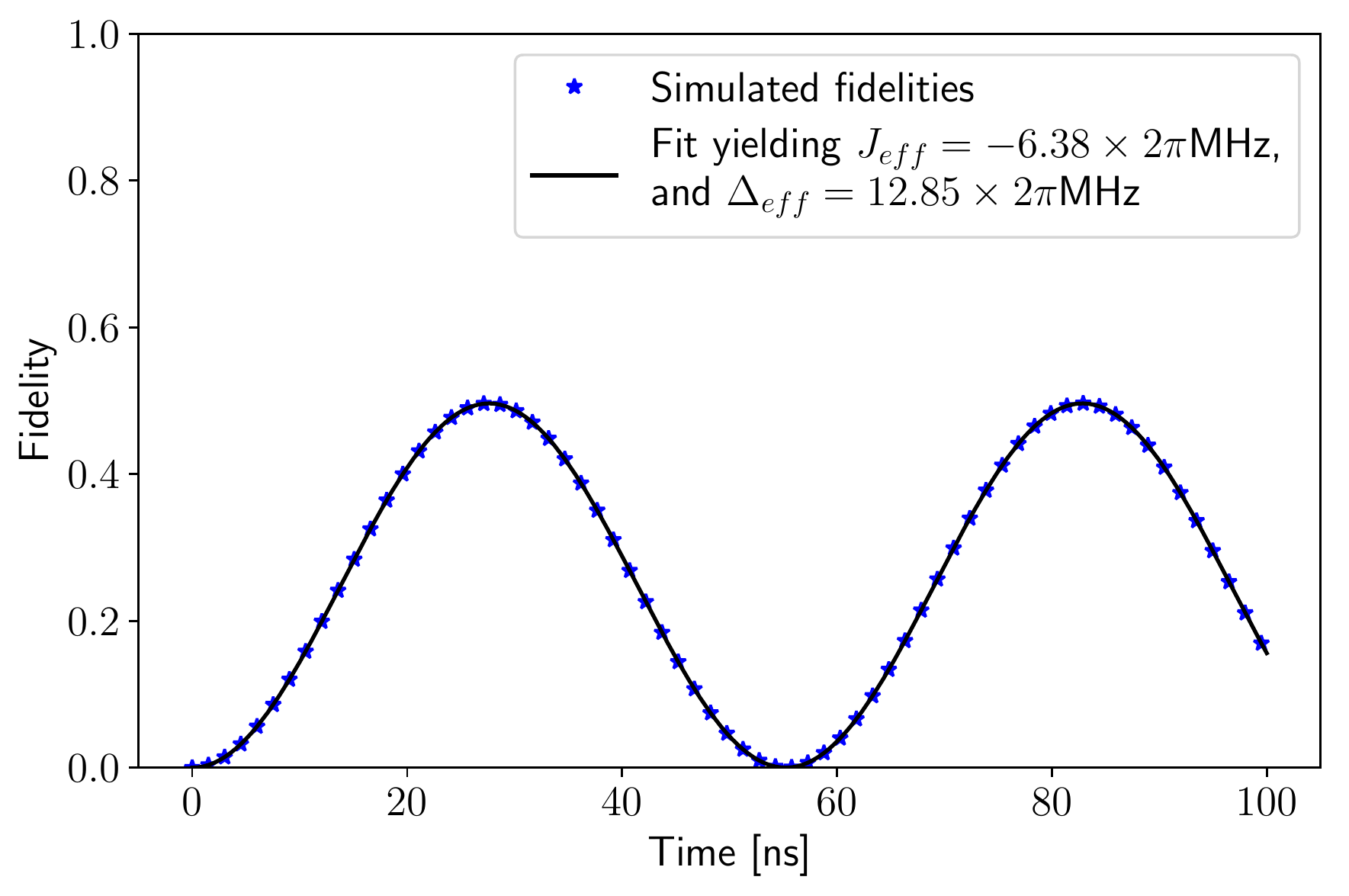}
	\caption{\label{fig:effParamsFit} }
\end{figure}
In the above section we found exact spin model parameters, taking two levels in each mode into account. For the detuning and XXX-coupling strength, however, we need to numerically find effective parameters, as the above bare spin model parameters are renormalized by virtual processes involving the higher levels of the system. We consider only the effective detuning and XXX-coupling strength this way, because they need to be accurate, while anharmonicities and other parameters do not have such exact demands imposed on them. The effective detuning we are looking for is the detuning between the states $ \ket{1_{0}1_{g}0_{1}} $ and $ \ket{0_{0}0_{g}1_{1}} $ with respect to the $ \sigma_{0}^{+}\sigma_{g}^{+}\sigma_{1}^{-} + \textup{H.c.} $ interaction. The effective coupling strength we are looking for is the strength of this interaction. To find these we simply simulate the dynamics of the circuit Hamiltonian truncated to four levels in each mode and initialized in $ \ket{1_{0}1_{g}0_{1}} $ (which is also a valid state, when each mode has four states). We then calculate the fidelity with the state $ \ket{0_{0}0_{g}1_{1}} $ at subsequent times. Ideally, these two levels behave as if they are only coupled to each other, and as such follow the evolution of a single two level system, with a certain detuning and coupling between the states, described by 
\begin{align}
H_{2} = -\frac{1}{2}\omega\sigma^{z} + J\sigma^{x} = \mat{-\frac{1}{2}\omega & J \\ J & \frac{1}{2}\omega}
\end{align}
Initializing in of the states, the evolution of the fidelity, $ F(t) $, of the other is then described by 
\begin{align}
F(t) = \frac{J^2}{\omega^2/4 + J^2}\sin^2\left(\sqrt{\frac{\omega^2}{4} + J^2}t\right). \label{eq:fidelity2level}
\end{align}
Hence, fitting the simulated data with a sine curve gives us information about $ \Delta_{\textup{eff}} $ between $ \ket{1_{0}1_{g}0_{1}} $ and $ \ket{0_{0}0_{g}1_{1}} $, and also the effective strength, $ J_{\textup{eff}} $, of the coupling between these states. \cref{fig:effParamsFit} shows an example of such a fit yielding the effective parameters we discussed in the main text. Clearly the fidelity does follow a sine, as if these two states are all alone in the world. This is effectively the case because all other interactions are suppressed by detuning, but as explained it is virtual processes resulting from these suppressed interactions that renormalizes the detuning and interaction strength such that we can not directly use the bare expressions found above.

\subsection{Optimizing circuit parameters}\label{app:opt}
Higher order contributions from interactions with states outside the spin-$ \frac{1}{2} $ subspace mean that we must consider effective spin model parameters when optimizing the circuit parameters. We are in particular interested in the detuning between $ \ket{1_{0}1_{g}0_{1}} $ and $ \ket{0_{0}0_{g}1_{1}} $, and the coupling strength of the XXX-coupling. Furthermore, we want the numerical value of the anharmonicities of the modes to be about $ \SI{100}{\times2\pi\mega\hertz} $ or larger in order to be able to address the spins for initialization and control \cite{Krantz2019}. The effective detuning, $ \Delta_{\textup{eff}} $, and the effective coupling strength, $ J_{\textup{eff}} $, are found as described in the previous section.

We want to show that our circuit can be used to realize the quench dynamics studied in the first part of the main text. In this case, the effective detuning should not be zero, but rather it defines the staggered mass $ m $. Considering the rotated Hamiltonian $ H_{R} $ in Eq. (5) of the main text, it can be seen that we will have $ m = \Delta_{\textup{eff}}/2 $. Likewise, if $ J $ is the desired strength of the matter-gauge coupling in Eq. (1) of the main text, then $ J = 2J_{\textup{eff}} $, because of the factor $ 1/2 $ in the interaction term in Eq. (1). We will thus be tuning $ J/m = 4J_{\textup{eff}}/\Delta_{\textup{eff}} $ to find good circuit parameters producing spin model parameters that would be interesting for the analog simulation of the dynamics we found in the main text. The anharmonicities are obtained with the expressions given above. These too will be renormalized to some extent, but since we do not need them to have some specific value, it is sufficient to calculate their bare value. 

Since only the Josephson energies of a SQC can be tuned \emph{in situ}, it is difficult to actually perform the appropriate quench of the circuit. Instead we intend for the circuit to be constructed with the post-quench parameters. The quench will then be implemented by initializing the system in the ground state of the pre-quench Hamiltonian. Whether we have the system in its pre-quench setup, go into its ground state, and then quench to the post-setup, or simply start with the system in the post-quench setup, and then quickly initialize in the ground state of the pre-quench Hamiltonian, we will see the same resulting dynamics. This moves the difficulty from performing a fast quench to performing a fast initialization. 

With this in mind we tune the circuit parameters to yield a negative $ J/m $ (as the quench is to a negative mass, $ m \rightarrow -m $), corresponding to post-quench parameters. After finding appropriate circuit parameters we do a check of the overall behaviour of the circuit, ensuring that it works as intended, including no disturbing interactions with higher levels. To do this we use average fidelity \cite{Nielsen2002}. Average fidelity is a measure of how well a certain process implements a desired operation. In our case the process is the time evolution of the circuit, determined by $ H_{c} $ in Eq. (3) of the main text in a frame rotating such that the resulting Hamiltonian is $ H_{R} $ in Eq. (5), where $ m = \Delta_{\textup{eff}}/2 $ and the coupling strength is $ J_{\textup{eff}} $ (rather than the bare $ J_{0g1}^{x} $). To take contributions from higher level interactions into account we again truncate to the lowest \emph{four} levels. We only rotate the spin-$ 1/2 $ states, i.e. we use a four-level version of $ H_{0} + \frac{1}{2}m(\sigma_{0} - \sigma_{1}) $, where all entries pertaining to levels higher than the spin-$ 1/2 $ states are just zero. The operation we compare this with is the time evolution according to the target Hamiltonian, i.e. $ H $ from Eq. (1) of the main text with two matter sites and the gauge link between them. We compare only the dynamics of the spin-$ 1/2 $ states, i.e. time-evolution of the circuit takes place with four levels included for each mode, and the result is then projected down to the spin-$ 1/2 $ subspace, before comparing with the time-evolution of $ H $. The average fidelity will thus be comparing the very time evolution operators themselves for the circuit and the target system. The mass and coupling strength of the target Hamiltonian are chosen to be $ \Delta_{\textup{eff}}/2 $ and $ 2J_{\textup{eff}} $. The fidelity would thus \emph{a priori} be expected to be quite high, but since this is all done with four levels included in each anharmonic mode, the fidelity will be a measure of how much the higher levels affect the dynamics of the circuit beyond just the renormalization of the mass and coupling strength. In particular, some population will be lost to the higher levels, and just as virtual processes contribute to the strength of the XXX-coupling, they will also to some extent induce other effective interactions. These will disturb the desired dynamics and might be gauge-variant, resulting in population moving outside of the $ G_{n} = 0 $ gauge sector of the spin-$ 1/2 $ subspace. 

The average fidelity can be calculated from \cite{Nielsen2002}
\begin{align}
\overline{F}(U,\mathcal{E}) = \frac{\sum_{j}\tr{UU_{j}^{\dagger}U^{\dagger}\mathcal{E}(U_{J})} + d^2}{d^2(d + 1)}
\end{align}
where $ U $ is the target operation, and $ \mathcal{E} $ is a quantum map representing the process used to implement the target operation. The $ U_{j} $ are a unitary basis for operators on the relevant space of states, which has dimension $ d $. In our case the target operation is time evolution $ U = e^{-iHt} $, with $ H = H(\Delta_{\textup{eff}}/2,2J_{\textup{eff}}) $ from Eq. (1). This is an operator on the spin-$ 1/2 $ subspace, and so the $ U_{j} $ are a unitary basis for operators on that subspace. The quantum map $ \mathcal{E} $ takes an operator $ \rho $ pertaining to the spin-$ 1/2 $ subspace, and returns $ \mathcal{E}(\rho) = Pe^{iH_{0,\textup{eff}}t}e^{-iH_{c}t}P^{\dagger}\rho P e^{iH_{c}t}e^{-iH_{0,\textup{eff}}t}P^{\dagger} $, where $ P $ is the projection operator from the four-level subspace to the spin-$ 1/2 $ subspace. Hence, this map first projects the operator $ \rho $ into the four-level space, performs time evolution according to the four-level operators, and then projects the result back into the spin-$ 1/2 $ subspace. Here $ H_{0,\textup{eff}} $ is an effective four-level equivalent of $ H_{0} $, taking energy shifts caused by virtual processes into account. $ H_{c} $ is the circuit Hamiltonian with four levels in each anharmonic oscillator mode. The construction $ e^{iH_{0,\textup{eff}}t}e^{-iH_{c}t} $ implements time evolution according to $ H_{c} $ in the frame rotating with respect to $ H_{0,\textup{eff}} $ by first time evolving forwards with $ H_{c} $ and then backwards with $ H_{0,\textup{eff}} $. See the next section for details on this. We determine $ H_{0,\textup{eff}} $ numerically by turning off the XXX-coupling and time evolving each of the eight spin-$ 1/2 $ states in a frame rotating with respect to $ H_{0} $ (expanded to the four-level Hilbert space). As the Hamiltonian only involves Z-type interactions, and other, very suppressed interactions caused by the virtual processes, the spin-$ 1/2 $ states will almost be eigenstates of the system, and their phase will thus be $ e^{-iE_{\textup{eff}}t} $, where $ E_{\textup{eff}} $ is their effective energy. With these $ H_{0,\textup{eff}} $ can be constructed.

\begin{figure}
	\includegraphics[width=0.9\textwidth]{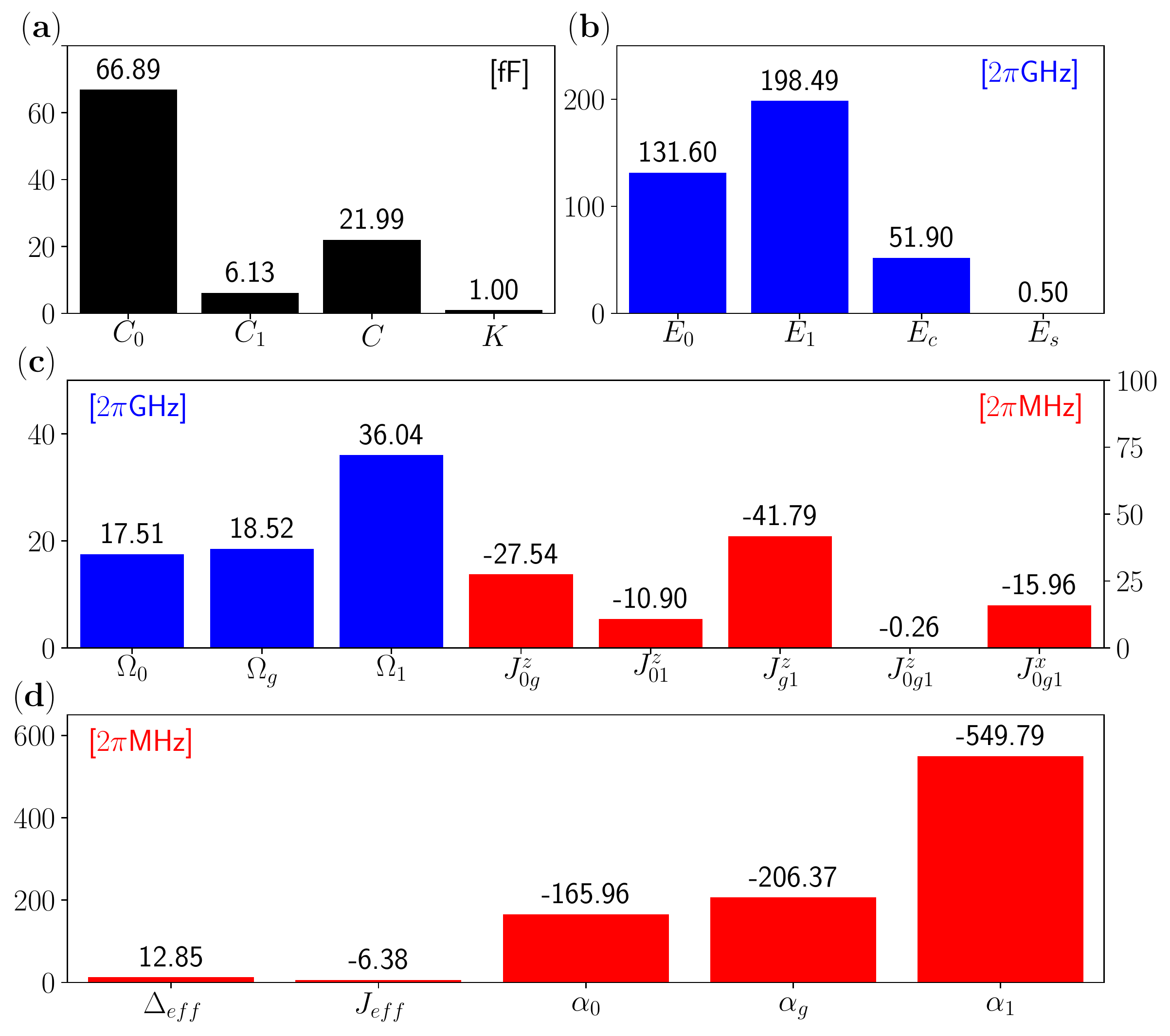}
	\caption{\label{fig:params} Circuit and spin model parameters. Notice that some parameters are negative, despite being plotted with positive bars. The bars are for graphical comparison of sizes, while the numbers above the bars give the exact value. (a),(b) Optimized circuit parameters implementing $ J/m = -2.0 $ and good anharmonicities. $ K $ has been fixed at $ \SI{1}{\femto\farad} $ as want it to be as low as possibly. $ E_{s} $ has been fixed at $ \SI{0.5}{\times2\pi\giga\hertz} $ to ensure that the odd interactions of the triple sine-term are weak enough to be suppressed by the anharmonicities. (c) The resulting bare spin model parameters. The Z-type coupling strengths are much smaller than the spin transition frequencies as required. (d) The effective detuning and XXX-coupling strength, as well as the anharmonicities. The anharmonicities are much larger than the effective XXX-coupling strength as required, and we can see that $ J/m = 4J_{\textup{eff}}/\Delta_{\textup{eff}} = -2.0 $.}
\end{figure}

In \cref{fig:params}a,b we show a set of circuit parameters, which yield $ J/m = -2.0 $ and good anharmonicities. In the parameter optimization we have fixed the capacitance to ground at $ K = \SI{1}{\femto\farad} $. As mentioned, $ K $ would ideally be zero, which is why we fixed it at this low value, but an experimental realization where it is completely negligible would be even better. $ E_{s} $, which controls the strength of $ J_{0g1}^{x} $, is likewise fixed at the small value of $ E_{s} = \SI{0.5}{\times2\pi\giga\hertz} $ to ensure that odd interactions with higher levels are suppressed by the anharmonicities. The other Josephson energies have large values, but there are also good sets of circuit parameters where they are smaller, but at the cost of $ E_{s} $ also being smaller, or interactions with higher levels being less suppressed. In \cref{fig:params}c the resulting bare spin model parameters can be seen. Notice that some of the parameters are negative, despite being plotted with bars pointing in the positive direction. The bars should be used as graphical comparison of absolute sizes, while the numbers indicate the exact values. Clearly the Z-type coupling strengths are much smaller than the spin transition frequencies, as we need. In \cref{fig:params}d the effective detuning and XXX-coupling strength, and the anharmonicities can be seen. We see that $ J_{\textup{eff}} $ is noticeably smaller than the bare $ J_{0g1}^{x} $. Furthermore, it is much smaller than the anharmonicities, as required to make the dynamics remain in the spin-$ 1/2 $ subspace. This effective detuning and XXX-coupling strength result in $ J/m = 4J_{\textup{eff}}/\Delta_{\textup{eff}} = -2.0 $, which according to Figs. 5 and 6 would result in interesting dynamics of the order parameter and Loschmidt amplitude within a time of $ tm = 2 $, corresponding to $ t = \SI{49.5}{\nano\second} $. 

\begin{figure}
	\includegraphics[width=0.9\textwidth]{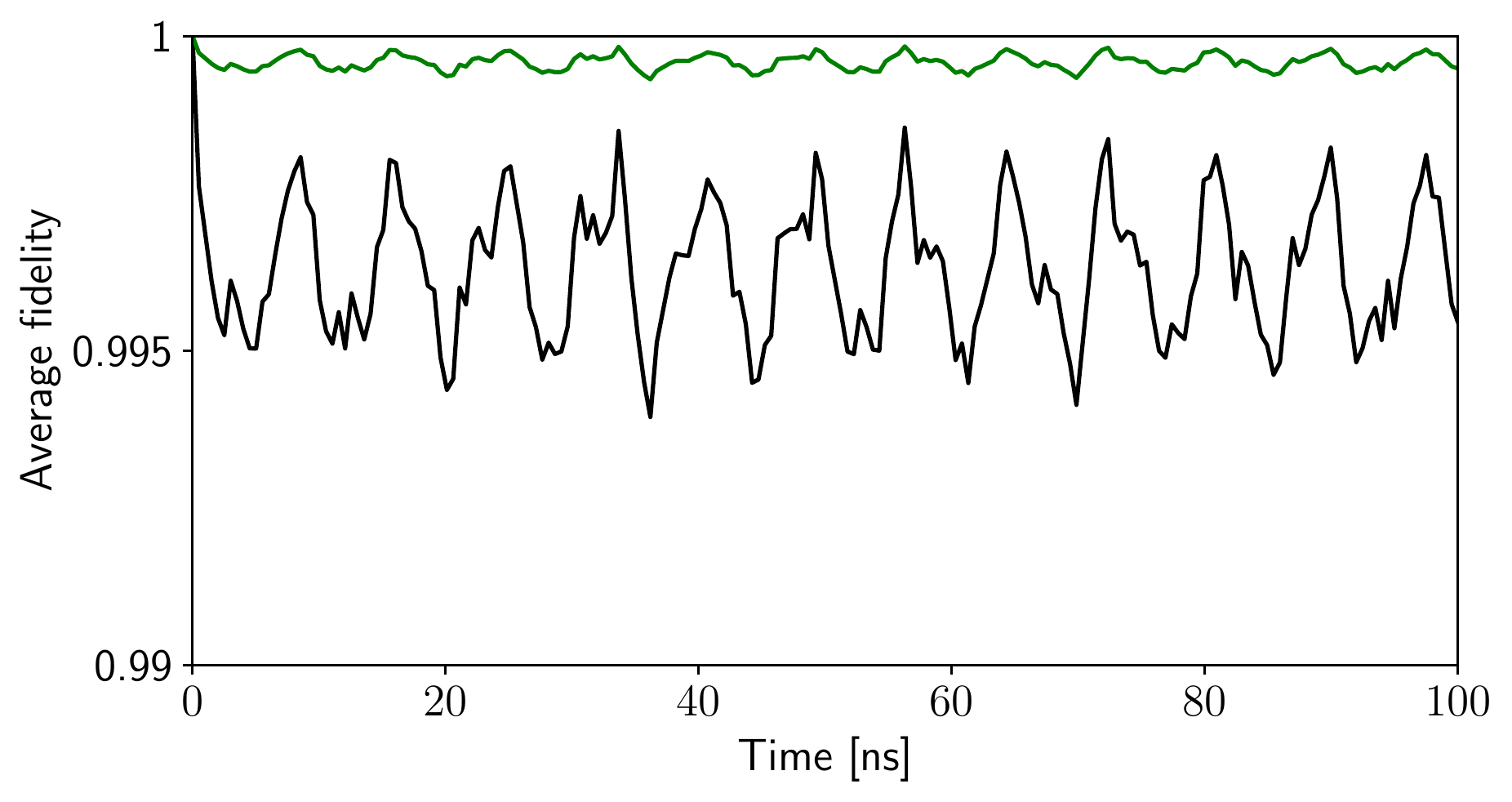}
	\caption{\label{fig:fids} In black: The average fidelity of the circuit's implementation of the target dynamics in a rotating frame. The fidelities are close to or above $ 99.5\% $ and keep steady over long times, with an oscillation on a short time scale. In green: The fidelity without taking leakage to the higher levels into account. From this we see that the largest part, $ 0.2-0.45\% $, of the lost fidelity is due to population immediately leaking into the higher levels and then partly oscillating back and forth. A smaller contribution, about $ 0.05\% $, is from the effective interactions induced by virtual processes involving the higher levels.}
\end{figure}

All calculations of dynamics are performed without including noise. However, with present superconducting qubit life times \cite{Wang2019,Touzard2019} we do not believe noise would significantly disturb the results presented here. Furthermore, while we have explained how the XXX-coupling in a RWA yields the desired U(1) interaction term, we do not actually use this approximation, but instead retain all terms to show that they do indeed not disturb the desired dynamics significantly. In \cref{fig:fids} the calculated average fidelity of the circuit's implementation of the target dynamics using the circuit parameters presented above can be seen in black. The fidelity is about or above $ 99.5\% $ at all times, and while it oscillates on a short timescale, it seems to keep steady over the plotted interval of $ \SI{100}{\nano\second} $. Hence, the implementation of the desired dynamics is good, and stable in the sense that we are not accumulating error or continuously losing population to the higher levels. We seem to lose a small fraction of the population immediately, which then partly oscillates back and forth. In green is plotted the average fidelity plus the leakage to higher levels, i.e. this plot shows the fidelity if we do \emph{not} take leakage into account. Hence, we can see that about $ 0.2-0.45\% $ fidelity is lost because of population leaking to the higher levels of the circuit, while about $ 0.05\% $ is lost due to effective interactions induced by virtual processes involving the higher levels.

In our work with tuning the circuit we have found that it is well capable of implementing the interval of $ J/m $ considered in the first part of this paper. The tuning effectively takes place via two options. Either $ J_{\textup{eff}} = J/2 $ is tuned separately by changing $ E_{s} $, or $ \Delta_{\textup{eff}} = 2m $ is tuned by changing all parameters but $ E_{s} $ (which does also change $ J_{\textup{eff}} $, but only slightly). While tuning $ J_{\textup{eff}} $ is easy because it is proportional to $ E_{s} $, $ \Delta_{\textup{eff}} $ is quite sensitive to the circuit parameters. This is essentially because in the parameter regime we are interested in, $ \Delta_{\textup{eff}} $ is about the same size as $ J_{\textup{eff}} $, which itself has to be much smaller than the anharmonicities, say a factor of $ 10 $, and finally the anharmonicities are about $ 1-2\% $ of the spin transition frequencies. Hence, the effective detuning $ \Delta_{\textup{eff}} $ of the spins must be about $ 0.1-0.2\% $ of the spin frequencies. They must therefore be tuned carefully. If a circuit is made that implements an interesting value of $ J/m $, other nearby values could be achieved by varying just the Josephson energies, making it possible to use the same circuit to study different values of $ J/m $.

An important thing to note is that the Loschmidt amplitude in the circuit's own frame will not be the same as in the rotating frame. Expectation values like $ \mel{\psi(t)}{A}{\psi(t)} $ or, in the Heisenberg picture, $ \mel{\psi}{A_{0}(t_{0})A_{1}(t_{1})...A_{n}(t_{n})}{\psi} $ (e.g. two-time correlators) are invariant under unitary transformations like $ e^{-iHt} \rightarrow Ue^{-iHt} $ for some appropriate Hamiltonian $ H $, where the operators are simultaneously transformed as $ A \rightarrow UAU^{\dagger} $. However, expectation values like $ \mel{\psi(0)}{A}{\psi(t)} $ are not. This is essentially because while the time-evolution operator gets a single factor of $ U $ in the transformed frame, any other operator gets two. Thus if there is an odd number of time-evolution operators, the resulting odd number of factors of $ U $ can not cancel (they would always cancel pairwise). Hence, the Loschmidt amplitude $ \mathcal{G}(t) = \braket{\psi(0)}{\psi(t)} = \mel{\psi(0)}{e^{-iH_{c}t}}{\psi(0)} $ is not invariant under rotations like $ U = e^{iH_{0,\textup{eff}}t} $. Likewise, the order parameter $ g(k,t) = \mel{\psi(0)}{\mathfrak{g}(k)}{\psi(t)} $ is not invariant. However, the Loschmidt amplitude is often measured by performing state tomography \cite{Qin2017,Baur2012,Filipp2009,Dicarlo2009,Steffen2006} and then calculating $ \mathcal{G}(t) $ from the results \cite{Heyl2018,Xu2019,Guo2019,Xu2020,Jurcevic2017,Flaschner2018}. With full information about the state of the circuit at any time, the Loschmidt amplitude of the rotated system can easily be calculated as this amounts to calculating $ \mel{\psi(0)}{e^{iH_{0,\textup{eff}}t}}{\psi(t)} $. The necessary Hamiltonian $ H_{0,\textup{eff}} $ describing the effective energies could be determined in a separate experiment prior to studying the quench dynamics in a fashion similar to what we have done numerically. The XXX-coupling is turned off by setting $ E_{s} = 0 $ via flux tuning, and by initializing in each of the spin-$ 1/2 $ states their phase over time, and thus their effective energy, could be measured. An alternative to performing full state tomography is to have multiple copies of the circuit, perform the quench experiment in just one of them while initializing the other in the appropriate initial state, and then connect the circuits using some appropriate scheme to make their states interfere, potentially yielding information about the quantities we are interested in. Such an approach is proposed in the context of atoms in an optical lattice in Refs. \cite{Daley2012,Pichler2013} to measure the Rényi entropy, and has in fact been experimentally probed since \cite{Islam2015,Linke2018,Brydges2019}.

\subsection{Time evolution in rotating frame} \label{app:rotFrame}
To find an expression for the time evolution operator in a rotating frame, we essentially just have to take advantage of the fact that we can transform between the non-rotating and the rotating frame at any time using the same time-dependent operator. Imagine some state in the non-rotating frame $ \ket{\psi(t)} = U(t)\ket{\psi(0)} $, where $ U(t) $ is the time evolution operator in the non-rotating frame. We then define a unitary, time-dependent operator $ R(t) $, which we use to define the state in the rotating frame $ \ket{\psi(t)}_{R} = R(t)\ket{\psi(t)} $, and which satisfies that the two frames are identical at $ t = 0 $, i.e. $ R(0) = \mathcal{1} $. Let $ U_{R}(t) $ be the time evolution operator of the rotating frame that we seek. We then quite simply have
\begin{align*}
U_{R}(t)\ket{\psi(0)} = U_{R}(t)\ket{\psi(0)}_{R} = \ket{\psi(t)}_{R} = R(t)\ket{\psi(t)} = R(t)U(t)\ket{\psi(0)}
\end{align*}
From this we conclude that $ U_{R}(t) = R(t)U(t) $. Hence, if for example we have a Hamiltonian $ H = H_{0} + H_{\textup{int}} $ in the non-rotating frame, i.e. $ U(t) = e^{-iHt} $, and we want to look at the frame rotating with respect to $ H_{0} $, i.e. $ R(t) = e^{iH_{0}t} $, then the time evolution operator in the rotating frame is $ U_{R}(t) = e^{iH_{0}t}e^{-iHt} $. This is the result we used in the main text.